\documentclass[useAMS,usenatbib]{mn2e} 
\usepackage{amsmath,graphics,color,subfigure,epsfig,amssymb}
\usepackage{mathptmx}
\usepackage{epsfig}
\usepackage{booktabs}
\usepackage{rotating,rotfloat}
\usepackage{longtable}
\usepackage{graphicx}
\usepackage{url} 
\usepackage{listings}
\usepackage{color}
\usepackage{pslatex} 
\usepackage{booktabs}
\usepackage{dcolumn}
\usepackage[hyperindex,breaklinks=true, colorlinks, citecolor=blue]{hyperref}

\let\la=\lesssim    
\def\reff@jnl#1{{\rm#1\/}}
\def\aj{\reff@jnl{AJ}}                  
\def\araa{\reff@jnl{ARA\&A}}            
\def\actaa{\reff@jnl{Acta. Astron}}     
\def\apj{\reff@jnl{ApJ}}                
\def\apjl{\reff@jnl{ApJ}}               
\def\apjs{\reff@jnl{ApJS}}              
\def\ao{\reff@jnl{Appl.Optics}}         
\def\apss{\reff@jnl{Ap\&SS}}            
\def\aap{\reff@jnl{A\&A}}               
\def\aapr{\reff@jnl{A\&A~Rev.}}         
\def\aaps{\reff@jnl{A\&AS}}             
\def\azh{\reff@jnl{AZh}}                
\def\baas{\reff@jnl{BAAS}}              
\def\jrasc{\reff@jnl{JRASC}}            
\def\memras{\reff@jnl{MmRAS}}           
\def\mnras{\reff@jnl{MNRAS}}            
\def\pra{\reff@jnl{Phys.Rev.A}}         
\def\prb{\reff@jnl{Phys.Rev.B}}         
\def\prc{\reff@jnl{Phys.Rev.C}}         
\def\prd{\reff@jnl{Phys.Rev.D}}         
\def\prl{\reff@jnl{Phys.Rev.Lett}}      
\def\pasp{\reff@jnl{PASP}}              
\def\pasj{\reff@jnl{PASJ}}              
\def\qjras{\reff@jnl{QJRAS}}            
\def\skytel{\reff@jnl{S\&T}}            
\def\solphys{\reff@jnl{Solar~Phys.}}    
\def\sovast{\reff@jnl{Soviet~Ast.}}     
\def\ssr{\reff@jnl{Space~Sci.Rev.}}     
\def\zap{\reff@jnl{ZAp}}                
\def\nat{\reff@jnl{Nature}}             
\def\pasa{\reff@jnl{Publ. Astron. Soc. Aust.}}            
\def\jcap{\reff@jnl{Journal of Cosmology and Astroparticle Physics}}     
\def\jqsrt{\reff@jnl{Journal of Quantitative Spectroscopy \& Radiative Transfer}}
\def\nar{\reff@jnl{New Astronomy Reviews}}

\newcolumntype{K}{D{.}{.}{2,2}}
\newcolumntype{H}{D{,}{\pm}{4.4}}

\newcommand{\wmap}{{WMAP}}
\newcommand{\planck}{{\em Planck}}

\newcommand{\spitzer}{{\em Spitzer}}
\newcommand{\cobe}{{COBE}}

\newcommand{\spdust}{{\sc SPDUST}}

\newcommand{\dg}{$^{\circ}$~}
\newcommand{\beq}{\begin{equation}}
\newcommand{\eeq}{\end{equation}}
\newcommand{\mic}{$\,\mu\mathrm{m}$}

\newcommand{\halpha}{H$\alpha$} 

\newcommand{\parcm}{'$\!\!.$}

\newcommand{\ldn}{LDN\,1780}
\newcommand{\hii}{H{\sc ii}}

\title[Spinning dust modelling in LDN\,1780]{Modelling the spinning
  dust emission from LDN\,1780}

\author[Vidal et al.]{Matias~Vidal$^{1}$\thanks{E-mail:
    mvidal@das.uchile.cl}, Clive~Dickinson$^2$, S.\,E. Harper$^2$,
  Simon~Casassus$^1$, A.\,N. Witt$^3$ \\ $^1$Departamento de
  Astronom\'ia, Universidad de Chile, Casilla 36-D Santiago, Chile
  \\ $^2$Jodrell Bank Centre for Astrophysics, Alan Turing Building,
  School of Physics and Astronomy, The University of
  Manchester,\\ Oxford Road, Manchester M13 9PL, UK.  \\ $^3$Ritter
  Astrophysical Research Center, University of Toledo, Toledo, OH
  43606, USA } \voffset=-0.6in
\begin{document}
   

\pagerange{\pageref{firstpage}--\pageref{lastpage}} \pubyear{2019}

\maketitle

\label{firstpage}

\begin{abstract}

We study the anomalous microwave emission (AME) in the Lynds Dark
Nebula (LDN)\,1780 on two angular scales. Using available ancillary
data at an angular resolution of 1\,degree, we construct an SED
between 0.408\,GHz to 2997\,GHz. We show that there is a significant
amount of AME at these angular scales and the excess is compatible
with a physical spinning dust model. We find that LDN\,1780 is one of
the clearest examples of AME on 1\,degree scales. We detected AME with
a significance $> 20\sigma$. We also find at these angular scales that
the location of the peak of the emission at frequencies between
23--70\,GHz differs from the one on the 90--3000\,GHz map. In order to
investigate the origin of the AME in this cloud, we use data obtained
with the Combined Array for Research in Millimeter-wave Astronomy
(CARMA) that provides 2\,arcmin resolution at 30\,GHz. We study the
connection between the radio and IR emissions using morphological
correlations. The best correlation is found to be with MIPS
70\,$\mu$m, which traces warm dust ($T \sim 50$\,K). Finally, we study
the difference in radio emissivity between two locations within the
cloud. We measured a factor $\approx 6$ of difference in 30\,GHz
emissivity. We show that this variation can be explained, using the
spinning dust model, by a variation on the dust grain size
distribution across the cloud, particularly changing the carbon
fraction and hence the amount of PAHs.

\end{abstract}

\section{Introduction}

The \wmap\ \citep{Bennett2013} and \planck\ \citep{Planck2011_I}
satellites, as a byproduct of the making of Cosmic Microwave
Background (CMB) maps, have provided precise full-sky maps of the
different diffuse emission mechanisms on the Galaxy. Among them is the
anomalous microwave emission (AME), first detected by
\citet{Leitch1997} as a correlation between dust emission at
100\mic\ from IRAS and 14.5\,GHz radio emission toward the north
celestial pole, that could not be accounted for by synchrotron or
free-free emission.

In our Galaxy, AME can account for up to 30\% of the diffuse emission
at 30\,GHz \citep{Planck2016X,Planck2016XXV}.  AME has been observed
in different astrophysical environments, such as molecular clouds
\citep[][]{finkbeiner:02,watson:05,Casassus2006,casassus:08,scaife:09,dickinson:10},
translucent clouds \citep{Vidal2011}, reflection nebulae
\citep{Castellanos2011}, \hii~regions
\citep{dickinson:07,dickinson:09,todorovic:10} and in the galaxies
NGC\,6946 and NGC\,4725 \citep{murphy:10,Murphy2018}. AME may also be
important in compact objects like protoplanetary disks (PPDs).
\citet{Hoang2018} predicted that AME from spinning silicates or
polycyclic aromatic hydrocarbons (PAHs) dominates over thermal dust
emission at frequencies $<60$\,GHz in PPDs, even in the presence of
significant dust growth and \citet{Greaves2018} reproduced the EME
they detect in two disks using a model in which hydrogenated
nanodiamonds were the spinning carriers. For an up to date review on
AME, refer to \citet{Dickinson2018}.

AME is the least-understood emission mechanism in the 1\,--\,100\,GHz
range. It is difficult to study AME due to its diffuse nature, being
clearly detected by CMB experiments and telescopes at $\sim 1$\,degree
angular resolution but difficult to detect at higher angular
resolutions \citep[although there are well known regions were it has
  been observed with high resolution,
  e.g.][]{Scaife2010,Tibbs2011,Battistelli:2015}. This presents a
problem for the identification of the emitters and their physical
properties so, at the moment, we only know some general properties of
AME, like being associated with photodissociation regions (PDRs).  AME
is thought to be caused by dust grains, possessing electric dipole
moments, spinning at GHz frequencies. This is an old idea that was
first proposed by \citet{erickson:1957}.

The spinning dust (SD) hypothesis has been preferred by the
observations and the more convincing examples are the Perseus and
$\rho$ Ophiuchi molecular clouds
\citep{watson:05,casassus:08,planck_sd:11}.  Currently, detailed
theoretical models have been constructed that predict the SD spectrum
for different grain types and astrophysical environments
\citep{DL98b,ali:09,hoang:10,silsbee:11,ysard:11,hoang:12}. They
present an opportunity to study the ISM, in particular the smallest
dust grains, from a new window at GHz frequencies. Spinning dust
emission depends on factors such as the gas density, temperature,
ionization fraction and grain size distribution, so the detailed
comparison of the models with good observations will allow us to study
the ISM conditions in a variety of environments.

Nevertheless, some doubt has been cast on the SD paradigm by
\citep{HensleyDraine2016}, who found that the \planck\ AME map is
uncorrelated with a template of PAH emission. PAH are thought to be
one of he main carriers of AME in the SD model.  This shows that much
research is still needed in this area.

Here we present 30\,GHz data from the Combined Array for Research in
Millimeter-wave Astronomy (CARMA) of the Lynds Dark Nebula
(LDN)\,1780, a high Galactic latitude ({\em l} = 359$^{\circ}$\!\!.0,
{\em b} = 36$^{\circ}$\!\!.7) translucent region at a distance of
110$\pm$10\,pc \citep{franco:89}. LDN\,1780 has a moderate column
density (a few $\times 10^{21}$\,cm$^{-2}$) that corresponds to the
``translucent cloud'' type of object, i.e. interstellar clouds with
some protection from the radiation field, with optical extinctions in
the range $A_V \sim 1-4$\,mag \citep{snow:06}. Using an optical-depth
map constructed from {\em ISO}\,200\,$\mu$m observations,
\citet{Ridderstad2006} found a mass of $\sim$ 18~M$_{\odot}$ and
reported no young stellar objects based on the absence of colour
excess in point sources.

LDN\,1780 is a known source of AME. \citet{Vidal2011} detected AME from
this cloud through observations at 31\,GHz. They found that the AME at
30\,GHz correlates best with the IRAS\,60\,$\mu$m map, which traces
hot and small dust grains. This correlation was even tighter than with
an 8\,$\mu$m, which traces PAH. Here we revisit this cloud, using
archival data to study the spectral energy distribution (SED) at
1\,degree angular scales. We also use our CARMA data in addition to IR
and sub-mm templates to study and model the AME on angular scales of
2\,arcmin.

In Section \ref{sec:data} we describe the CARMA observations, as well
as the ancillary data used in the analysis. Section \ref{sec:onedeg}
describes the SED of the cloud on 1\,deg angular scales. Section
\ref{sec:dust_properties} correspond to the analysis at 2\,arcmin
resolution based on the CARMA data. Section \ref{sec:conclusions}
concludes.

\section{Data}
\label{sec:data}
\subsection{CARMA data}
\label{sec:carma_data}

We obtained 31\,GHz data from the Combined Array for Research in
Millimeter-wave Astronomy (CARMA). It consists of 8 antennas of 3.5\,m
diameter. Six ``inner'' telescopes are arranged in a compact
configuration, with baselines ranging from 4.5 to 11.5\,m. The two
other telescopes provide baselines of 56 and 78\,m. The receivers
observe the frequency range 26--36\,GHz in total intensity. The
primary beam corresponds to $\approx 11'$ at 31\,GHz.

\begin{figure}
  \centering
  \includegraphics[angle=-90,width=0.49\textwidth]{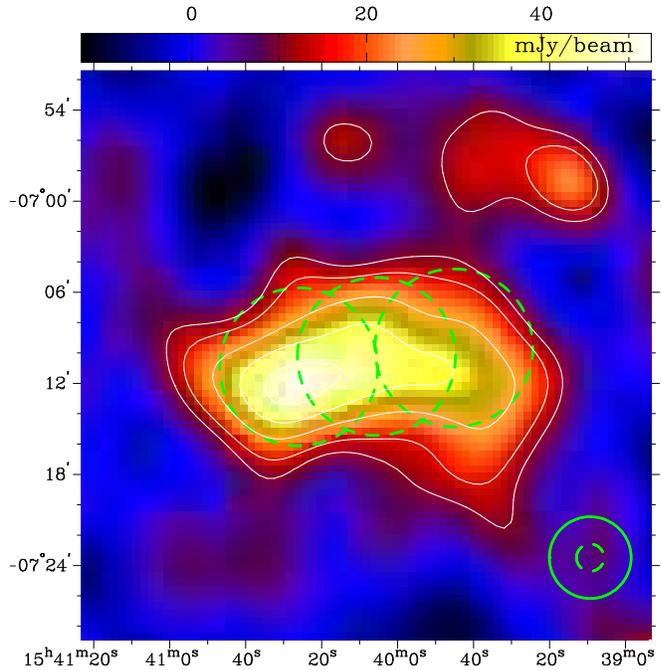}
   \caption{Mosaic pointings of the CARMA observations. The colour
     image shows the emission at 31\,GHz as seen by the CBI
     \citep{Vidal2011}, with a peak of 53\,mJy/beam and a synthesised
     beam of 5.4\,arcmin. The dashed circles along the cloud show the
     location of the three pointings where we observed the cloud with
     CARMA, with the size of them indicating the primary beam of the
     CARMA antennas of 10.5\,arcmin. At the bottom-right corner are
     shown the synthesised beam sizes for the CBI (5.4\,arcmin) and
     CARMA (inner, 1.8\,arcmin). }
  \label{fig:carma_mosaic}
\end{figure}

We prepared a three pointings mosaic observation centred at the peak
of the cloud at 31\,GHz as observed by the Cosmic Background Imager
(CBI) in \citet{Vidal2011}, and also, to include the ``gradient'' of
IR emission, i.e. regions where the morphology of the different IR
maps differs.  In Fig. \ref{fig:carma_mosaic} we show the three
pointing mosaic, overlaid on top of the CBI image of the cloud, from
\citet{Vidal2011}.

\subsection{Observations and calibration}

The observations were performed in two runs, first between 2012-06-09
and 2012-07-21 and, and between 19-05-2013 and 14-06-2013.  Each run
is divided into small observations blocks (OB). The total observing
time adds up to 25.2\,hours (or 4 sidereal passes) of telescope time.
During each one of the OB, the source is observed along with three
calibrators, namely flux calibrator (3C273), passband calibrator
(1337-129) and phase calibrator (1512-090). The OB consisted on
observations of the flux calibrator during 5\,min, then observation of
the passband calibrator during 5\,min, followed by the target cycle
where the phase calibrator is observed during 3\,m, followed by 15\,on
source.

We calibrated the data using the \texttt{Miriad} data-reduction
package \citep{sault:95}. We performed a small amount of flagging to
remove particularly noise combinations of baselines and spectral
channels.

\subsection{Imaging}

\begin{figure}
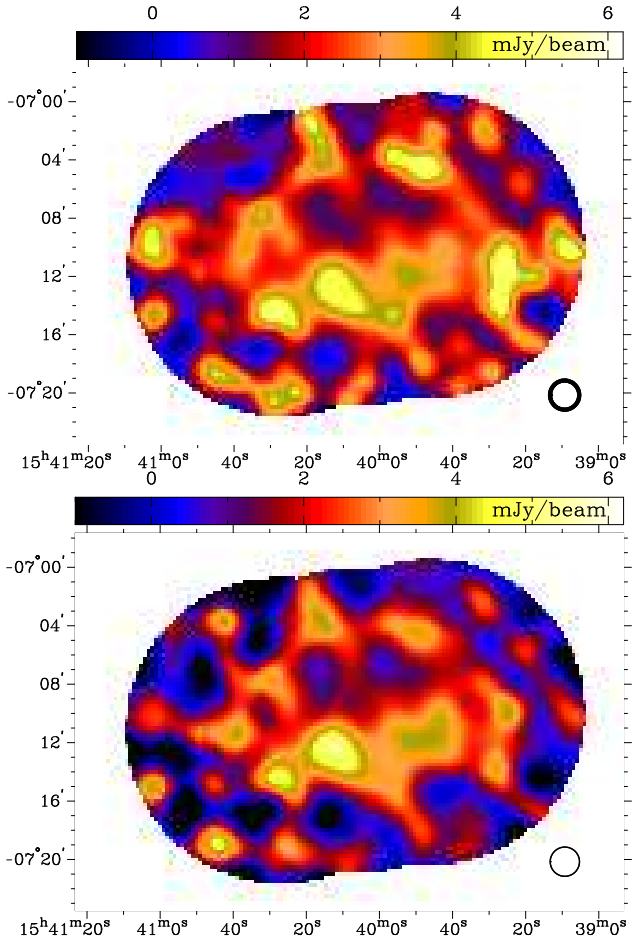

  \centering
  \includegraphics[angle=-90,width=0.47\textwidth]{figs/L1780.all.point_sub.restor.clean.ps}
  \includegraphics[angle=-90,width=0.47\textwidth]{figs/L1780.all.point_sub.restor.mem.ps}
  \caption{Clean ({\it top}) and MEM ({\it bottom}) mosaic
    reconstructions of the data after the $uv$ tapering, in order to
    increase the SNR of the extended emission. The angular resolution
    in this case is 2$'$ for both maps.  The r.m.s. noise value of the
    CLEAN map is 1.4\,mJy\,beam$^{-1}$\!, and the noise of the MEM map
    is 0.99\,mJy\,beam$^{-1}$\!. }
   \label{fig:ldn_final}
\end{figure}

To image the calibrated visibilities, we tried both CLEAN
\citep[e.g.][]{hogbom:74} and MEM \citep[e.g.][]{cornwell:85}
reconstructions. This was done to identify any possible imaging
artifact of the extended emission.  We used natural weights in order
to maximize the S/N in the restored image.

Two radio sources from the \citet{condon:98} catalogue are visible in
the maps. They are listed in Table \ref{tab:point_sources}.  We
subtracted these sources from the visibilities, as we are only
interested in the diffuse emission from the cloud. We did this by
first using CLEAN to obtain their flux density. An appropriate point
model is then subtracted from the visibilities. Table
\ref{tab:point_sources} also lists our measured coordinates and flux
densities at 31\,GHz. We inspected the subtracted maps after to check
for artifacts in case of a bad estimation of the flux. Any residual
from the source subtraction is smaller than the r.m.s noise of the
maps.

\begin{table}
  \centering
  \caption{Point sources subtracted from the visibilities. The
    location and 1.4\,GHz fluxes are from the \citet{condon:98}
    catalogue. The 31\,GHz fluxes were obtained from our CARMA
    observations.  }
   \vspace{3mm}
  \begin{tabular}{lccccc}
    \hline
    NVSS name &  $S_{1.4\,{\rm GHz}}$ & $S_{31\,{\rm GHz}}$  & $\alpha_{1.4/31}$ \\
    & mJy & mJy & \\
    \hline
    J154006-070442 & $25.8\pm1.3$ & $3.5\pm1.2$ & $-0.64\pm0.26$\\ 
    J154024-070858 & $13.3\pm0.6$ & $6.2\pm1.3$ & $-0.25\pm0.16$\\ 
    \hline
  \end{tabular}
  \label{tab:point_sources}
\end{table}

In order to increase the sensitivity to extended emission, we use
Gaussian $(u,v)$ tapering. This is the multiplication of the
visibilities with a Gaussian filter which has the effect of
down-weighting the longer baselines, degrading the final angular
resolution of the map and increasing the S/R of the more extended
emission. The original angular resolution of the map using natural
weights is $\sim 1\parcm6$. We used a filter size in the Fourier space
equivalent to a Gaussian smoothing kernel in the image plane which
produces a final map with 2$'$ resolution. An added advantage of
smoothing CARMA data to 2 arcmin resolution is that it symmetrises the
beam.

After imaging with both methods, CLEAN and MEM, the r.m.s. noise in
the CLEAN map is 40\% larger than the noise in the MEM map
(1.4\,mJy\,beam$^{-1}$ and 0.99\,mJy\,beam$^{-1}$ respectively).  In
Fig. \ref{fig:ldn_final} we show the two maps. The synthesised beam
size is plotted as an ellipse at the bottom-right corner. It has a
size of $2'$\,FWHM.  Both maps present a similar morphology, but the
MEM reconstruction seem to recover more of the diffuse and extended
flux. We use the MEM map for the rest of the analysis due to this and
its lower noise value.

\subsection{Ancillary data}

Besides the CARMA data, we also used ancillary data to study the
cloud.  We used radio and IR data to build an SED of LDN\,1780 from
0.408\,GHz to 2997\,GHz on a 1\dg scale. Table \ref{tab:anc_data}
lists all the data used.

\begin{table*}%
 \centering
  \caption{List of ancillary data used in the analysis.}
  \begin{tabular}{lrrl}
    \hline
    Telescope/Survey & Freq.\,[GHz] & Nominal Resolution & Reference \\
    \hline
      Haslam&         0.408& 56\parcm0&  \citet{haslam:82,Remazeilles2015}\cr
      Reich&          1.42& 35\parcm4&  \citet{reich:82,reich:86,reich:01}\cr
      Jonas&          2.3& 20\parcm0&  \citet{jonas:98}\cr
      \wmap\ ~9-year&   22.8& \,\,51\parcm3&  \citet{Bennett2013}\cr
      \planck\ &        28.4& 32\parcm3&  \cite{Planck_results2015}\cr
      \wmap\ ~9-year&   33.0& \,\,39\parcm1&  \citet{Bennett2013}\cr
      \wmap\ ~9-year&   40.7& \,\,30\parcm8&  \citet{Bennett2013}\cr
      \planck\ &        44.1& 27\parcm1&  \cite{Planck_results2015}\cr
      \wmap\ ~9-year&   60.7& \,\,21\parcm1&  \citet{Bennett2013}\cr
      \planck\ &        70.4& 13\parcm3&  \cite{Planck_results2015}\cr
      \wmap\ ~9-year&   93.5& \,\,14\parcm8&  \citet{Bennett2013}\cr
      \planck\ &        100& 9\parcm7&  \cite{Planck_results2015}\cr
      \planck\ &        143& 7\parcm3&  \cite{Planck_results2015}\cr
      \planck\ &        217& 5\parcm0&  \cite{Planck_results2015}\cr
      \planck\ &        353& 4\parcm8&  \cite{Planck_results2015}\cr
      \planck\ &        545& 4\parcm7&  \cite{Planck_results2015}\cr
      \planck\ &        857& 4\parcm3&  \cite{Planck_results2015}\cr
      \cobe-DIRBE&    1249& 37\parcm1&  \citet{hauser:98}\cr
      \cobe-DIRBE&    2141& 38\parcm0&  \citet{hauser:98}\cr
      \cobe-DIRBE&    2997& 38\parcm6&  \citet{hauser:98}\cr
      \hline
  \end{tabular} 
  \label{tab:anc_data}
\end{table*}

We used the re-processed version of the 0.408\,GHz map of
\citet{haslam:82} by \citet{Remazeilles2015} available at the LAMBDA
website\footnote{\url{http://lambda.gsfc.nasa.gov/}}.  It has an
effective resolution of 56$'$ and includes all the point sources.  At
1.42\,GHz the Reich et al. map \citep{reich:82,reich:86,reich:01} has
an angular resolution of 36$'$. The 2.326\,GHz map from
\citet{jonas:98} has an angular resolution of 20\parcm~ We assumed a
10\% uncertainty in these three data sets. An additional uncertainty
of 0.8\,K is added to the 0.408\,GHz map, in order to account for the
striations on the map as measured in \citet{Remazeilles2015}.

We included the five \wmap\ 9-yr maps \citep{Bennett2013}, from 23 to
94\,GHz, smoothed to 1\dg\!\!. We assumed a conservative 4\%
uncertainty \citep[the calibration uncertainty quoted in][is
  0.2\%]{Bennett2013}. This is to account for any additional
uncertainty due to non-symmetric beams, and any colour correction
effect, as the spectral index of the source is not equal to the CMB
spectrum.

We also include \planck\ data. \planck\ observes the full sky in nine
frequency bands between 28 and 857\,GHz. We used the temperature maps
which were released in 2015 (PR2), described in
\citet{Planck_results2015} and are available in the \planck\ Legacy
Archive\footnote{\url{http://pla.esac.esa.int/pla/}}.

\section{SED at 1$^{\circ}$ resolution}
\label{sec:onedeg}

\subsection{Flux densities measurement}

To obtain the flux densities of the cloud at the different
frequencies, the maps that are originally in antenna temperature
units, are expressed in flux units (Jy\,pixel$^{-1}$) using the
relation,
\begin{equation}
  S= \frac{2 \,k\,T_{RJ}\,\nu^2\,\Omega_{\mathrm{pix}}}{c^2},
\end{equation}
where $\Omega_{\mathrm{pix}}$ is the solid angle of each pixel,
$T_{RJ}$ the brightness temperature, $\nu$ the observing frequency,
$k$ the Boltzmann constant and $c$ the speed of light \citep[see ][for
  a similar analysis on other sources]{planck_sd:11,planck_sd:13}.

The flux densities are measured by integrating the flux density over a
2\dg diameter aperture, centred at the position of the cloud. We
subtract the background level using the median value of the pixels
that lie in an annular aperture between 80$'$ and 100$'$ from the
position of the source. The r.m.s. variations in this ring are used to
estimate the uncertainty in the measured fluxes, including noise, CMB
and background variations.

\begin{figure*}
  \centering \newcommand{\widthfig}{0.2} \newcommand{\angfig}{0}
  \includegraphics[angle=\angfig,width=\widthfig\textwidth]{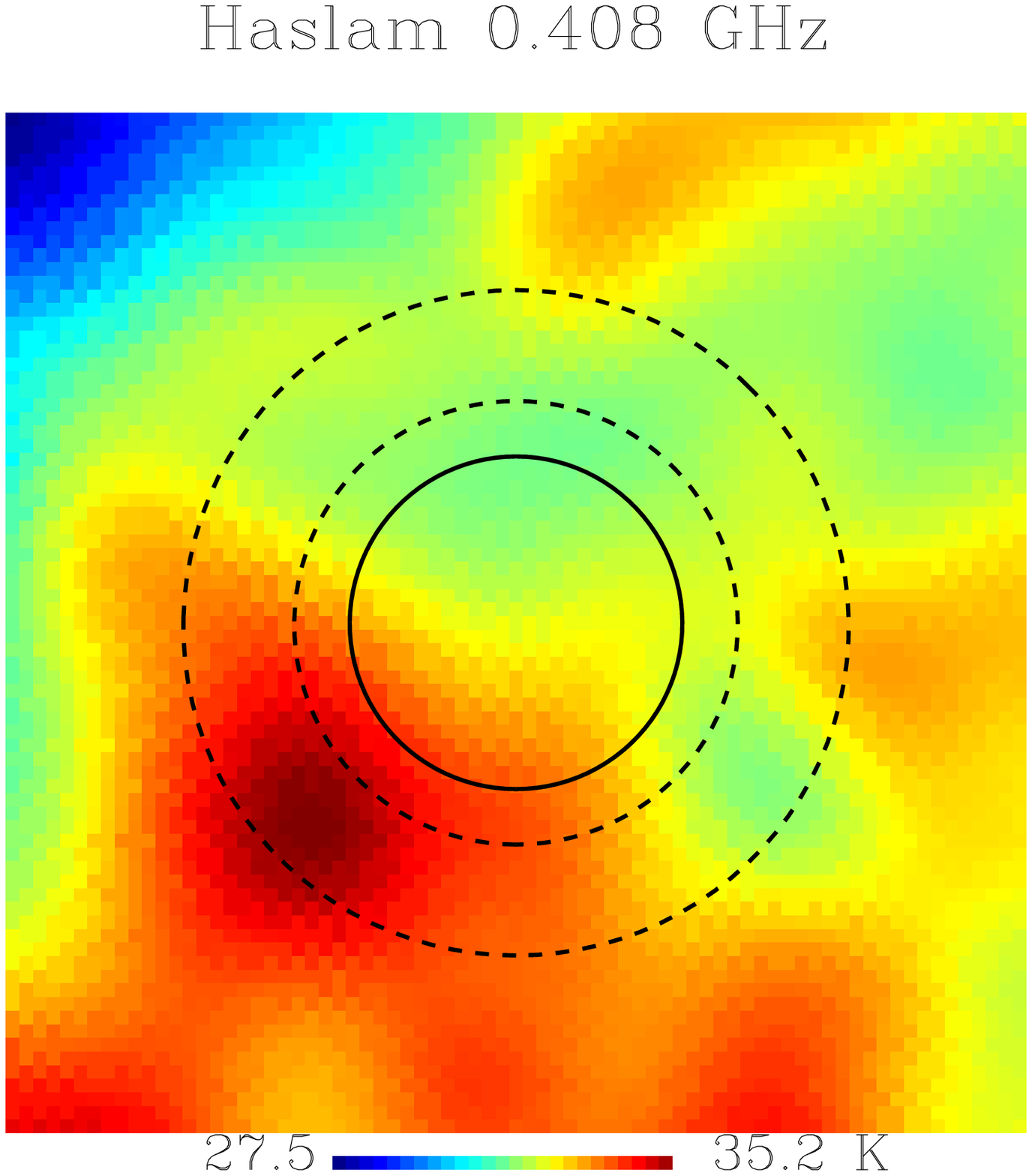}
  \includegraphics[angle=\angfig,width=\widthfig\textwidth]{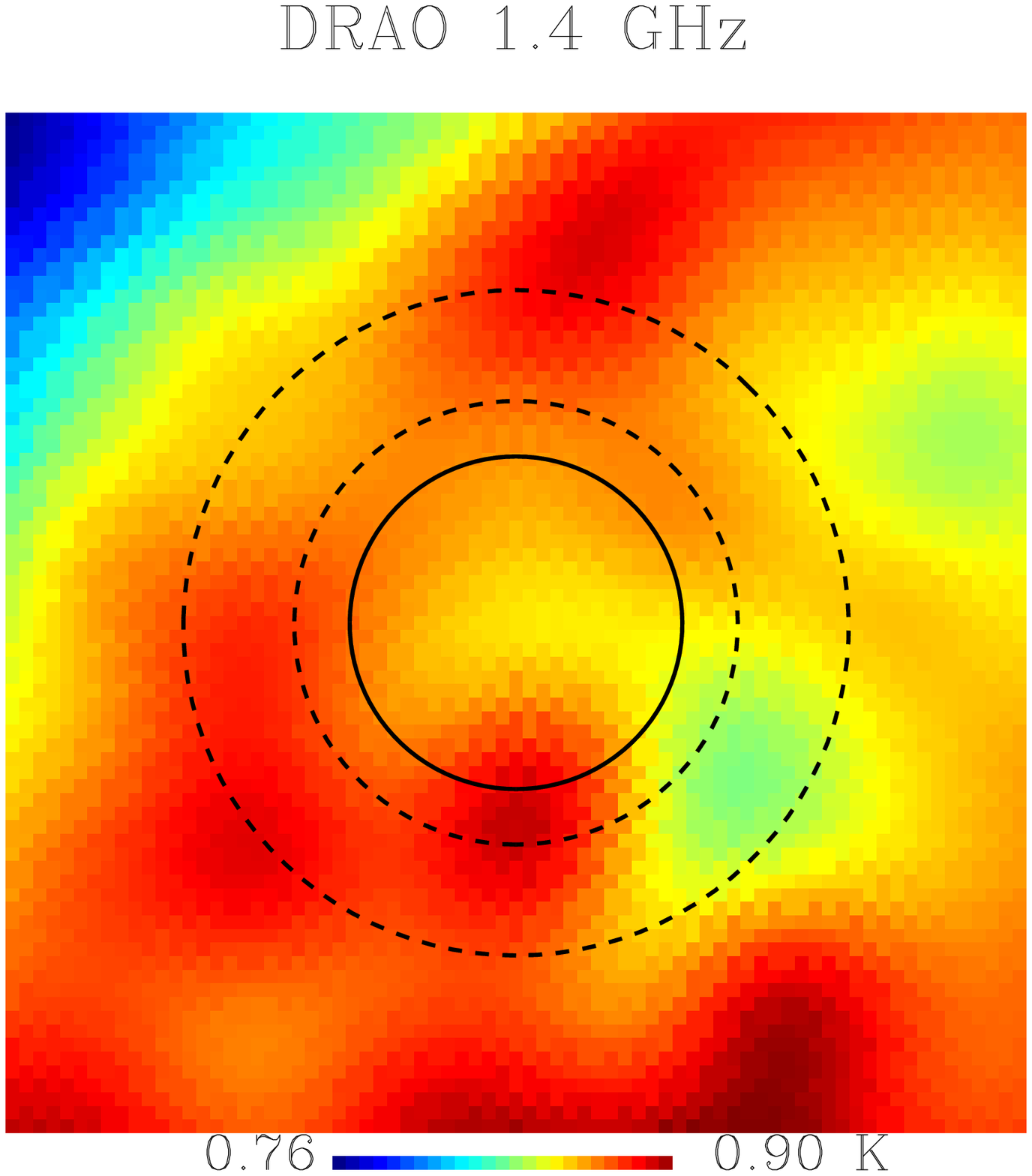}
  \includegraphics[angle=\angfig,width=\widthfig\textwidth]{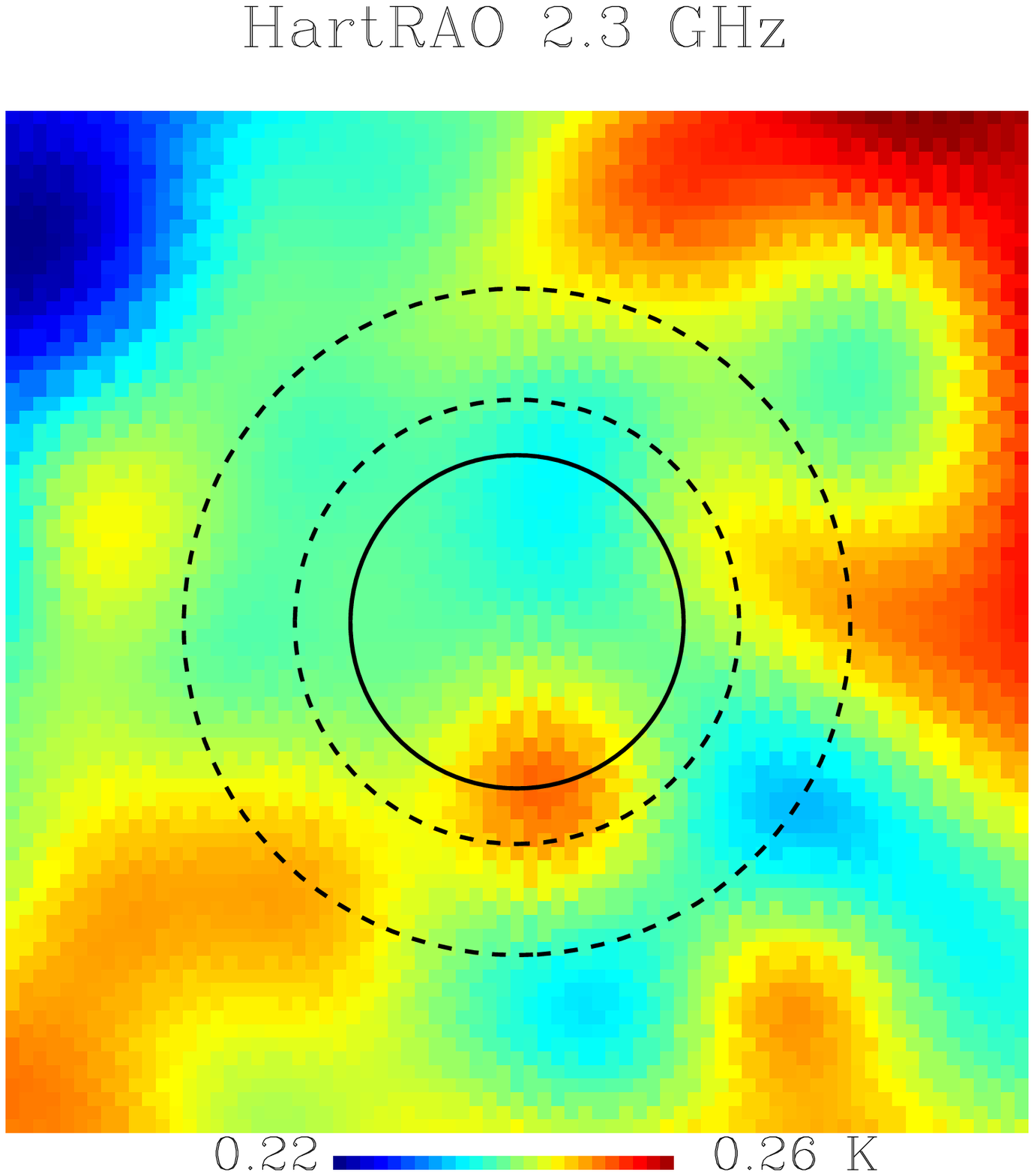}
  \includegraphics[angle=\angfig,width=\widthfig\textwidth]{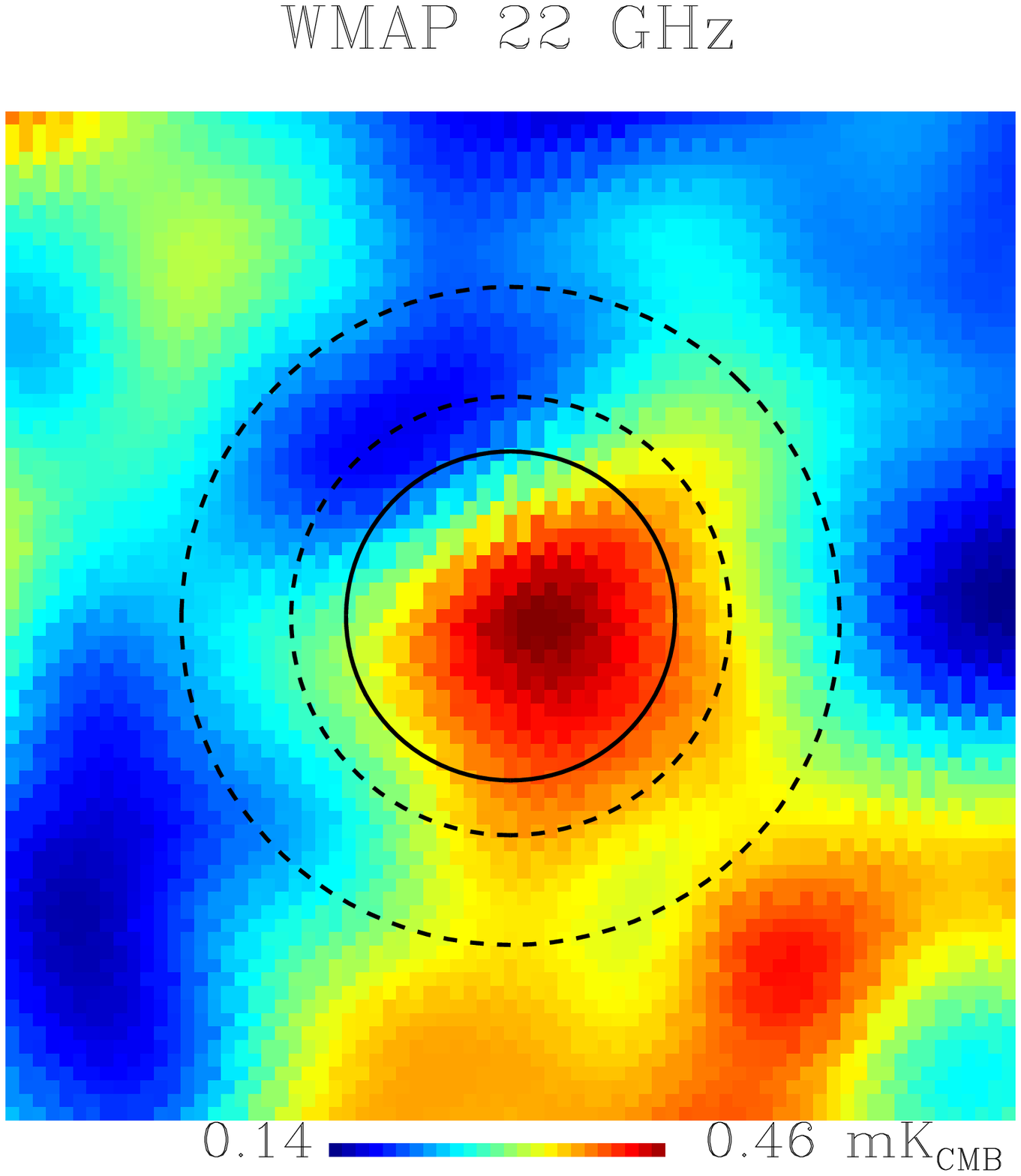}

  \includegraphics[angle=\angfig,width=\widthfig\textwidth]{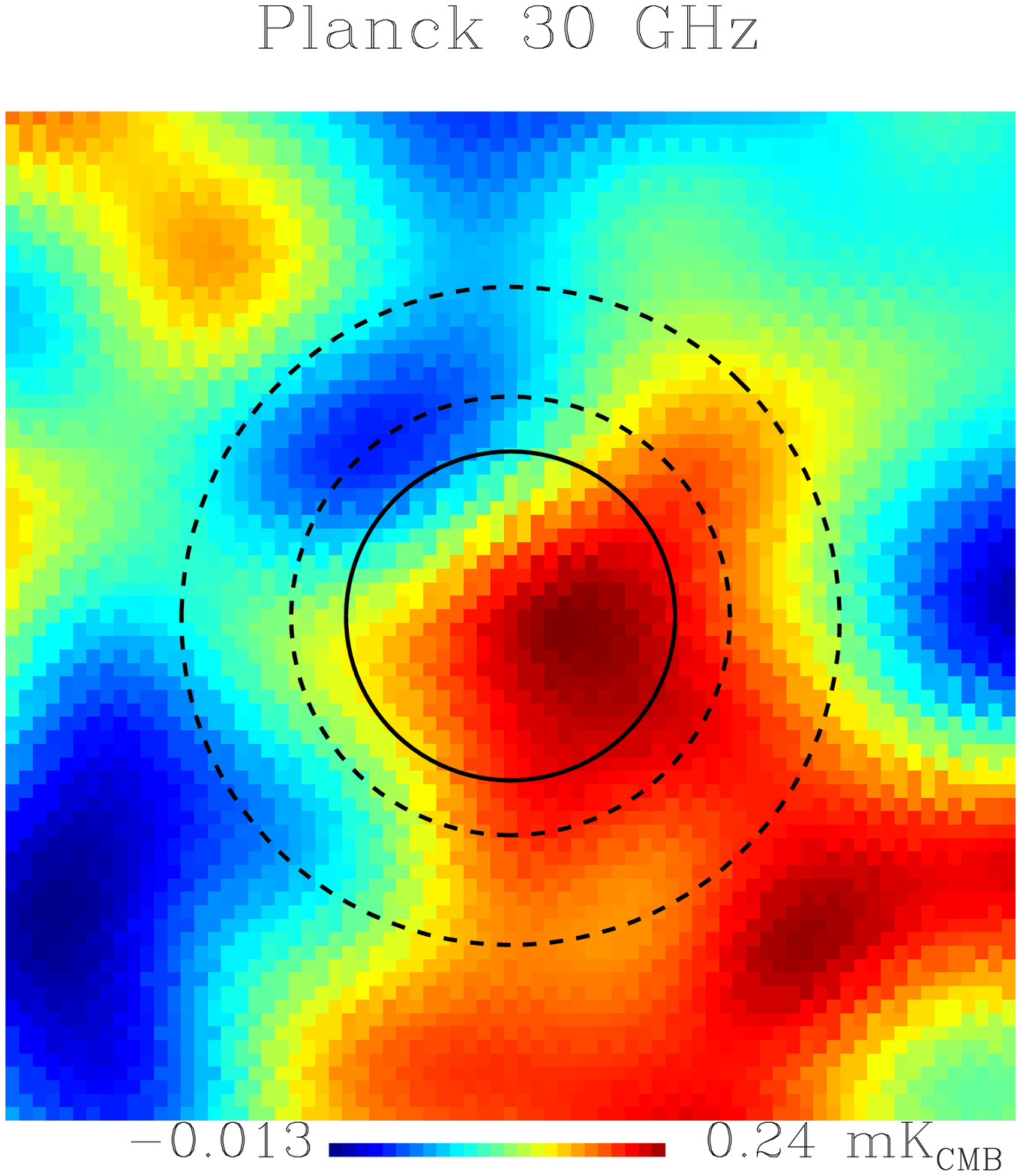}
  \includegraphics[angle=\angfig,width=\widthfig\textwidth]{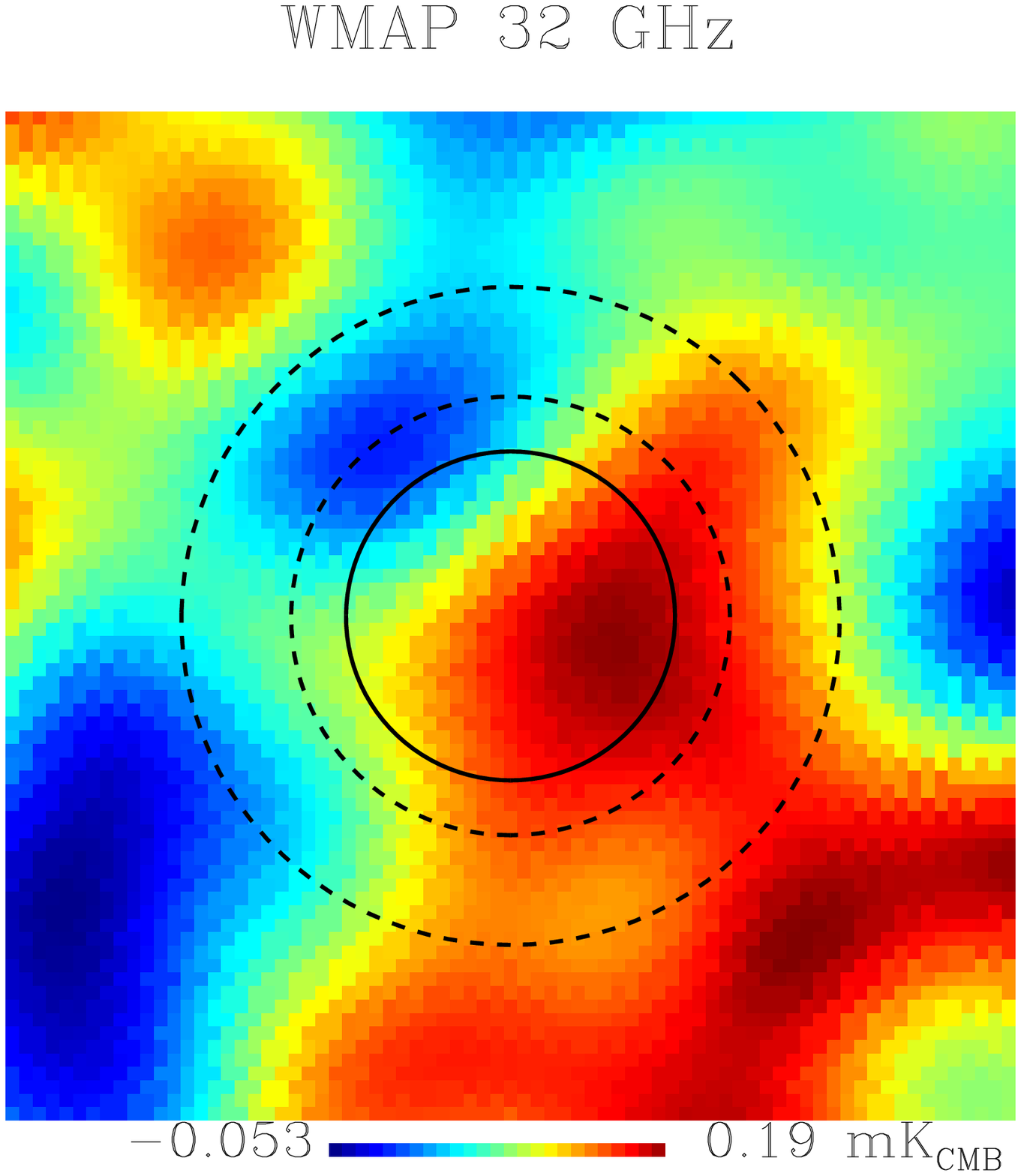}
  \includegraphics[angle=\angfig,width=\widthfig\textwidth]{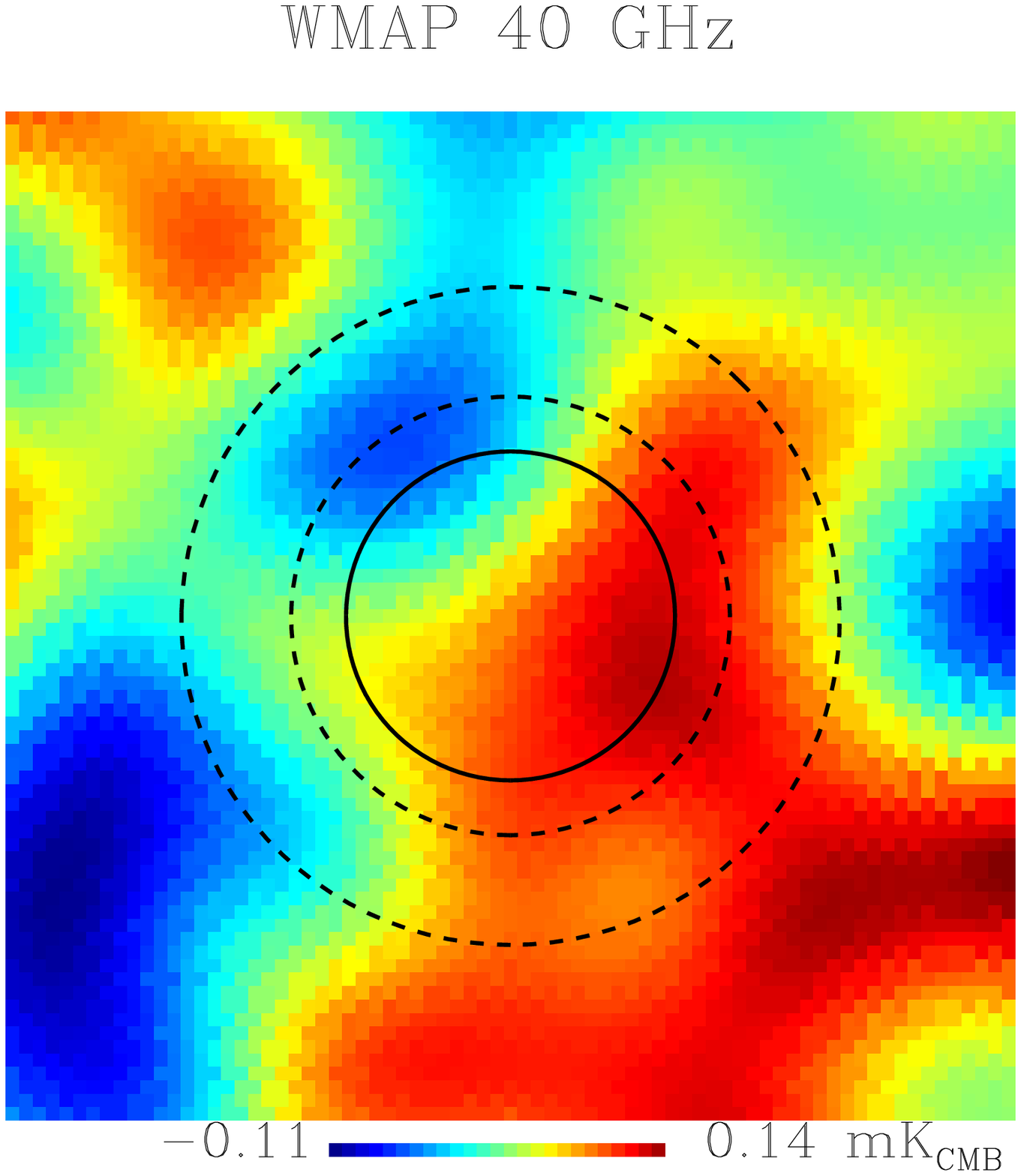}
  \includegraphics[angle=\angfig,width=\widthfig\textwidth]{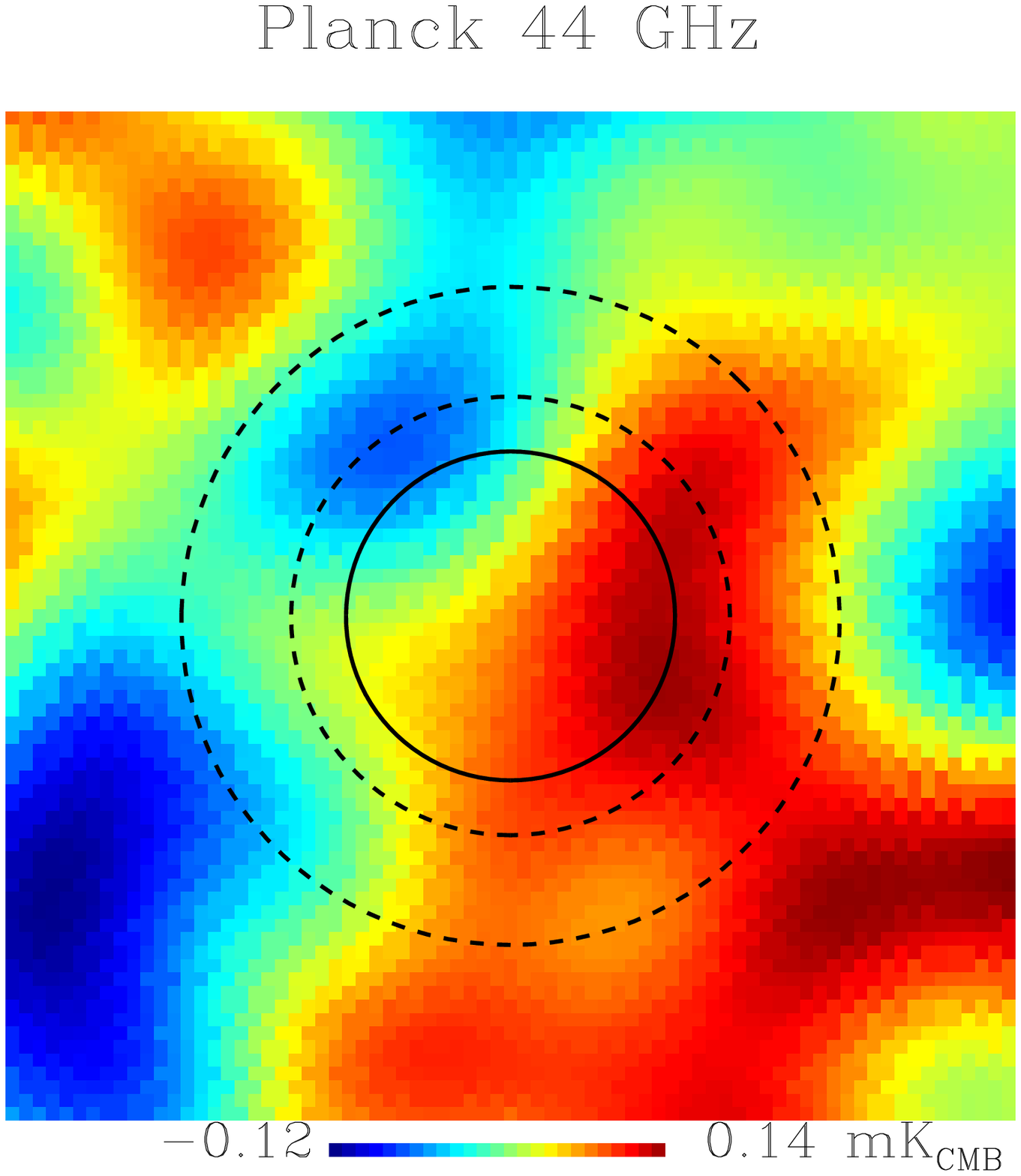}
  
  \includegraphics[angle=\angfig,width=\widthfig\textwidth]{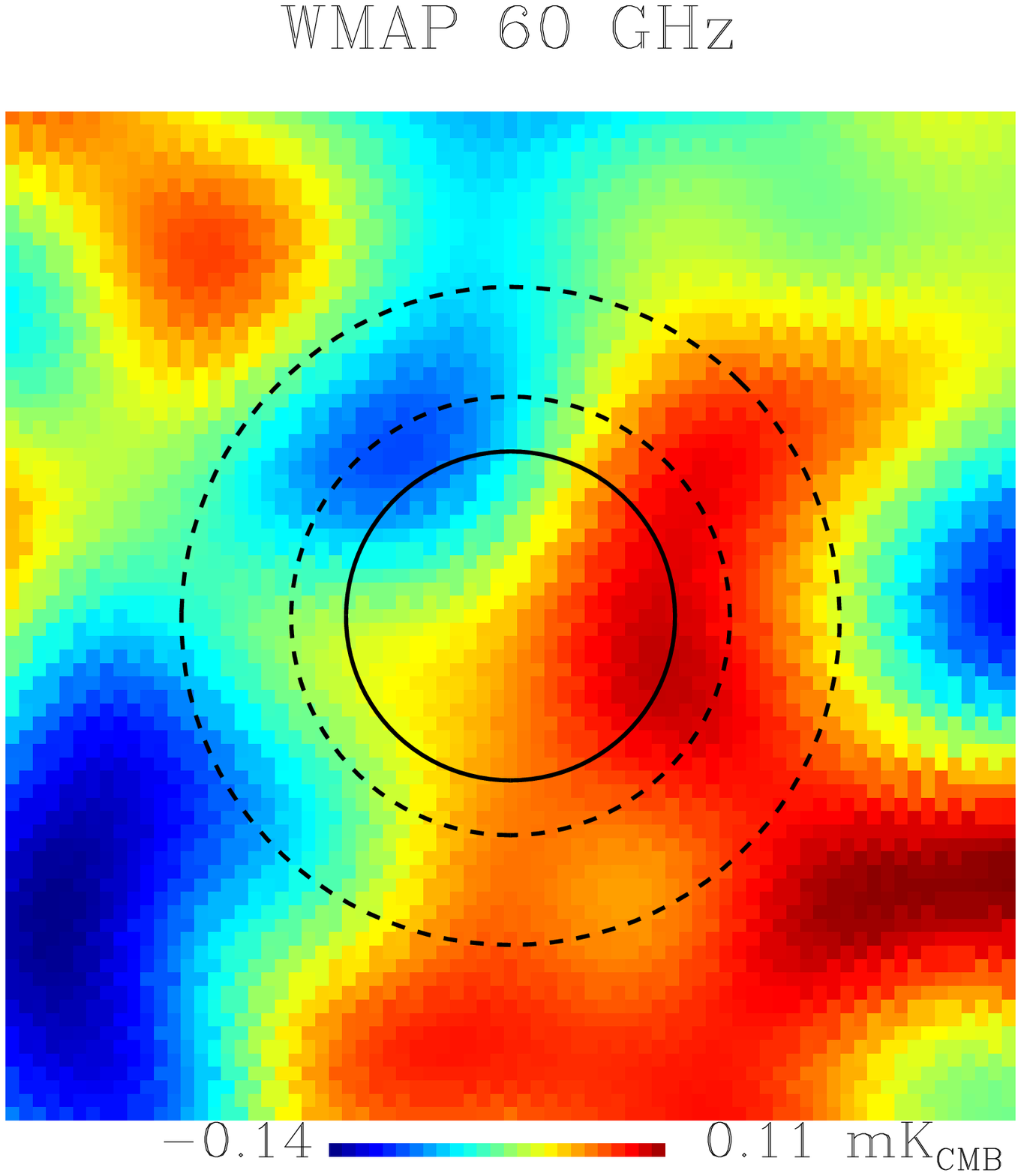}
  \includegraphics[angle=\angfig,width=\widthfig\textwidth]{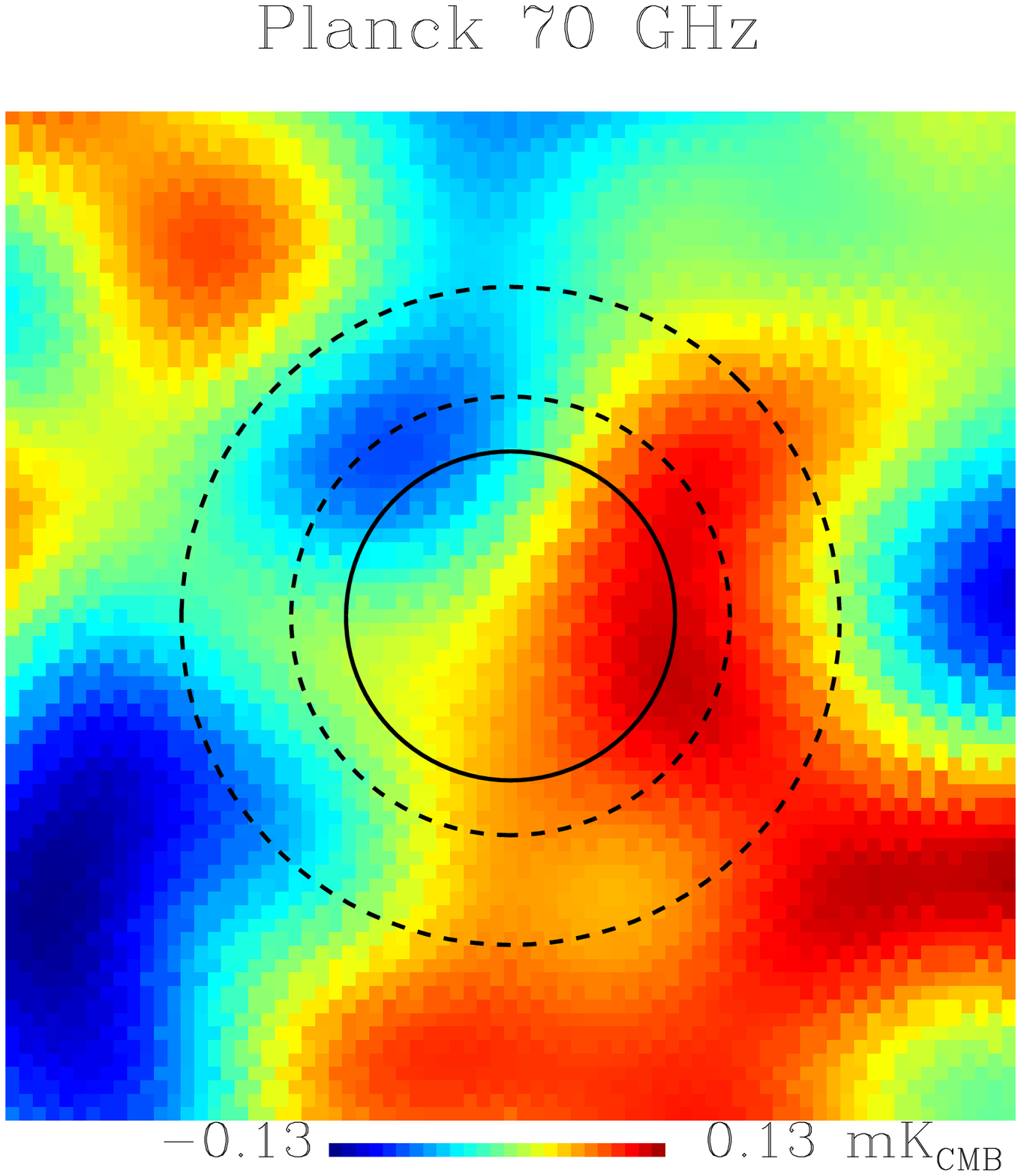}
  \includegraphics[angle=\angfig,width=\widthfig\textwidth]{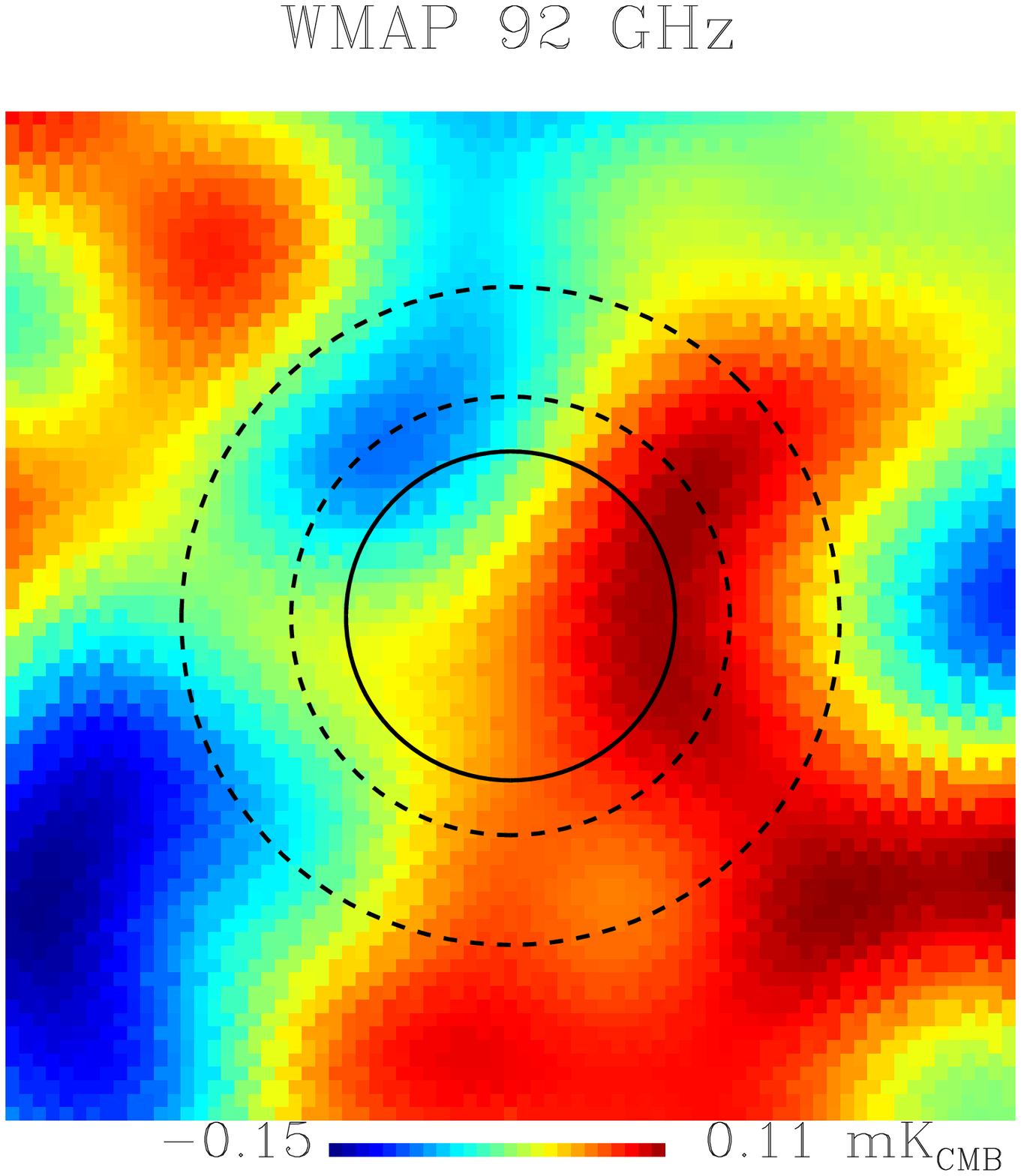}
  \includegraphics[angle=\angfig,width=\widthfig\textwidth]{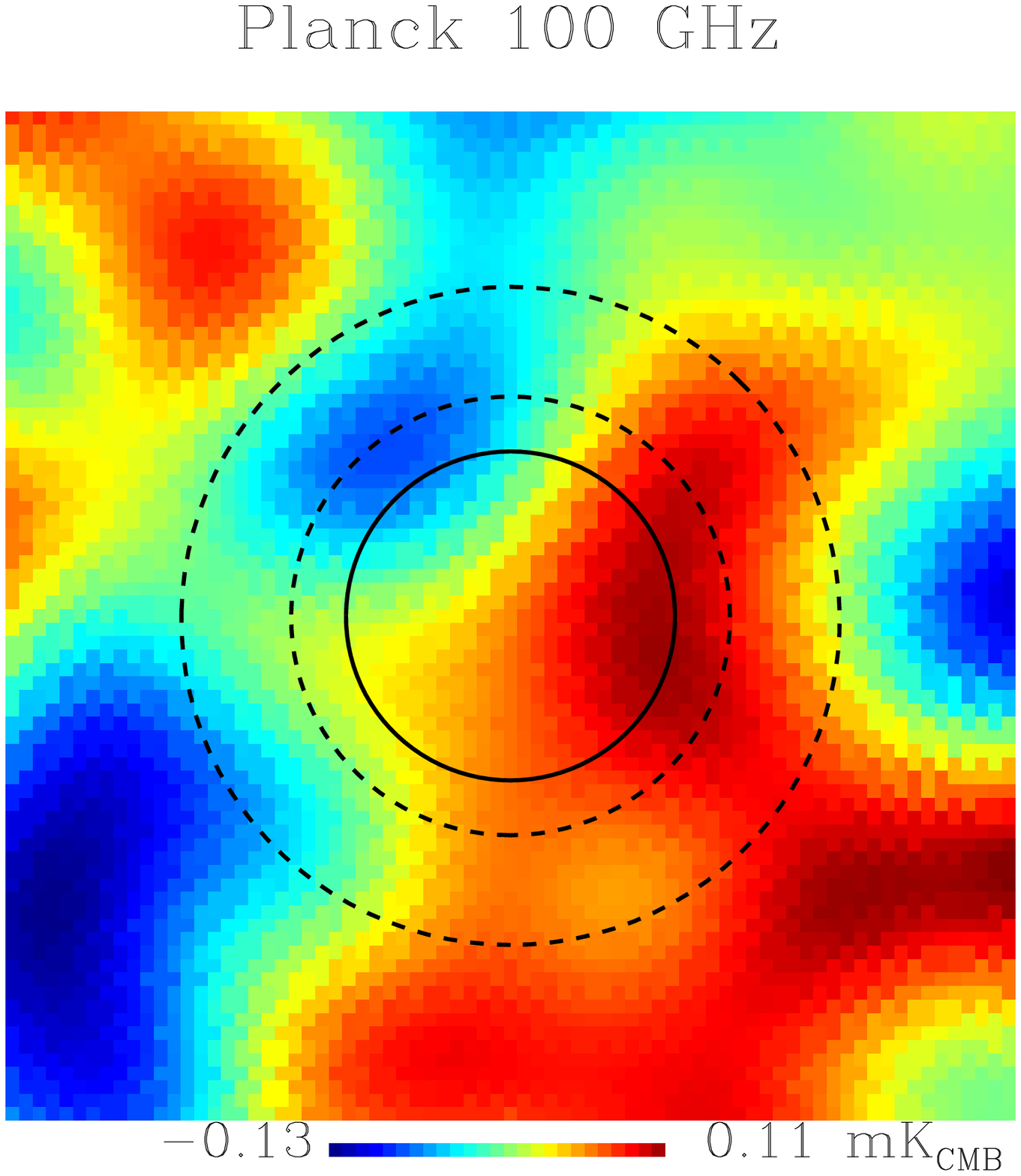}
  
  \includegraphics[angle=\angfig,width=\widthfig\textwidth]{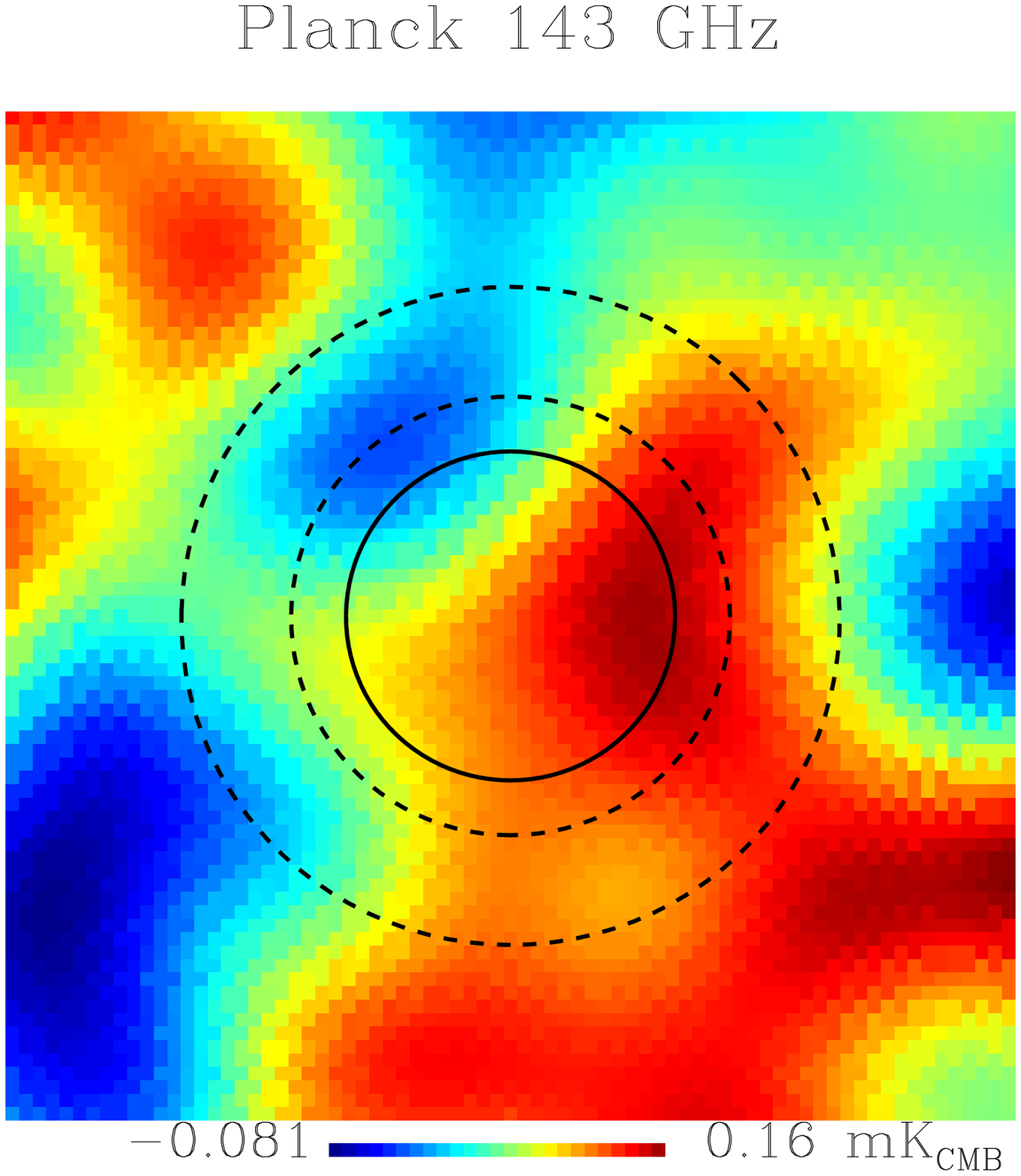}
  \includegraphics[angle=\angfig,width=\widthfig\textwidth]{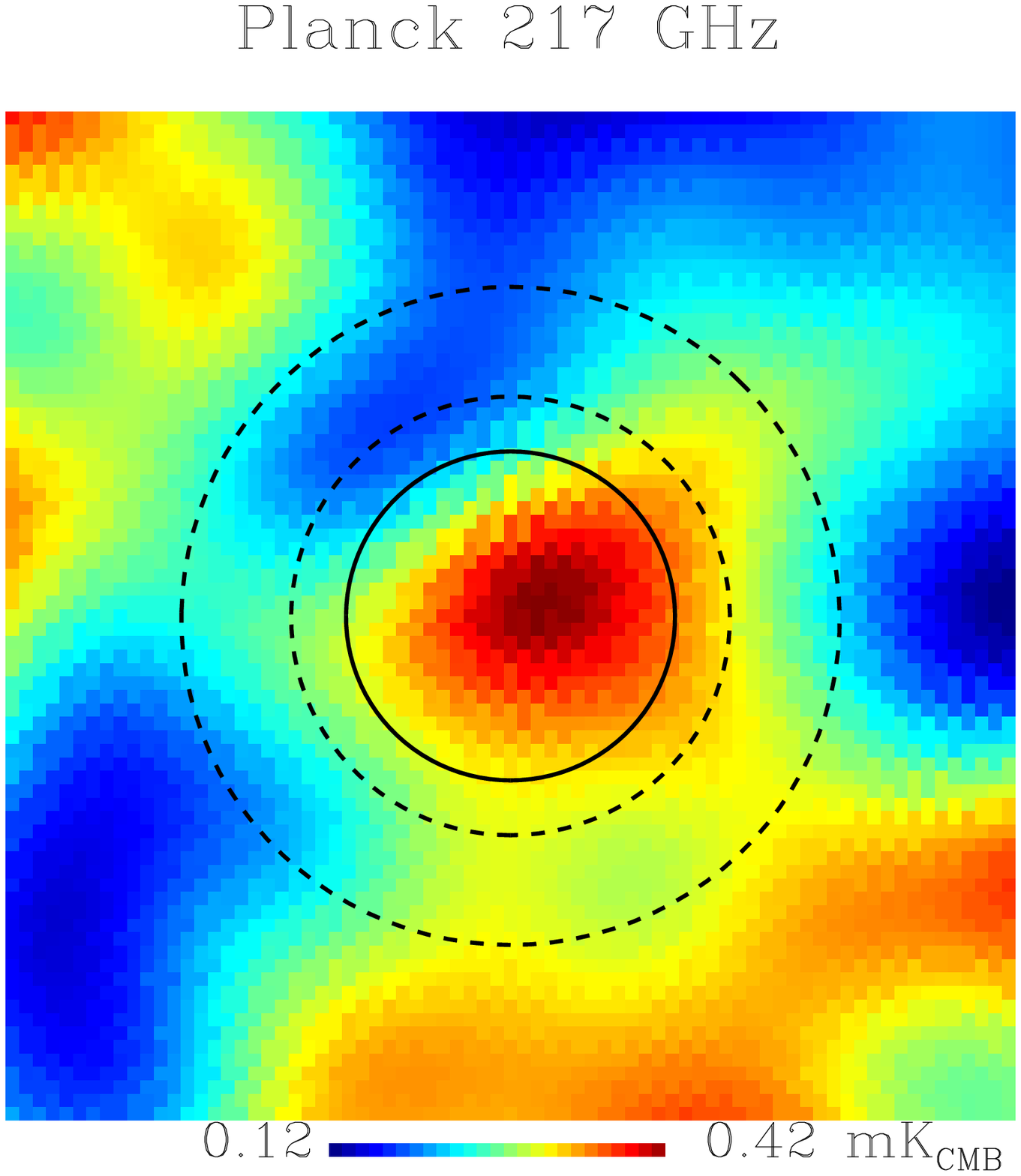}
  \includegraphics[angle=\angfig,width=\widthfig\textwidth]{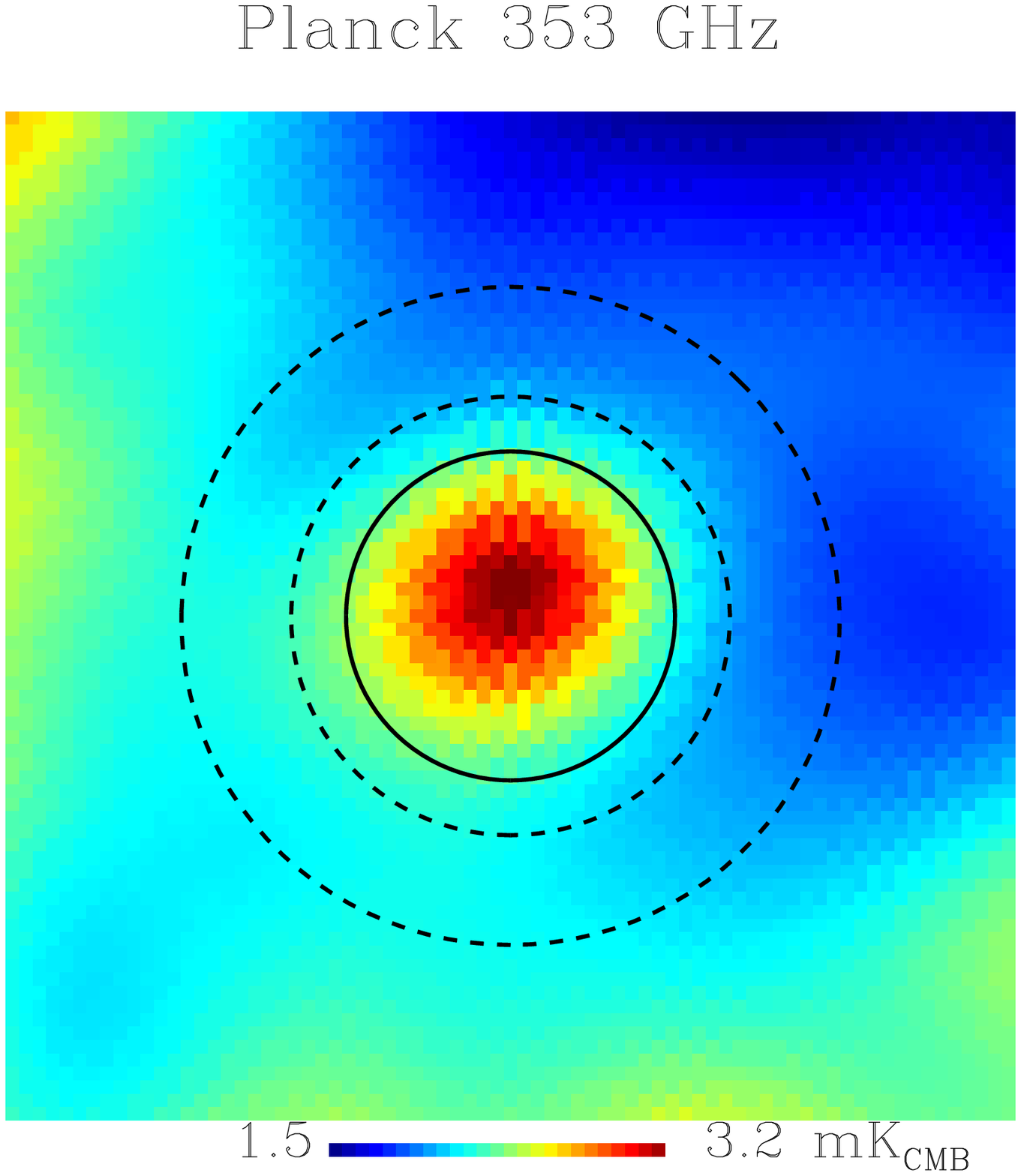}
  \includegraphics[angle=\angfig,width=\widthfig\textwidth]{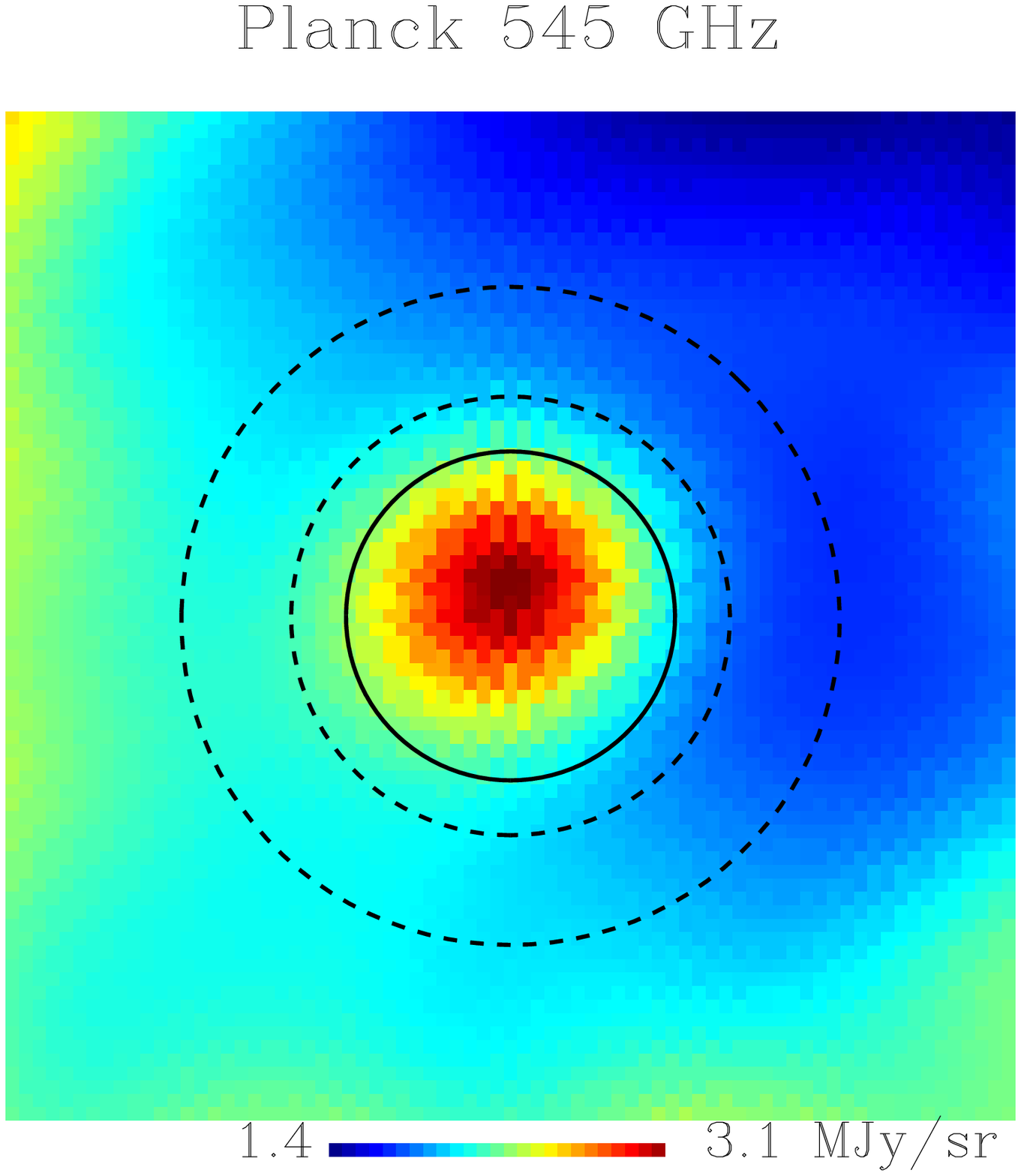}

  \includegraphics[angle=\angfig,width=\widthfig\textwidth]{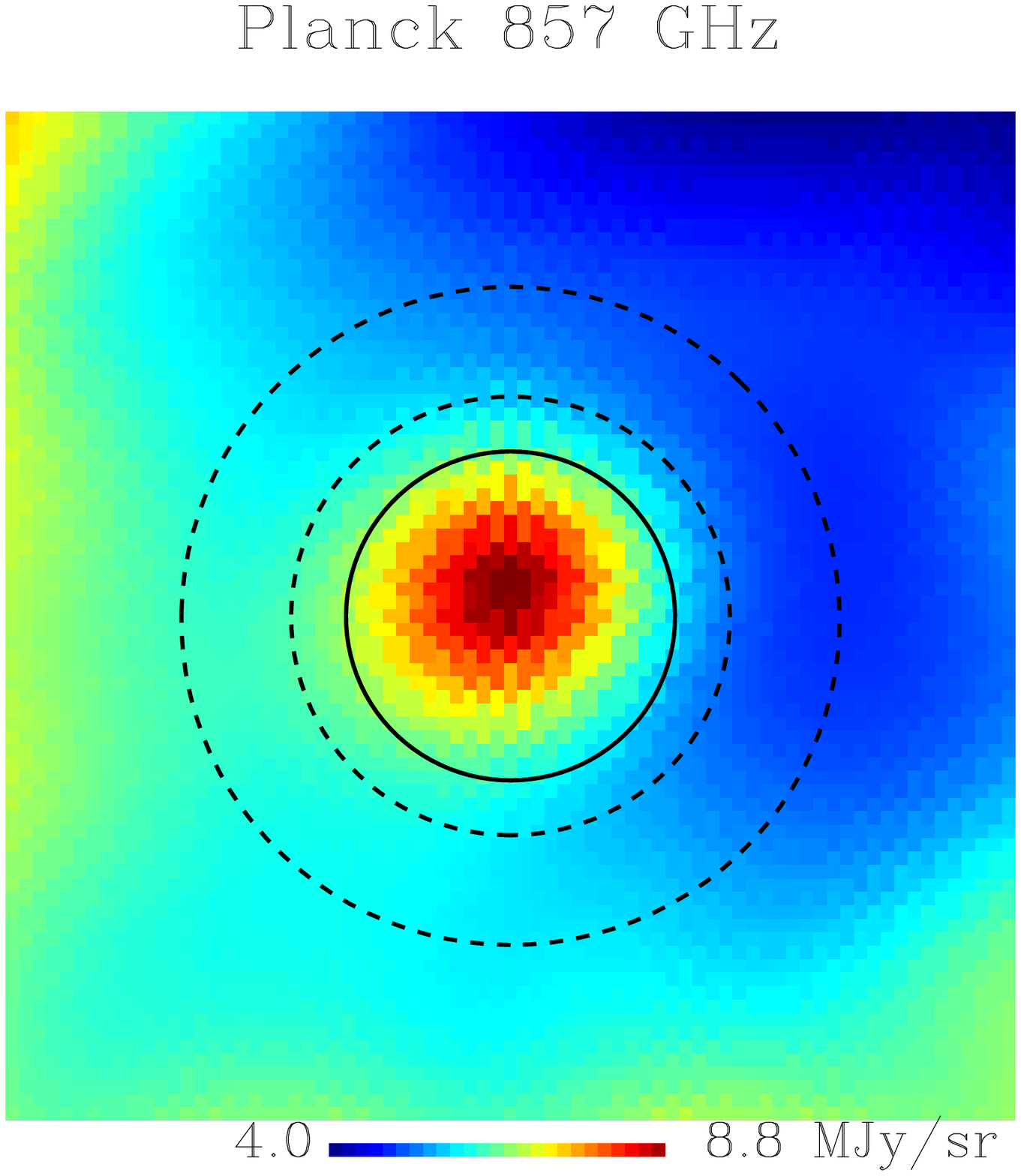}
  \includegraphics[angle=\angfig,width=\widthfig\textwidth]{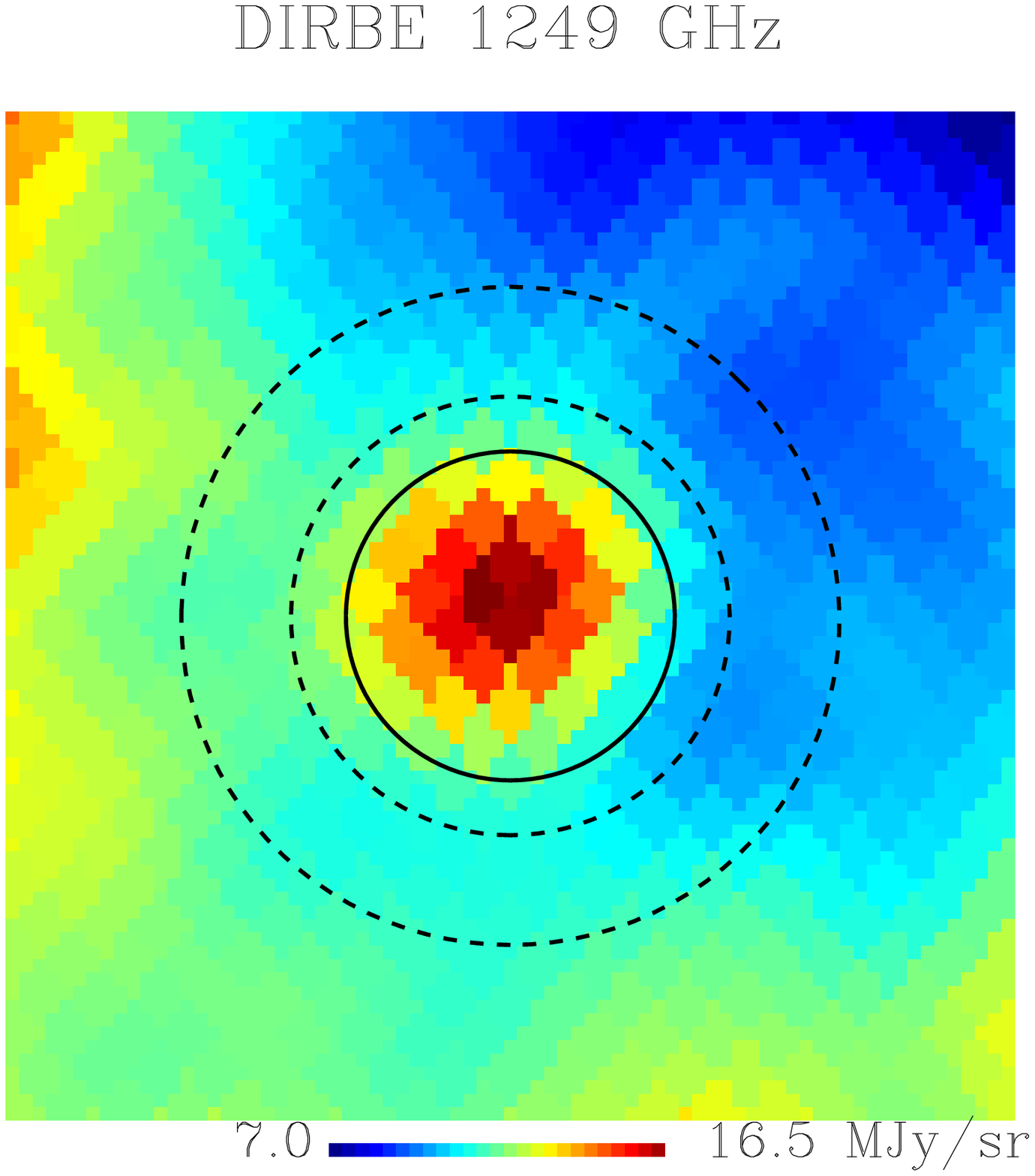}
  \includegraphics[angle=\angfig,width=\widthfig\textwidth]{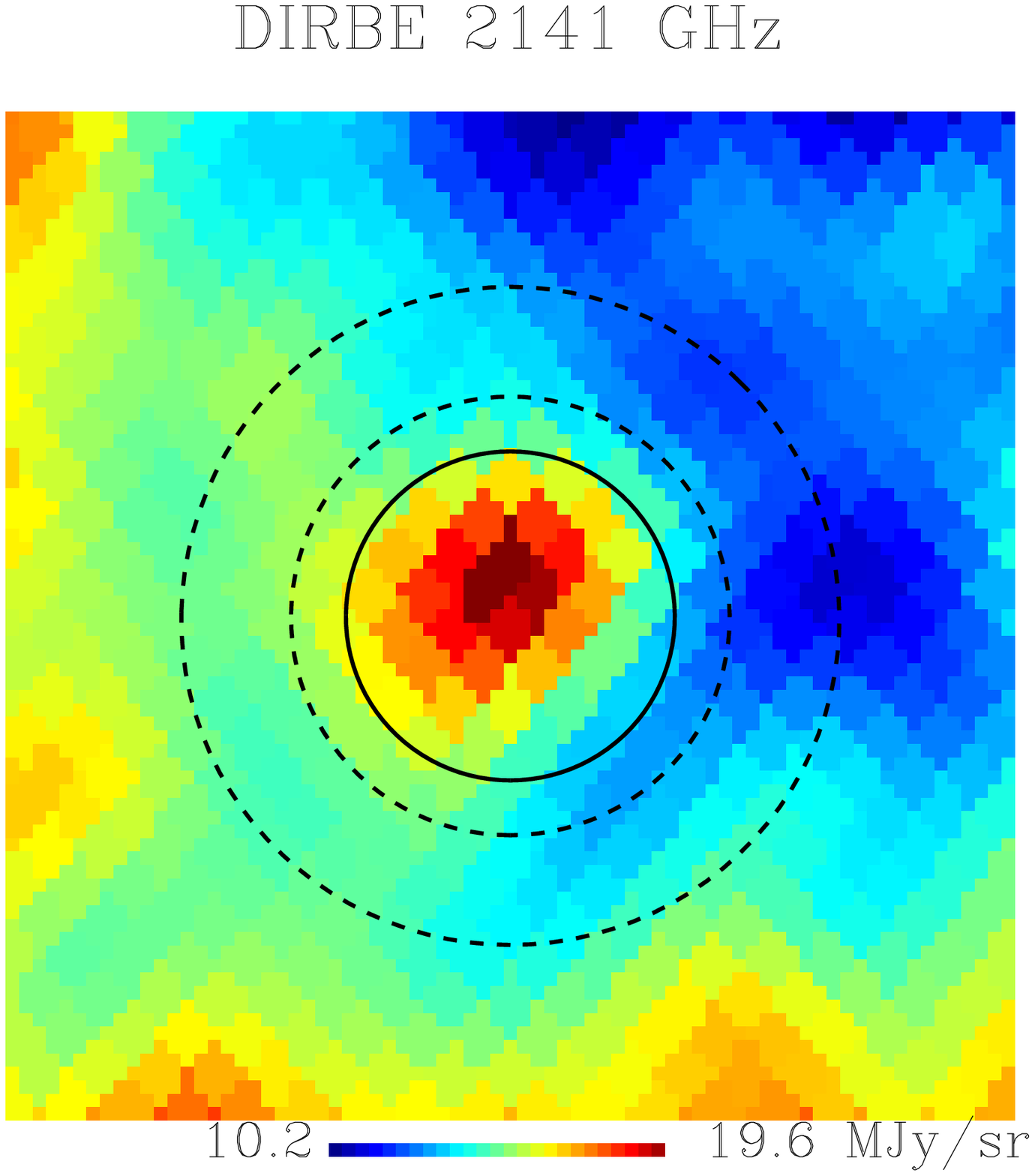}
  \includegraphics[angle=\angfig,width=\widthfig\textwidth]{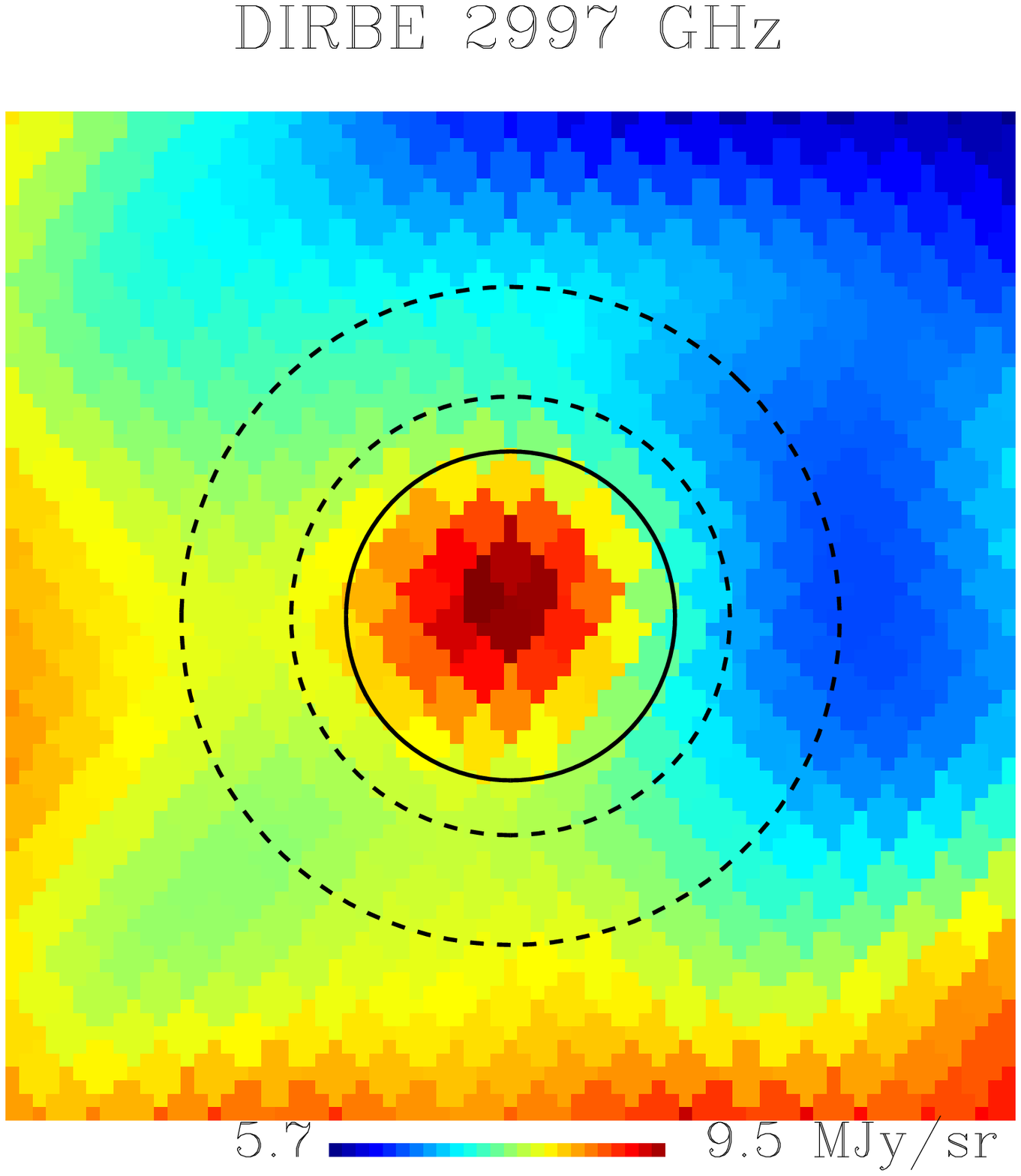}   
  \caption{Maps of LDN\,1780 at different frequencies ranging from the
    Haslam et al. map at 0.408\,GHz to the DIRBE image at
    2997\,GHz. The square maps have 5\dg in side and all have been
    smooth to a common angular resolution of 1\dg FWHM. The colour
    scale is linear, ranging from the minimum to the maximum of each
    map. The inner circle shows the aperture where we measured the
    flux and the dashed larger circles show the annulus used to
    estimate the background emission and noise around the aperture. }
  \label{fig:l1780_maps}
\end{figure*}

In Fig. \ref{fig:l1780_maps} we present the 20 maps we used of LDN\,1780,
from 0.408\,GHz up to 2997\,GHz. All of them have been smoothed to a
common 1\dg resolution. Each image is 5\dg on a side and the circular
aperture and ring used for the photometry are indicated. The cloud is
clearly visible in the high frequency maps, from 217\,GHz, where the
thermal dust emission dominates above the diffuse background. At lower
frequencies, between 23 and 143\,GHz, all the maps show a similar
structure, the CMB fluctuations that predominate over this frequency
range and angular scales. At even lower frequencies, in the 0.408, 1.4
and 2.3\,GHz maps, there is no emission from the region of the cloud
visible above the background.

\begin{figure*}
  \centering
   \newcommand{\widthfig}{0.2}
   \newcommand{\angfig}{0}
  \includegraphics[angle=\angfig,width=\widthfig\textwidth]{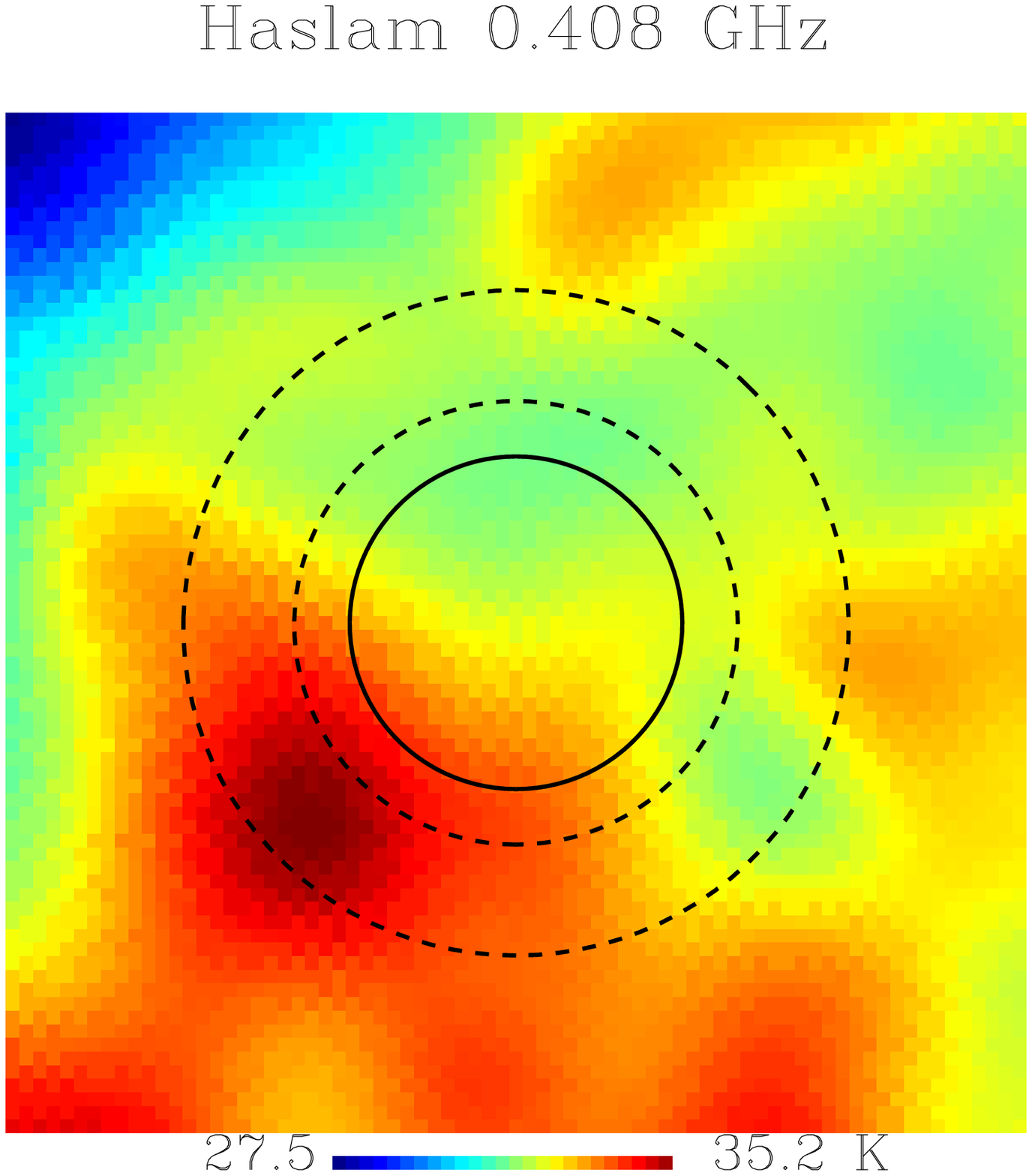}
  \includegraphics[angle=\angfig,width=\widthfig\textwidth]{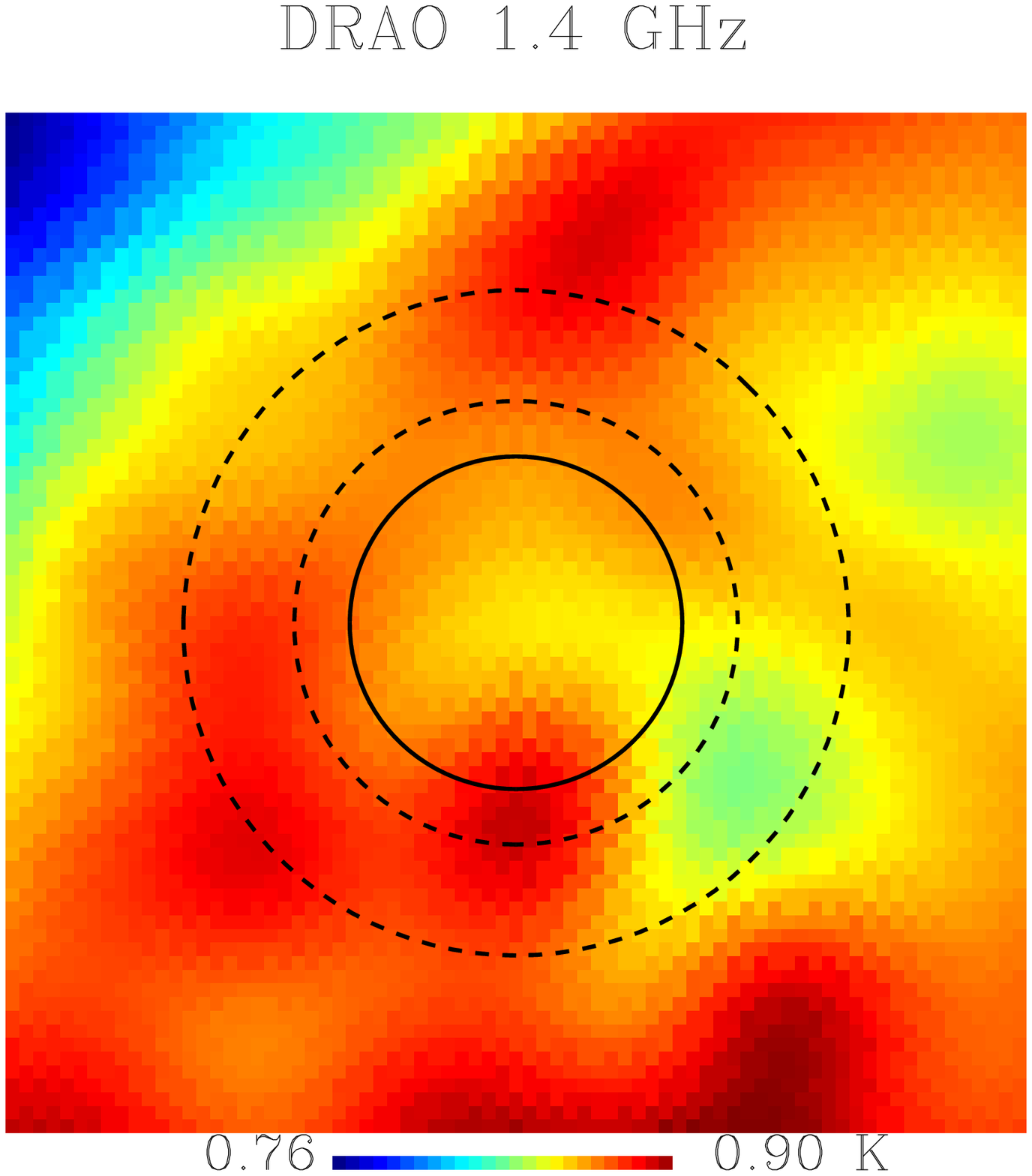}
  \includegraphics[angle=\angfig,width=\widthfig\textwidth]{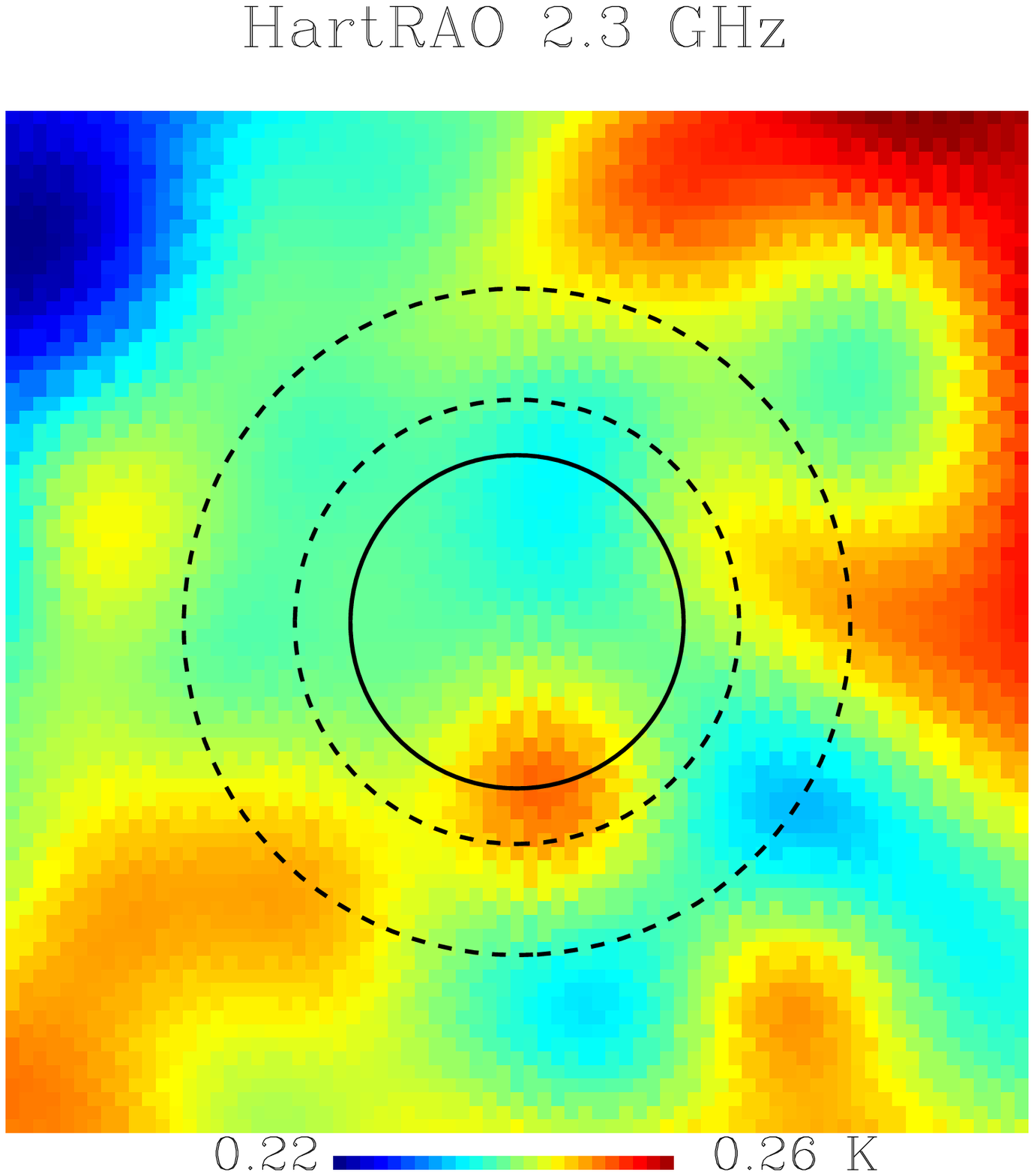}
  \includegraphics[angle=\angfig,width=\widthfig\textwidth]{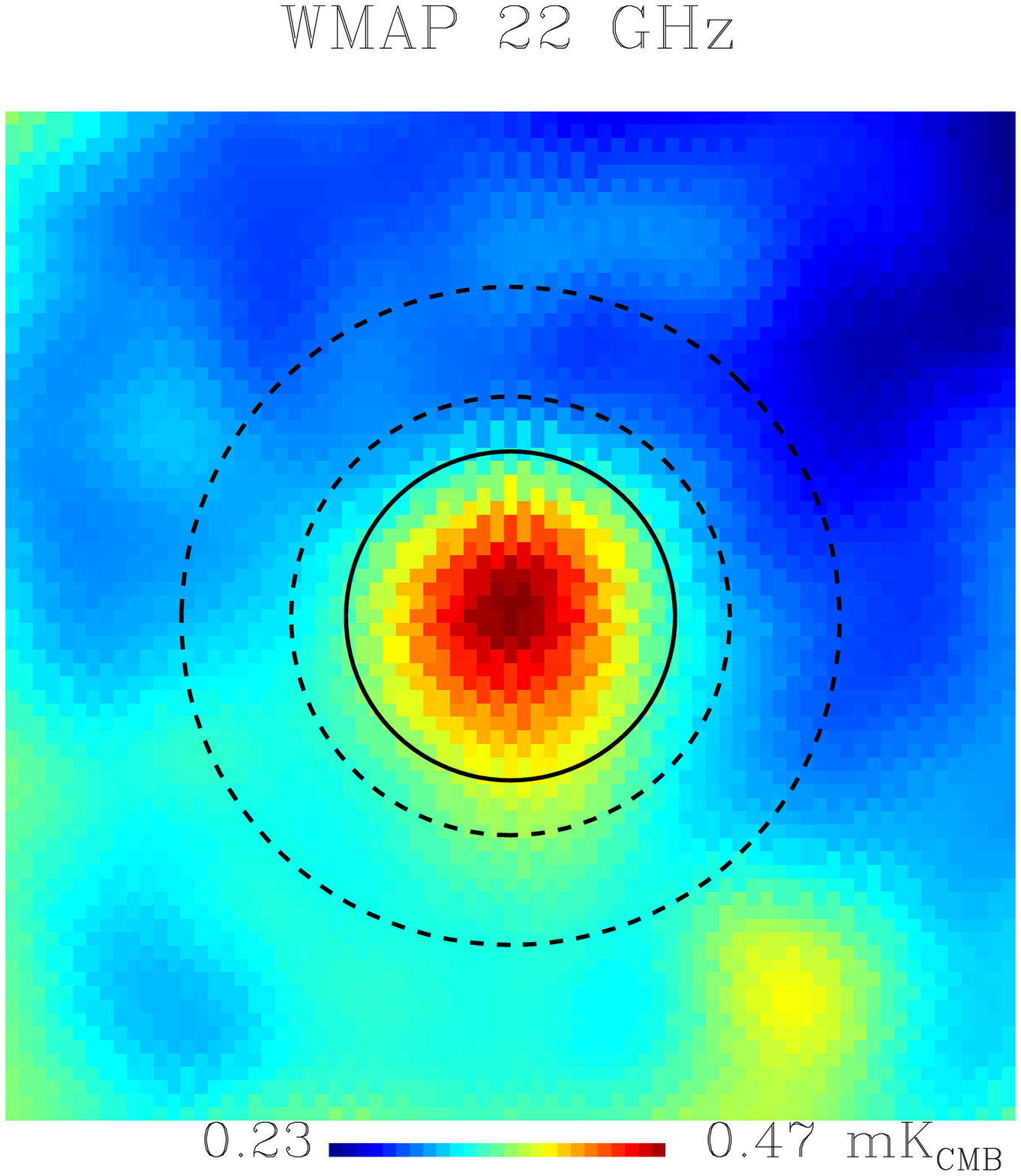} 

  \includegraphics[angle=\angfig,width=\widthfig\textwidth]{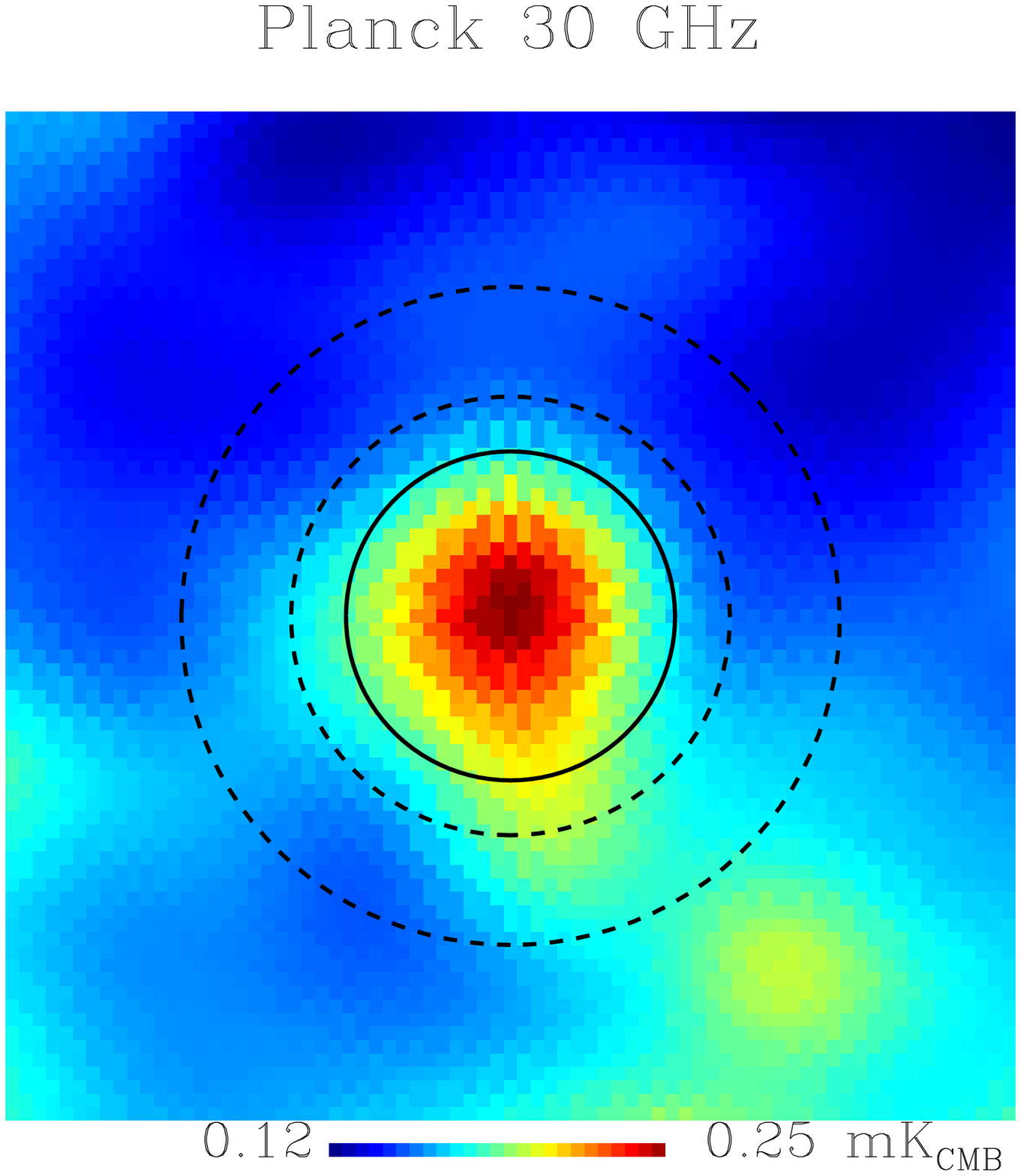}
  \includegraphics[angle=\angfig,width=\widthfig\textwidth]{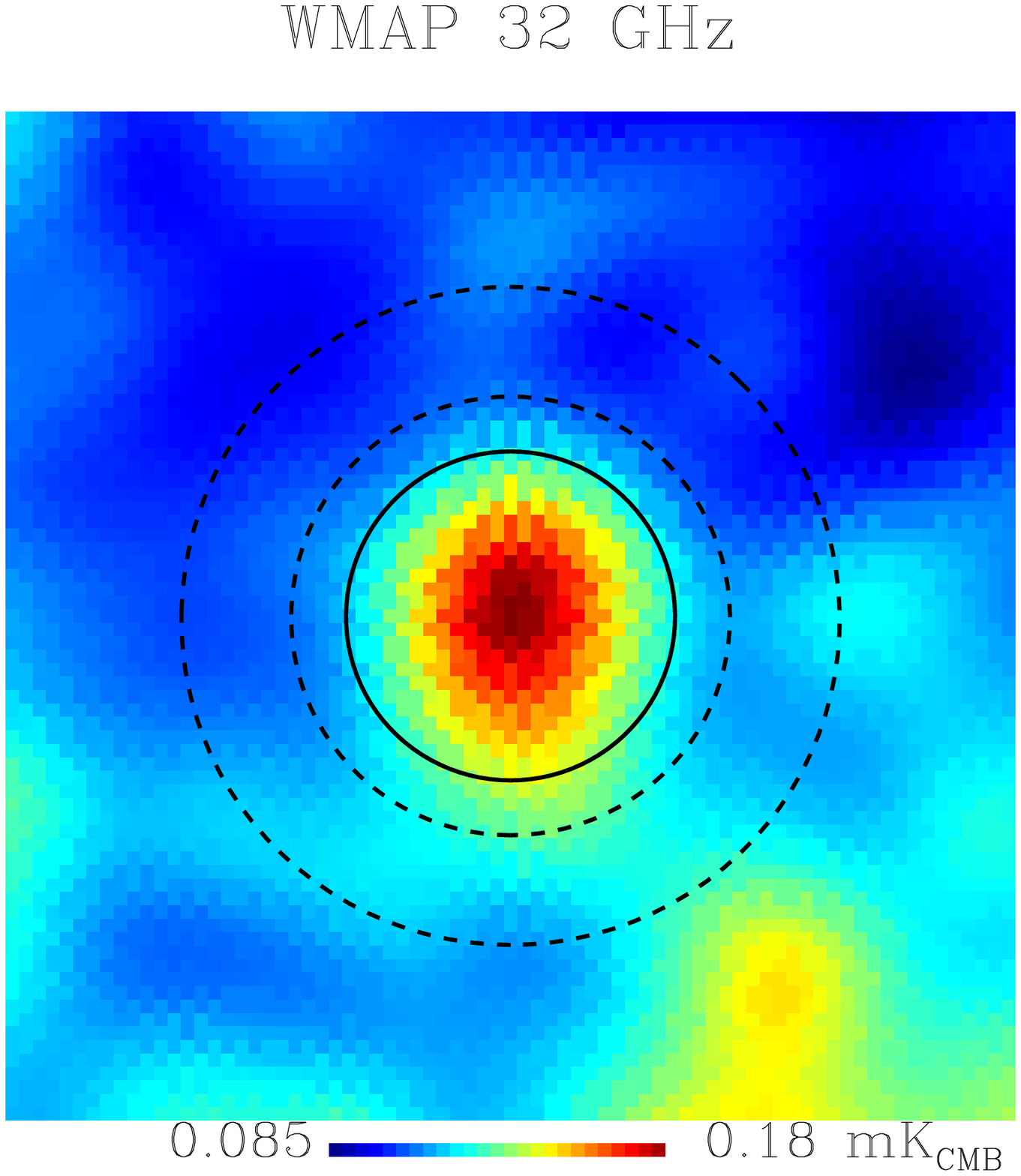}
  \includegraphics[angle=\angfig,width=\widthfig\textwidth]{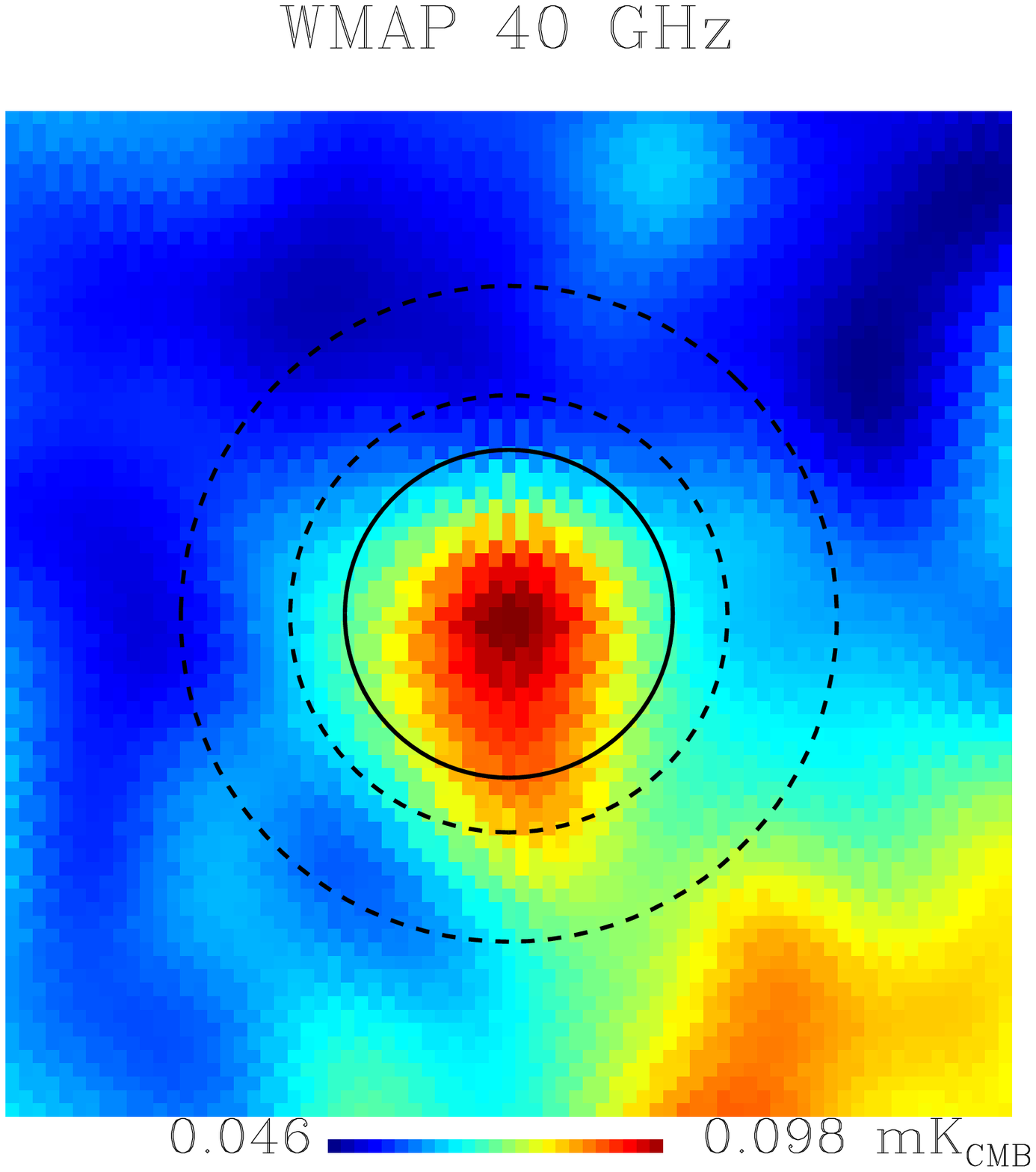}
  \includegraphics[angle=\angfig,width=\widthfig\textwidth]{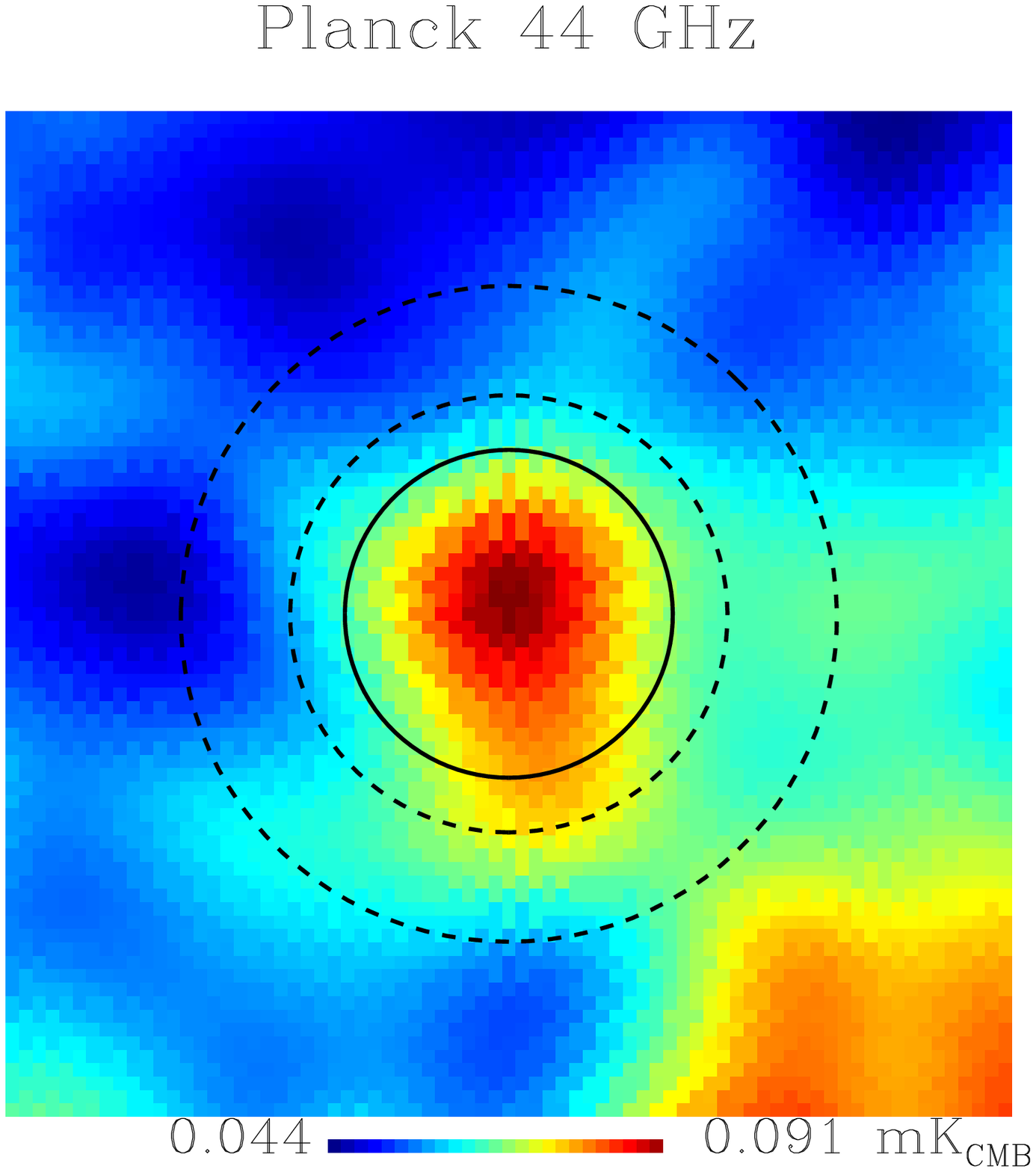}
  
  \includegraphics[angle=\angfig,width=\widthfig\textwidth]{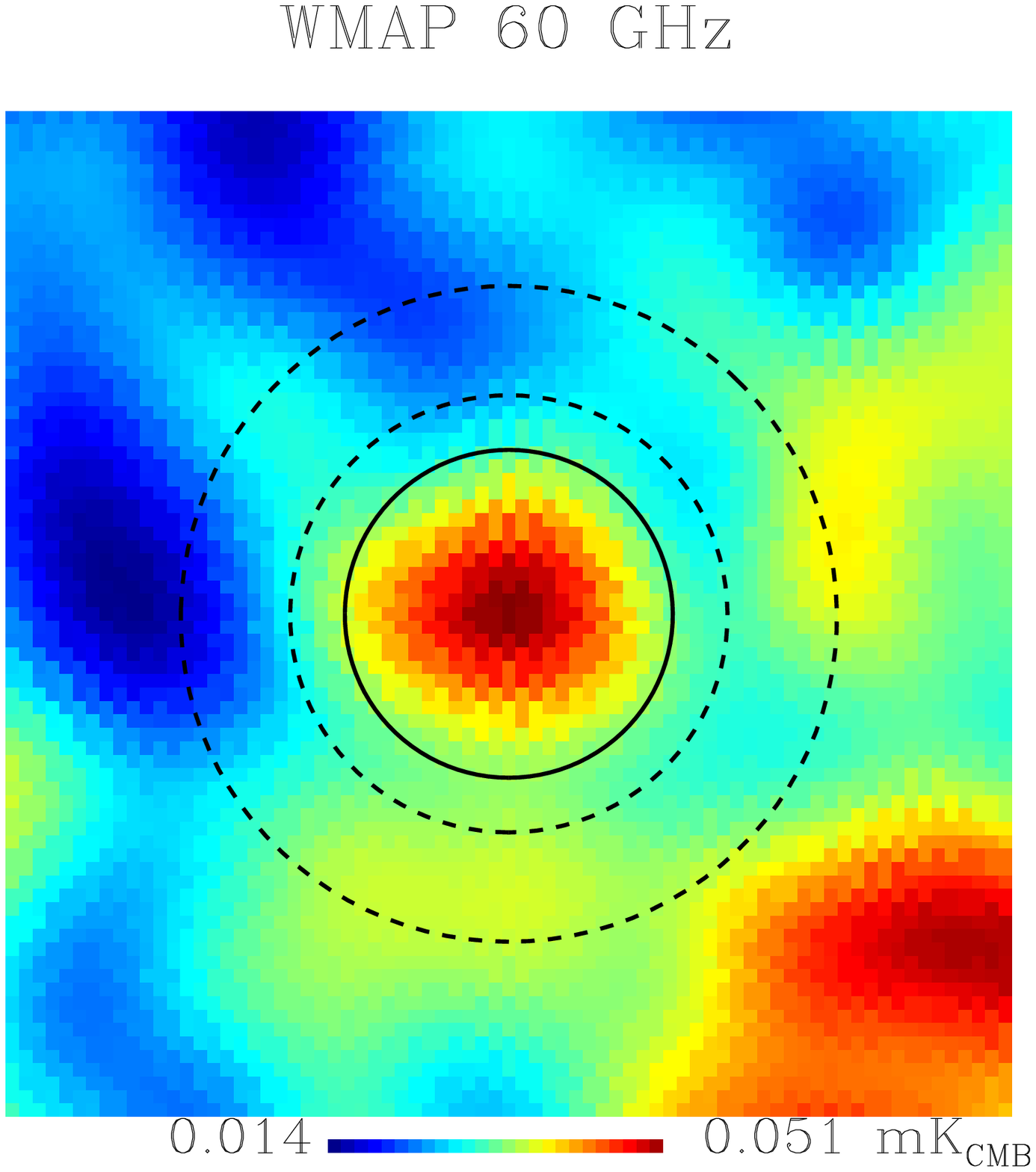}
  \includegraphics[angle=\angfig,width=\widthfig\textwidth]{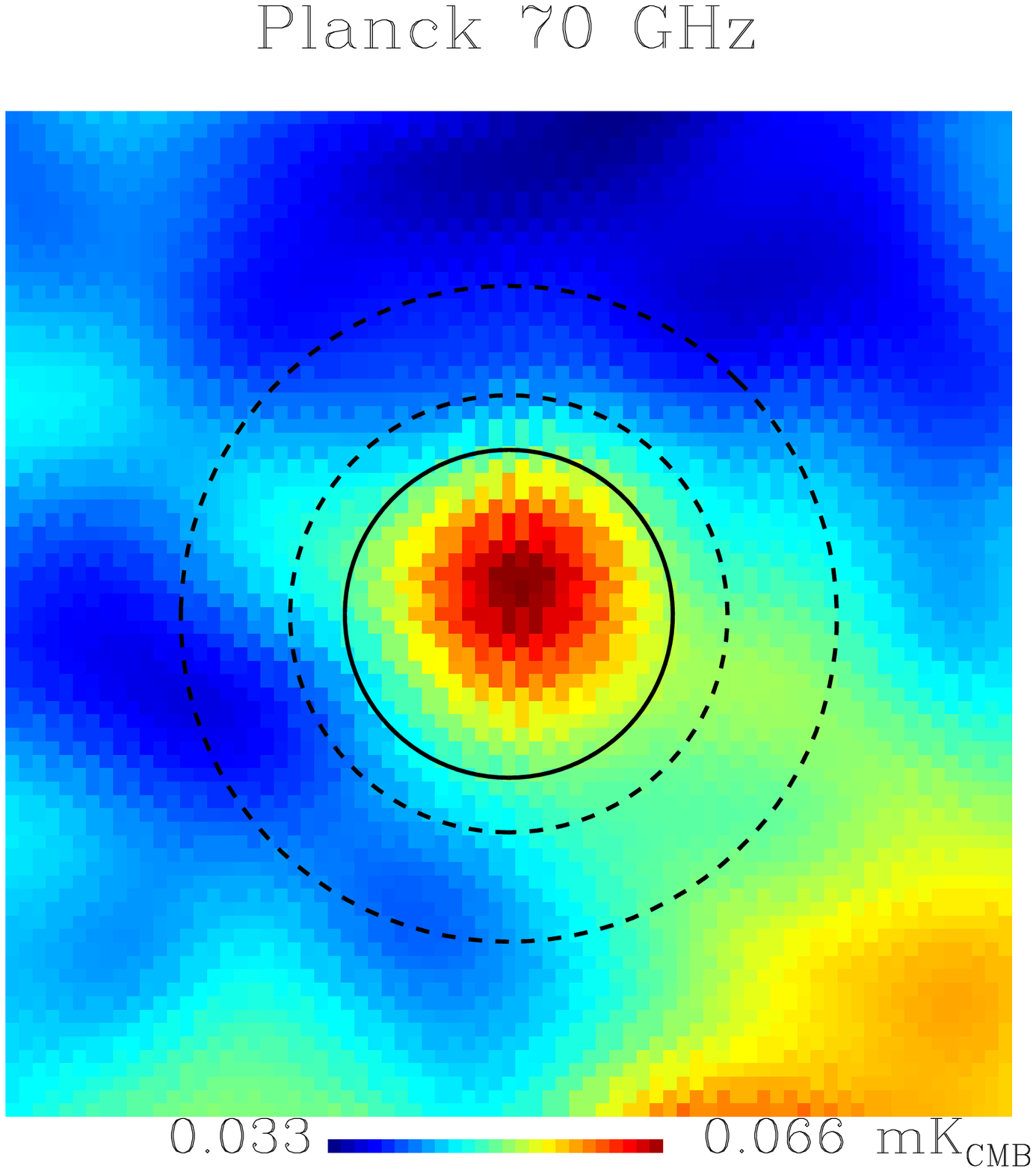}
  \includegraphics[angle=\angfig,width=\widthfig\textwidth]{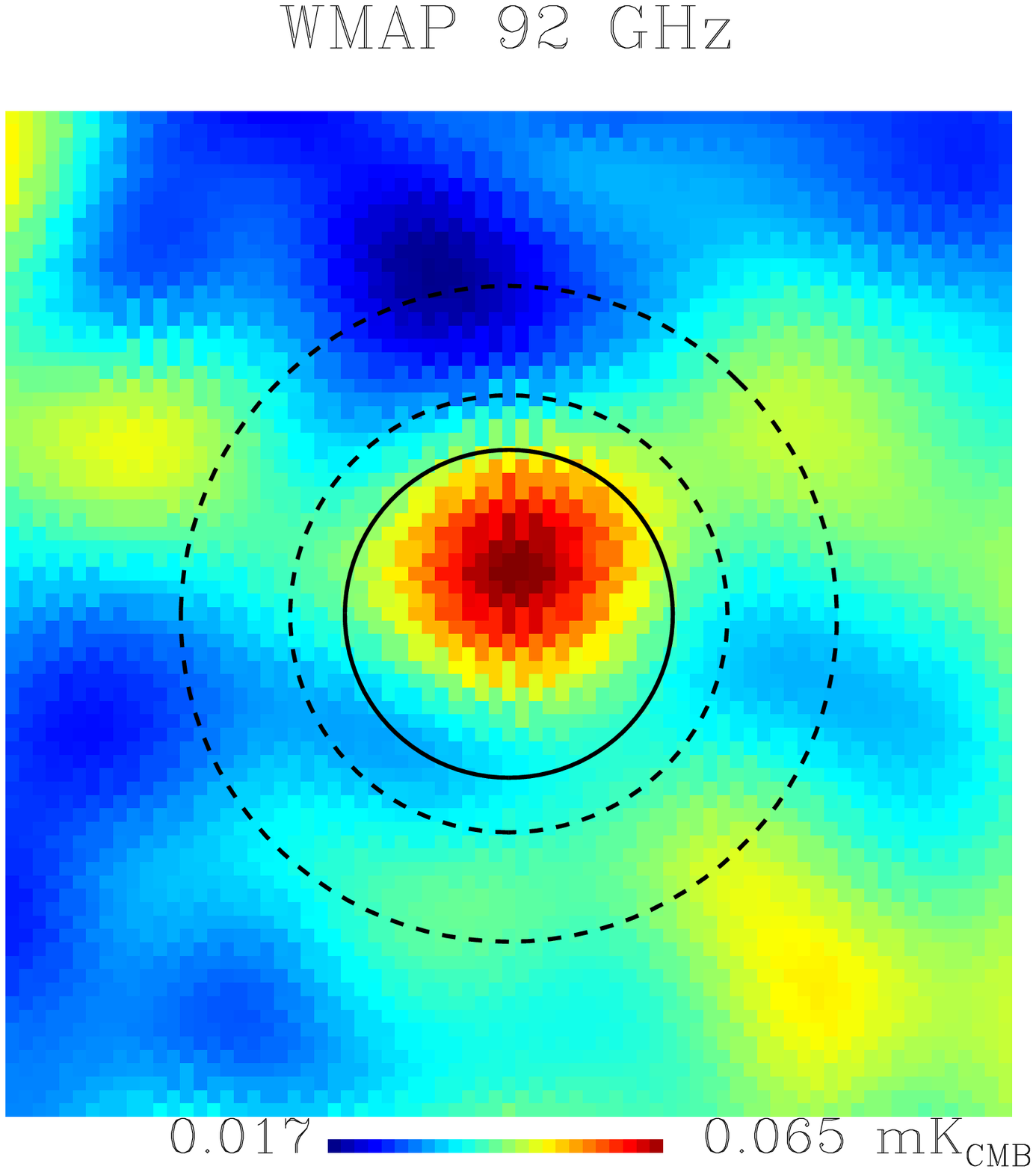}
  \includegraphics[angle=\angfig,width=\widthfig\textwidth]{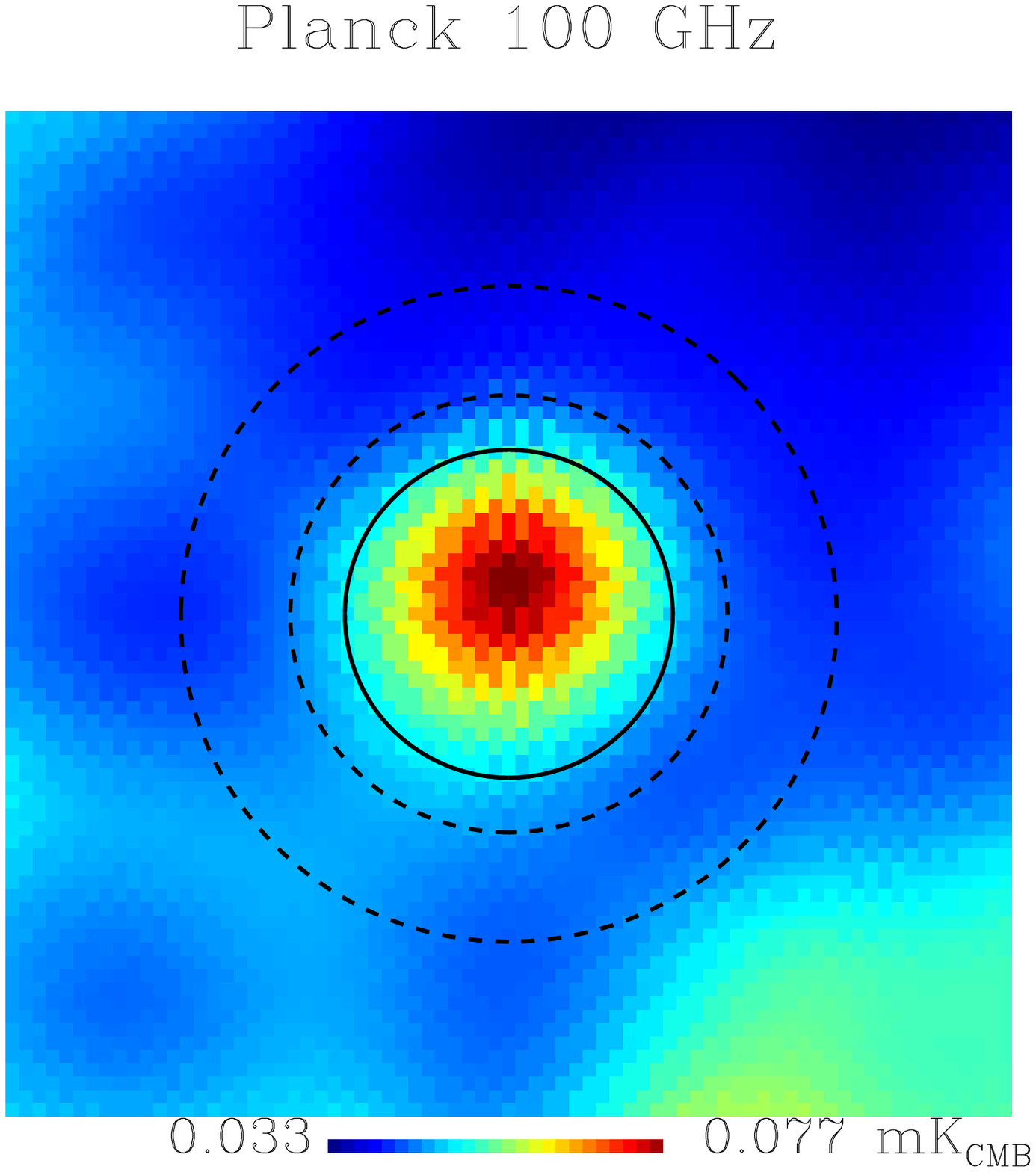}
  
  \includegraphics[angle=\angfig,width=\widthfig\textwidth]{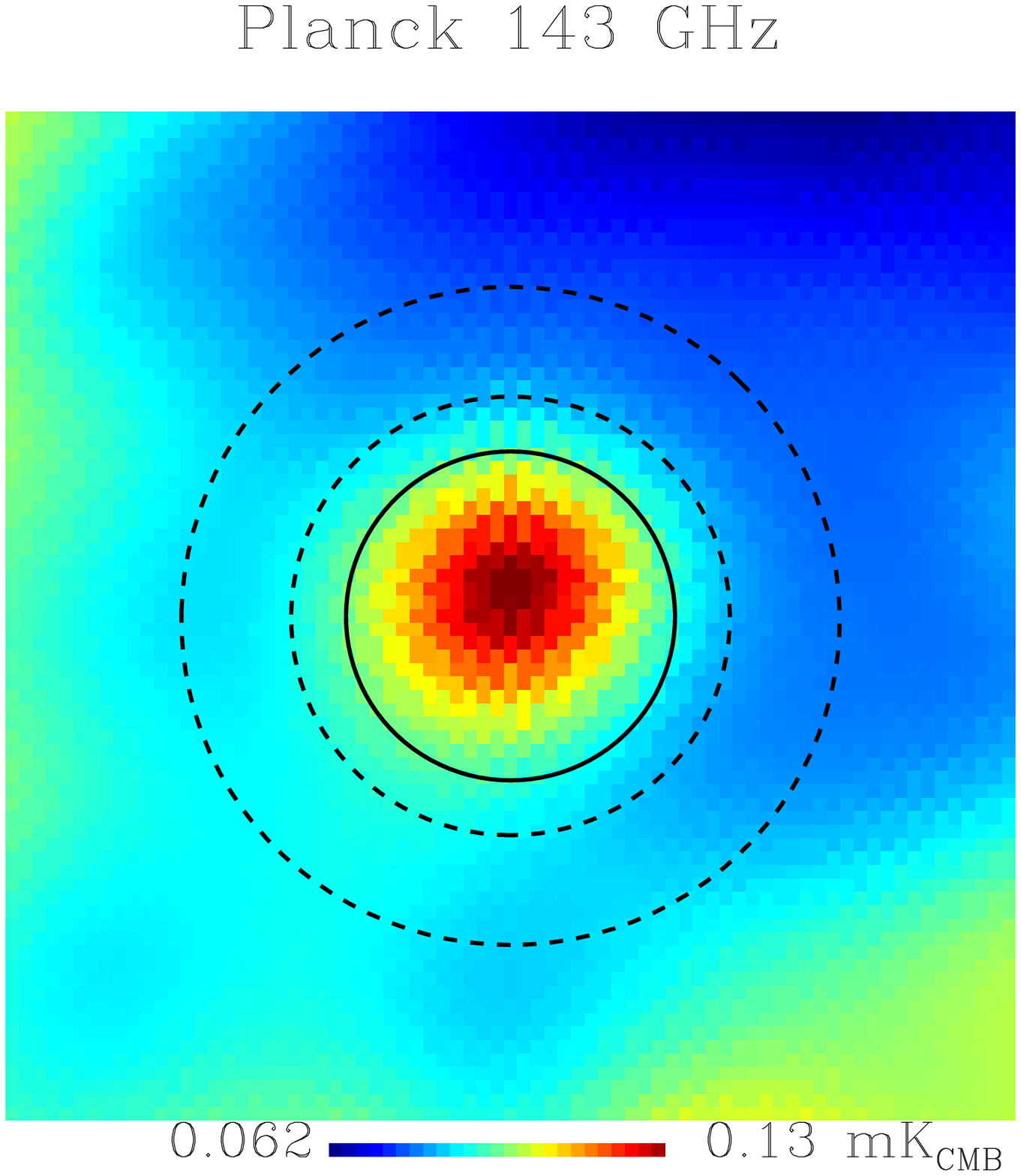}
  \includegraphics[angle=\angfig,width=\widthfig\textwidth]{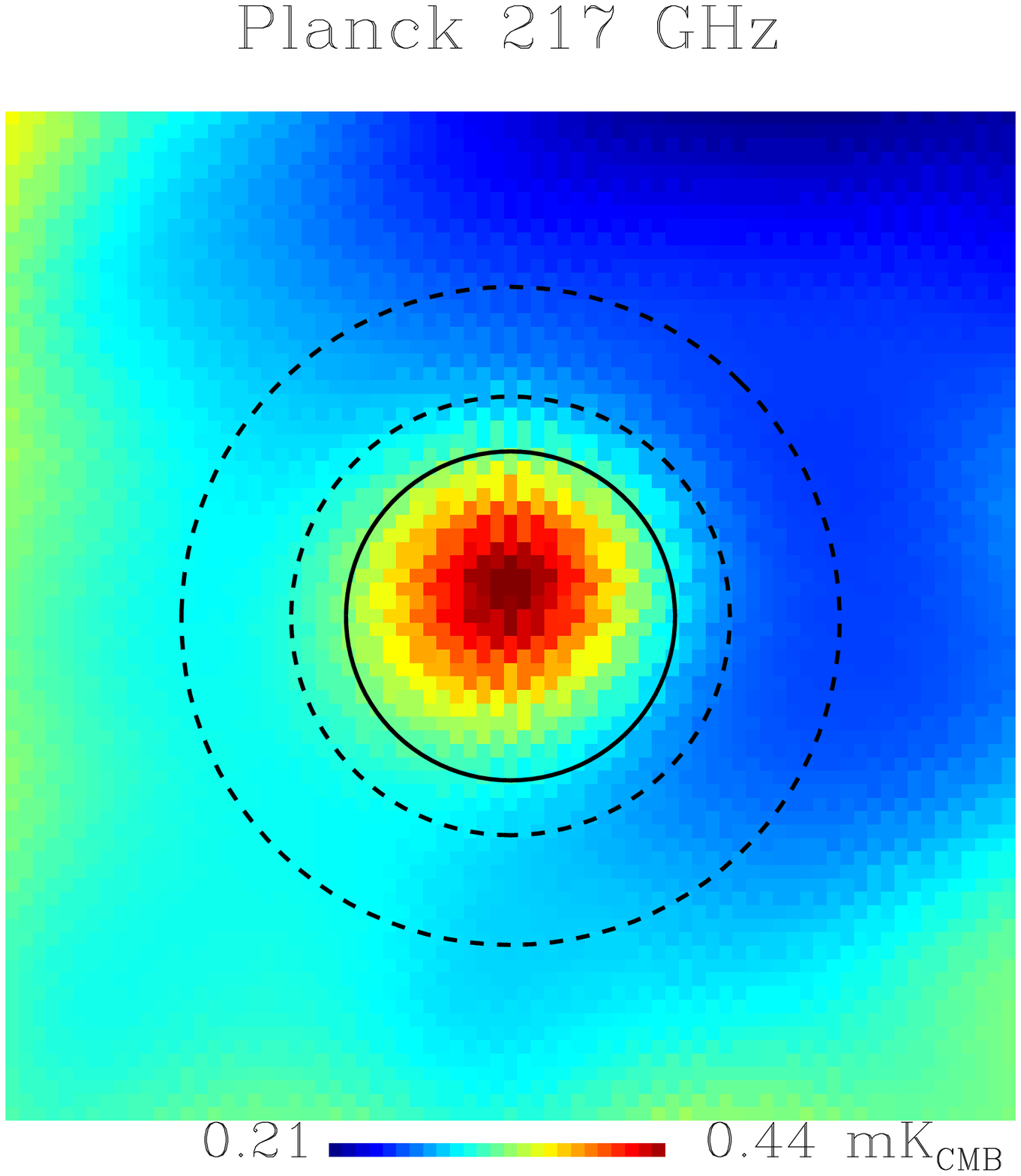}
  \includegraphics[angle=\angfig,width=\widthfig\textwidth]{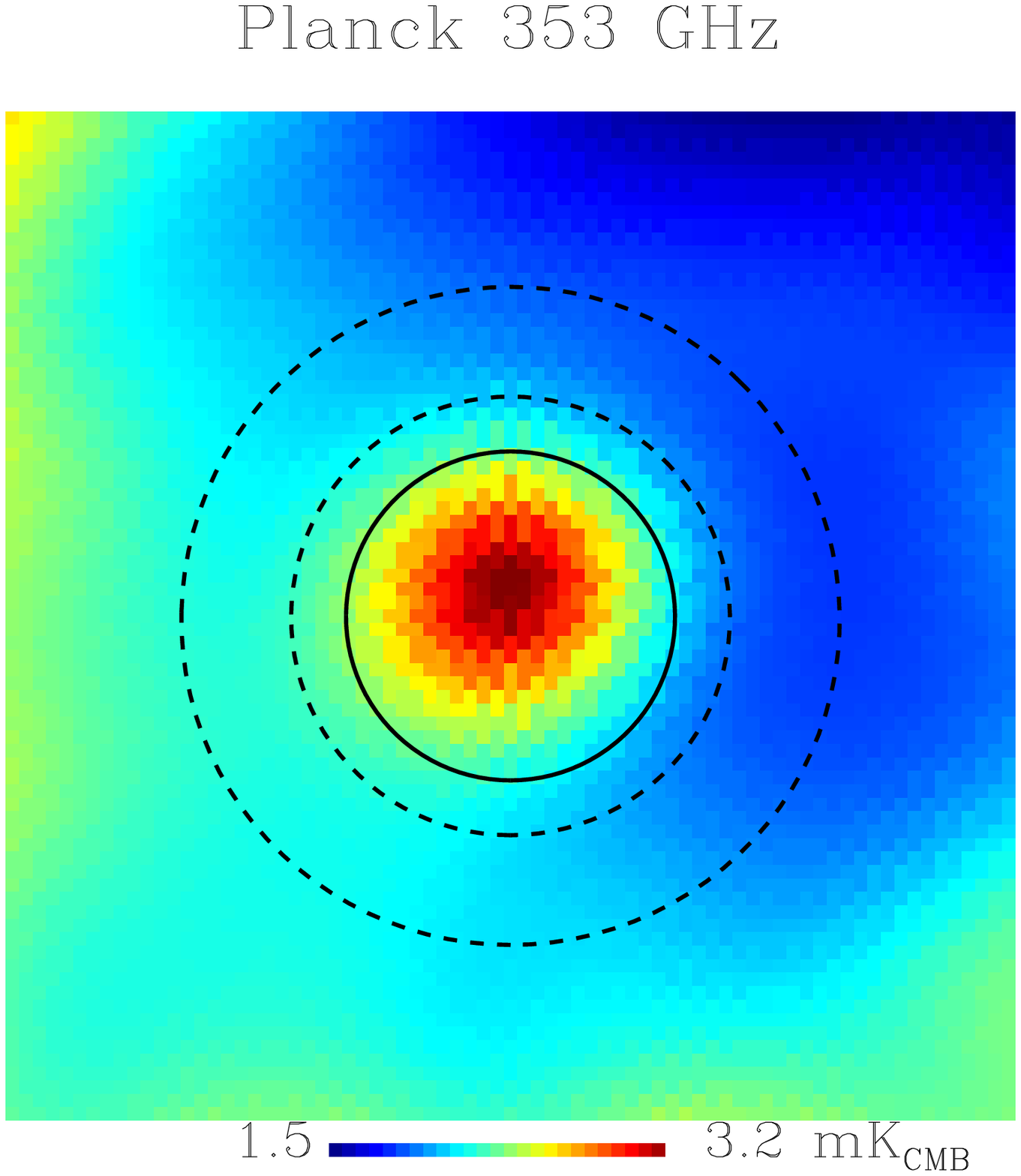}
  \includegraphics[angle=\angfig,width=\widthfig\textwidth]{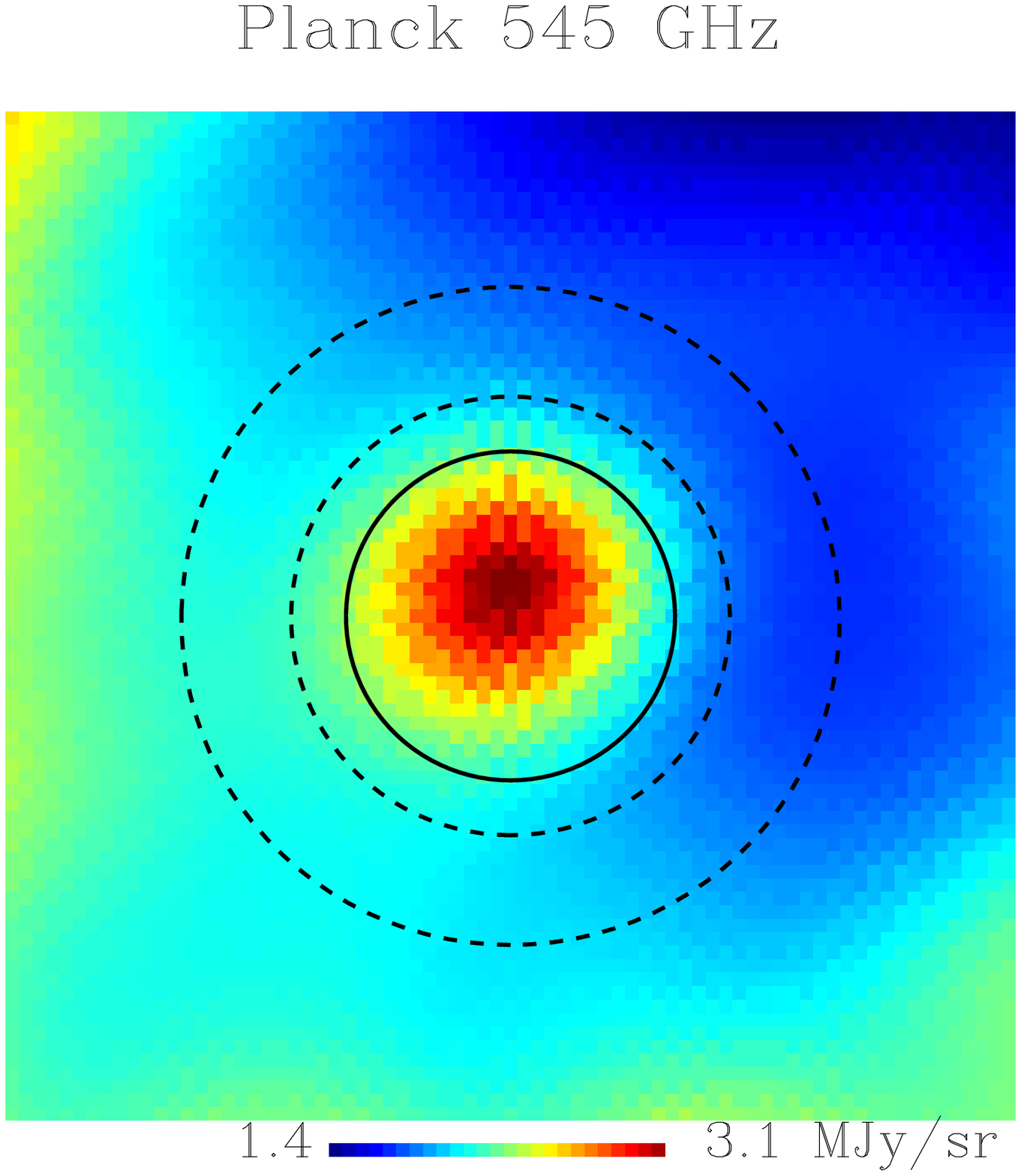}

  \includegraphics[angle=\angfig,width=\widthfig\textwidth]{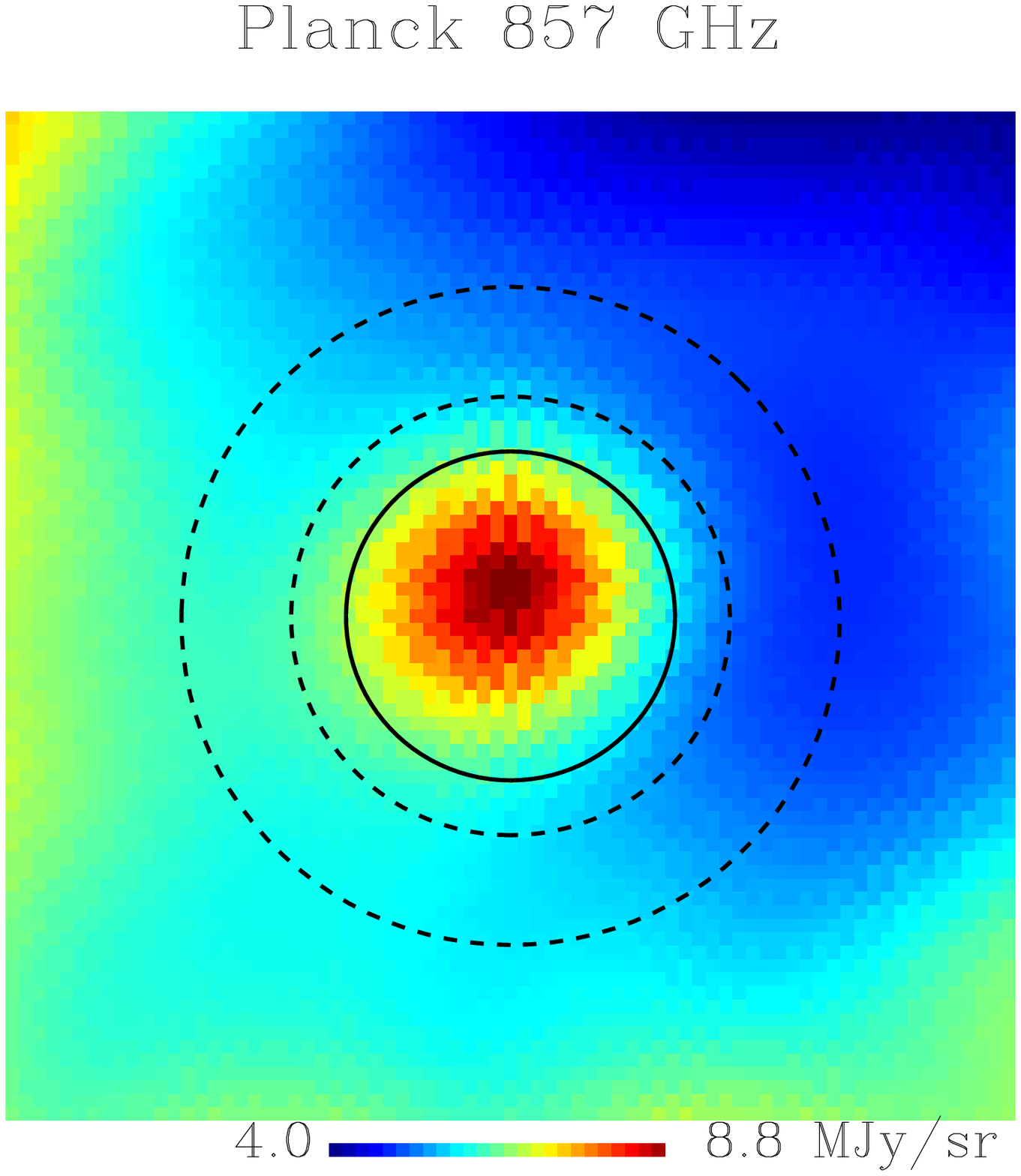}
  \includegraphics[angle=\angfig,width=\widthfig\textwidth]{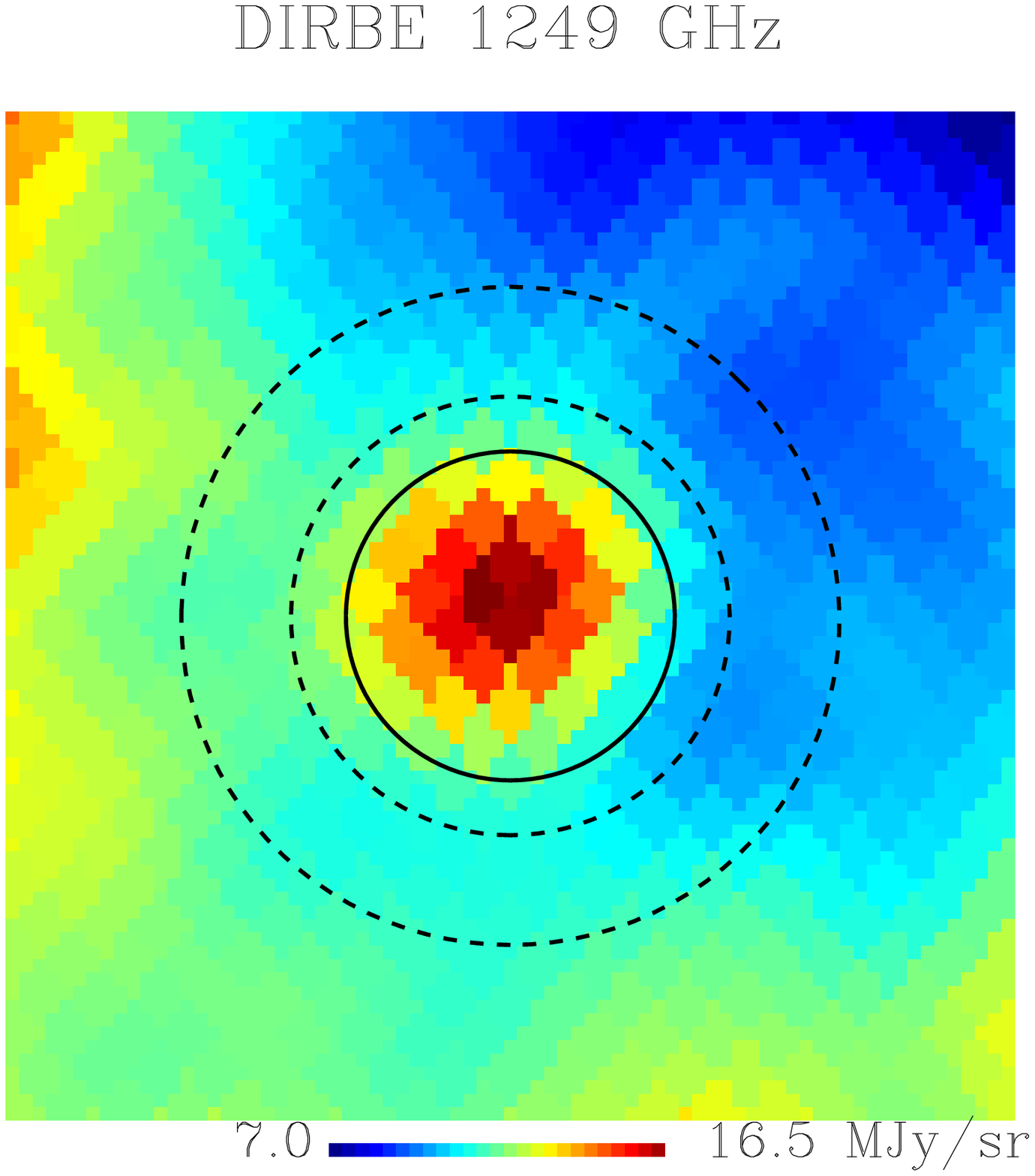}
  \includegraphics[angle=\angfig,width=\widthfig\textwidth]{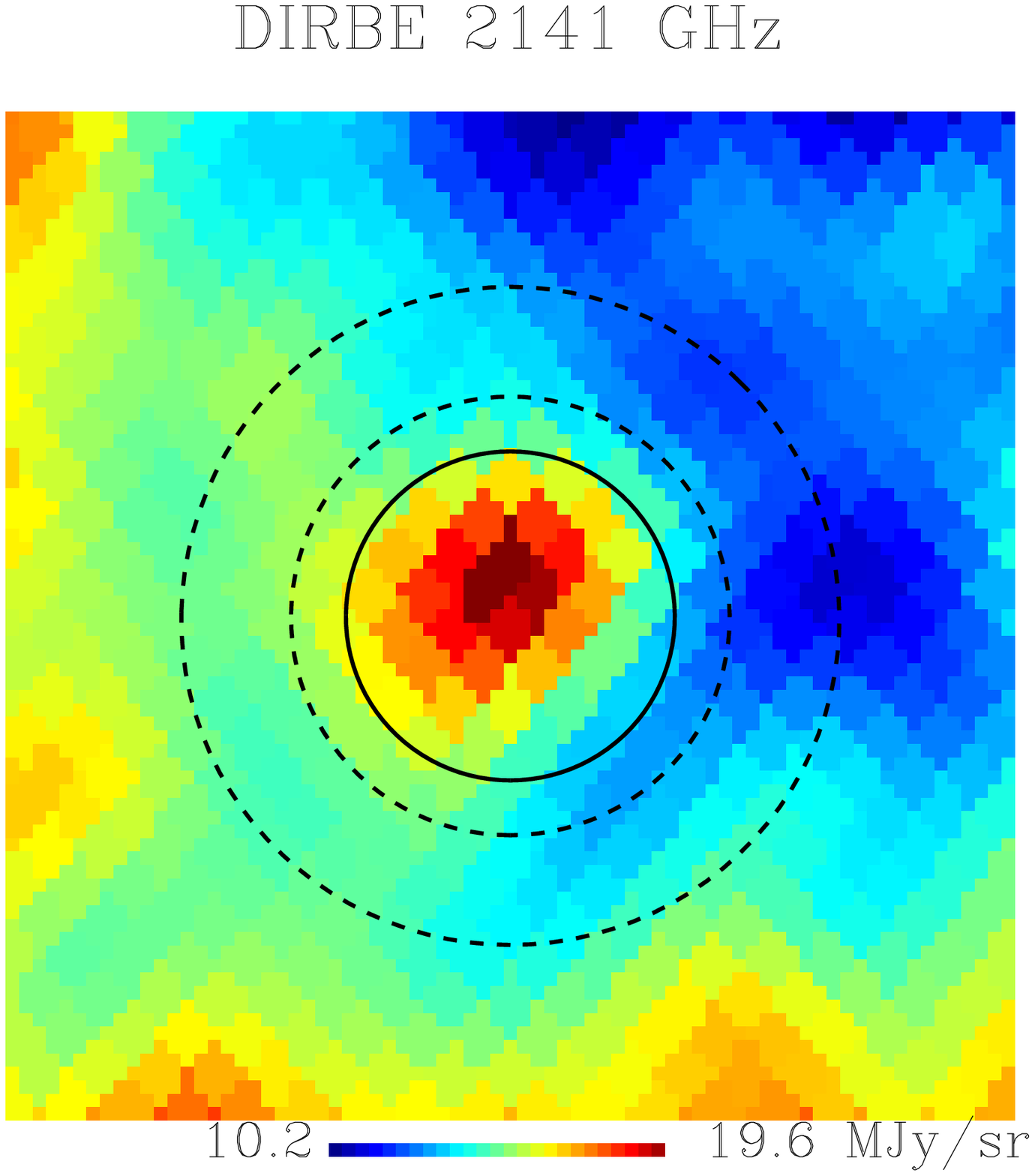}
  \includegraphics[angle=\angfig,width=\widthfig\textwidth]{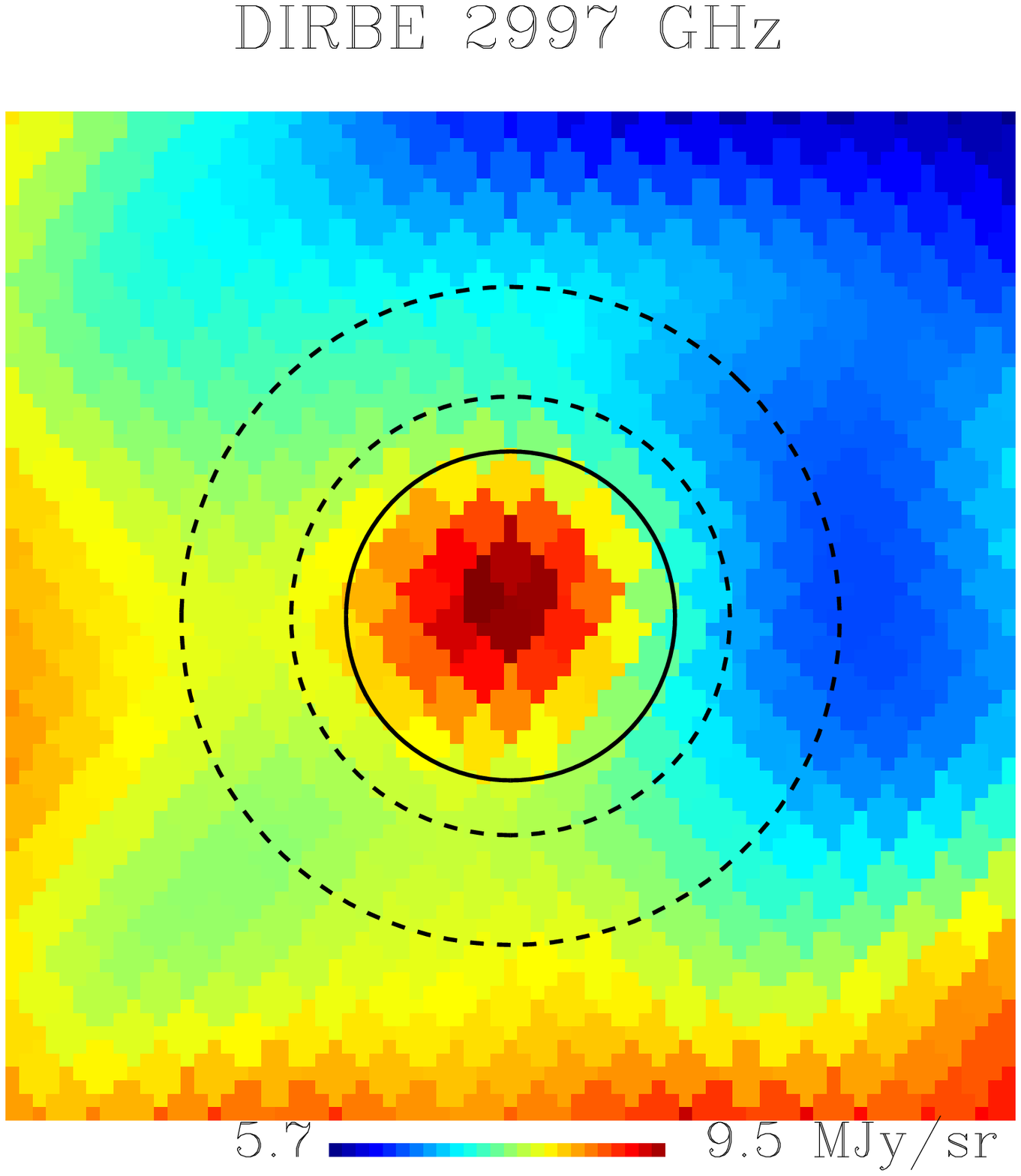}
 
\caption[CMB-subtracted maps of LDN\,1780 at different frequencies
  ranging from the Haslam et al. map at 0.408\,GHz to the DIRBE image at
  2997\,GHz.]{CMB-subtracted maps of LDN\,1780 at different frequencies
  ranging from the Haslam et al. map at 0.408\,GHz to the DIRBE image at
  2997\,GHz. The square maps have 5\dg in side and all have been
  smooth to a common angular resolution of 1\dg FWHM. The colour scale
  is linear, ranging from the minimum to the maximum of each map. The
  inner circle shows the aperture where we measured the flux and the
  dashed larger circles show the annulus used to estimate the
  background emission and noise around the aperture. Compare these 
  CMB-subtracted maps with the original ones shown in Fig. \ref{fig:l1780_maps}, 
  in particularly between 23 and 217\,GHz, where the CMB anisotropy dominates.}
  \label{fig:l1780_maps_cmbsub}
\end{figure*}

A main goal of the \planck\ and \wmap\ mission was to produce an
accurate map of the CMB fluctuations.  We use the {\sc SMICA} CMB map
\citep{planck_cmb_maps:2016} to subtract the CMB anisotropy from the
individual frequency maps. By doing this, LDN\,1780 cloud is
recognisable above the background in all the maps between 23 and
2997\,GHz. Fig. \ref{fig:l1780_maps_cmbsub} show these CMB-subtracted
maps. In them, LDN\,1780 is easily discernible. The peak position of
the source varies slightly among some of the maps
(e.g. \wmap\ 23\,GHz, \planck\ 545\,GHz).  We will discuss this in
Sect. \ref{sec:peak_location}.

Additional structure can be seen around LDN\,1780 at the 40--92\,GHz
CMB subtracted maps in Fig. \ref{fig:l1780_maps_cmbsub}. We measured
the standard deviation of these fluctuations around the cloud in the
three mentioned maps, using a ring with an inner radius of 1\dg\!
centred at the location of LDN\,1780, and a thickness of
3$^{\circ}$\!. We compared the standard deviation within the ring with
the r.m.s. noise of each map in the same ring. The r.m.s. noise was
calculated using 500 simulations of pure noise for each map,
constructed using the variance maps provided by the \wmap\ and
\planck\ collaborations. Table \ref{tab:stdev_noise_l1780} lists the
standard deviation values around LDN\,1780 in the \planck-44\,GHz,
\wmap-60\,GHz and \planck-70\,GHz CMB subtracted maps, as well as the
r.m.s. noise values of each of those data sets. The r.m.s. noise
values can account for at least 50\% of the measured fluctuations
around LDN\,1780. The additional residual fluctuations are a
combination between uncertainties in the CMB map and additional
diffuse foregrounds fluctuations. We measured the mean standard
deviation between four different CMB maps provided by the
\planck\ collaboration: Commander, , NILC, SEVEM and SMICA
\citep{planck_cmb_maps:2016}. In the same aperture, the fluctuations
between these maps average a value of 5.6\,$\mu$K. This value can be
used as a measurement of the uncertainty of the CMB map in this
region. This noise in the CMB map, in addition with the r.m.s. noise
value of the maps can account for the measured fluctuations around
LDN\,1780.

\begin{table}
  \centering
  \caption[]{Standard deviation of the fluctuations visible around
    LDN\,1780 in three of the CMB subtracted maps from
    Fig. \ref{fig:l1780_maps_cmbsub}. They are measured within a ring
    centred at the cloud, with an inner radius of 1\dg and an outer
    radius of 3\dg\!\!.  The second column shows the r.m.s. noise
    values measured in the same ring. }
  \begin{tabular}{l cD{.}{.}{-1} cD{.}{.}{-1}}
    \toprule
    \textit{Map} 
    & \multicolumn{1}{c}{Standard deviation}
    & \multicolumn{1}{c}{r.m.s. noise}\\ 
    & \multicolumn{1}{c}{$\mu$K}
    & \multicolumn{1}{c}{$\mu$K} \\
    \midrule
    \planck\  44\,GHz  & $ 12.0 $ &  5.9   \\
    \wmap\    60\,GHz  & $ 7.6  $ &  4.7   \\
    \planck\  70\,GHz  & $ 7.0  $ &  4.3   \\ 
    \bottomrule
  \end{tabular}
  \label{tab:stdev_noise_l1780}
\end{table}

In Table \ref{tab:fluxes_deg_scale} we list the flux densities
measured using the aperture photometry. We also list the values for
the flux densities measured in the CMB-subtracted map. The fluxes
measured at the three lowest frequencies were negative, i.e. the
background level in the ring is larger than the flux in the
aperture. Because of this, we give a $2\,\sigma$ upper limit for each
one of these points. In the next section we describe the SED fitting
to the values listed in Table \ref{tab:fluxes_deg_scale}.

\begin{table}
\centering
\caption{Flux densities of \ldn over a 2\dg diameter aperture.}
\begin{tabular}{lcHH}
\toprule
\textit{Survey}
   & \multicolumn{1}{c}{Frequency}
   & \multicolumn{1}{c}{Flux density} 
   & \multicolumn{1}{c}{CMB-sub flux density}\\
   & \multicolumn{1}{c}{[GHz]}
   & \multicolumn{1}{c}{[Jy]}
   & \multicolumn{1}{c}{[Jy]}\\
     \midrule
  Haslam  &      0.4 &  \multicolumn{1}{c}{$ < 2.2$}      &   \multicolumn{1}{c}{$ < 2.2$}     \\
  DRAO    &      1.4 &  \multicolumn{1}{c}{$ < 0.14$}       &  \multicolumn{1}{c}{$ < 0.14$}     \\
  HartRao &      2.3 &  \multicolumn{1}{c}{$ < 0.12$}  &  \multicolumn{1}{c}{$ < 0.12$}          \\
      WMAP &       23 &    1.4 ,  0.05   &    1.1 ,  0.06 \\    
  $Planck$ &       30 &    1.6 ,  0.09   &    1.1 ,  0.07 \\ 
      WMAP &       33 &    1.3 ,  0.10   &    0.8 ,  0.08 \\ 
      WMAP &       41 &    1.4 ,  0.16   &    0.8 ,  0.11 \\ 
  $Planck$ &       44 &    1.5 ,  0.18   &    0.8 ,  0.12 \\ 
      WMAP &       61 &    1.7 ,  0.33   &    0.7 ,  0.22 \\
  $Planck$ &       70 &    2.8 ,  0.42   &    1.4 ,  0.27 \\ 
      WMAP &       93 &    3.9 ,  0.70   &    1.7 ,  0.48 \\ 
  $Planck$ &      100 &    5.5 ,  0.72   &    3.0 ,  0.45 \\ 
  $Planck$ &      143 &    10.8 ,   1.1  &    6.5 ,  0.79 \\ 
  $Planck$ &      217 &    33.8 ,   1.4  &    27.8 ,   1.6 \\
  $Planck$ &      353 &     123 ,     2  &     124 ,     4 \\
  $Planck$ &      545 &     407 ,    10  &     407 ,    13 \\
  $Planck$ &      857 &    1151 ,    31  &    1151 ,    40 \\
     DIRBE &     1249 &    1860 ,    62  &    1860 ,    81 \\
     DIRBE &     2141 &    1639 ,    83  &    1639 ,   108 \\
     DIRBE &     2997 &     749 ,    35  &     749 ,    46 \\
\bottomrule
 \end{tabular}
\label{tab:fluxes_deg_scale}
\end{table}

\subsection{SED fitting}

SEDs in this frequency range are usually modeled using five
components, namely synchrotron, free-free, AME, CMB and thermal-dust
emission. In this case, we do not include a synchrotron component, due
to the small flux densities measured at the lower frequency bands. Our
model for the flux densities in a 2\dg aperture is therefore described
by four components,
\begin{equation}
  S = S_{ff} + S_{AME} + S_{CMB} + S_{TD}.
  \label{eq:sed_ldn}
\end{equation}

The free-free level in LDN\,1780 itself is very low. The H$\alpha$
line can be used as a tracer of free-free emission provided that the
line is the result of {\em in situ} recombination.  There is some
H$\alpha$ emission coming from the cloud but \citet{Witt:10} showed
that most of it consists of scattered light from the diffuse H$\alpha$
component of the galactic interstellar radiation field (ISRF). We
include a conservative upper limit for the free-free component in the
SED fitting, using the estimation at 31\,GHz over a 1\dg scale of
$S_{31} = 0.09$\,Jy from \citet{Vidal2011}, that was calculated using the
H$\alpha$ map from the SHASSAA survey \citep{Gaustad2001}.

We extrapolate this value to lower and higher frequencies using a
power law with the form,
\begin{equation}
  S_{ff} = S_{31}\,(\nu/31\,\mathrm{GHz})^{\beta_{ff}},
  \label{eq:ff_sed}
\end{equation}
where $\alpha_{ff} =-0.13$ is the free-free spectral index (for flux
density) valid for the diffuse ISM \citep{Draine_book}. 

The AME component is accounted for using a spinning dust model,
provided by the {\sc SPDUST} package \citep{ali:09,silsbee:11}. This
program calculates the emissivity, $j_{\nu}$, in terms of the hydrogen
column density of a population of spinning dust grains. It requires a
number of physical parameters to generate the spectrum. We used the
ideal description for the ``warm neutral medium'' (WNM) as defined by
\citet{DL98b}.  The generated spectrum produces peaks at 23.6\,GHz. We
fit for the amplitude of this generic spectrum, so this component in
the SED has only one free parameter, $A_{sd}$. In Section
\ref{sec:spdust_modelling} we describe in more detail the {\sc SPDUST}
modelling.

A CMB component is included, using the differential form of a
blackbody at $T_{\mathrm{CMB}} =2.726\,\mathrm{K}$
\citep{fixsen:09}. The flux density of this component has the form
\begin{equation}
  S_{\mathrm{CMB}} = \left( \frac{2 \,k\,\nu^2\,\Omega}{c^2} \right)
  \Delta \mathrm{T_{\mathrm{CMB}}},
\end{equation}
where $\Delta \mathrm{T_{\mathrm{CMB}}}$ is the CMB anisotropy
temperature, in thermodynamics units.
 
The dust emission at wave lengths $\lambda > 60$\,$\mu$m is usually
described using a modified blackbody model. The flux density measured
in a solid angle $\Omega$ is,
\begin{equation}
  S_{TD} = 2\,h\,\frac{\nu^3}{c^2}\frac{1}{e^{h\nu / k T_d}
    -1}\,\tau_{250}(\nu/1.2\,\mathrm{THz})^{\beta_d}\,\Omega,
  \label{eq:mod_bb}
\end{equation}
where $k$, $c$ and $h$ are the Boltzmann constant, the speed of light
and the Planck constant respectively; $T_d$ is the dust temperature
and $\tau_{250}$ is the optical depth at $250\,\mu$m.

We used the {\sc MPFIT} IDL package \citep{markwardt:09}, which uses
the Levenberg--Marquardt algorithm to calculate a non-linear
least-squares fit. There are two \planck\ bands, centered at 100\,GHz
and at 217\,GHz, which can include a significant amount of CO line
emission \citep{planck_XIII_CO:13}, corresponding to the transitions
$J$=1$\rightarrow$0 at 115\,GHz and $J$=2$\rightarrow$1 at
230\,GHz. LDN\,1780 is known to have a molecular component
\citep{laureijs:95}. To avoid contamination from these lines in the
fluxes measured at these bands, we did not include these two channels
in our fit.

\begin{figure}
  \centering
  \includegraphics[angle=90,width=0.47\textwidth]{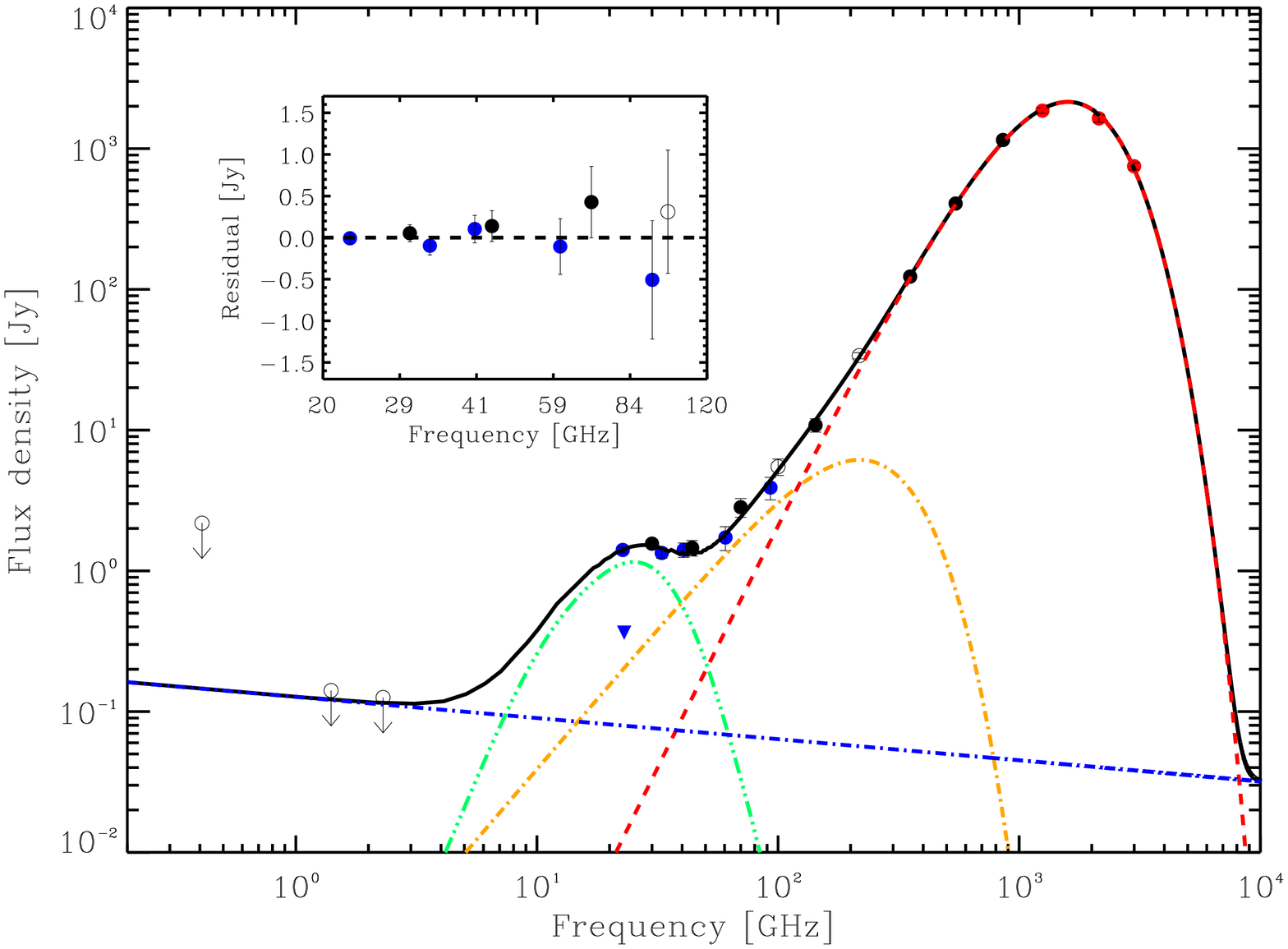}
  \includegraphics[angle=90,width=0.47\textwidth]{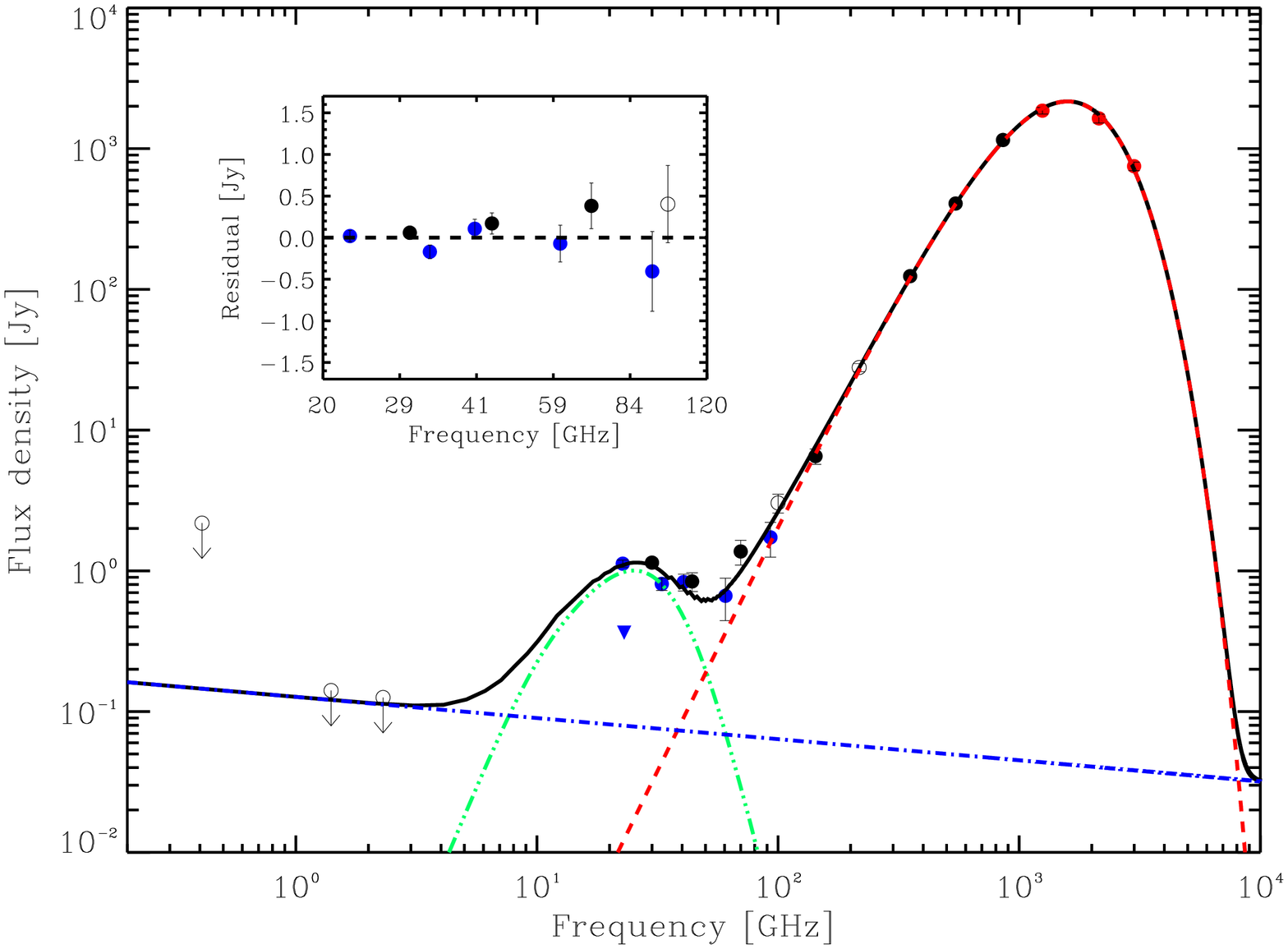}
  \caption[SED of LDN\,1780 in a 1\dg aperture]{Spectra of LDN\,1780
    including low frequency data, \wmap\ (blue dots), \planck\ (black
    dots) and \cobe-DIRBE~data (red dots). The black line is the best
    fit to the data, which includes four components. Thermal dust
    emission is represented with a red dashed line. A CMB component is
    shown in orange. The blue line represent an upper limit for the
    free-free emission as expected from the \halpha~map. The blue
    triangle shows the free-free emission expected from the
    \wmap\ 9-yr MEM template. The green line represents the spinning
    dust model. The two points that can be contaminated by CO line
    emission, at 100\,GHz and at 217\,GHz, are shown as a empty black
    circle.  The fits do not include a synchrotron component. The
    insert shows the residuals (data--model) around the region where
    the spinning dust component is important. The residuals are
    consistent with zero. The {\it bottom } panel shows the SED
    constructed using the CMB-subtracted maps. }
  \label{fig:l1780_sed_large}
\end{figure}

In the top panel of Fig. \ref{fig:l1780_sed_large} we show the best
fit to the data while the bottom panel shows the best fit to the
CMB-subtracted data.  In both plots, the low frequency data are
represented with $2\sigma$ upper limits. The largest uncertainty at
0.408\,GHz comes from the $\pm0.8$\,K striations measured by
\citet{Remazeilles2015}. The blue triangle at 23\,GHz represents the
expected free-free level predicted by the \wmap\ MEM map
\citep{Bennett2013}. A small CO contribution can be seen at 100\,GHz
and at 217\,GHz in the CMB-subtracted SED, however its flux is less
than 10\% at 100\,GHz. Being such at small effect at 100 and 217\,GHz
means that it will be negligible at 353\,GHz, so it is safe that we
have use the 353\,GHz map in our SED.  In Table \ref{tab:sed_fit_pars}
we list the parameters and uncertainties derived from the fit.

\begin{table}
\centering
\caption{Fitted parameters for the SED of \ldn. Also listed is the
  reduced $\chi^2$ of the fit. The second column lists the parameters
  of the fit using the CMB-subtracted maps.  }
\begin{tabular}{lcc}
\hline
\hline
Parameter & Normal & No--CMB \\
\hline
$\tau_{250} [\times 10^{-5}]$          & $  2.1 \pm  0.2$     &   $  2.2 \pm   0.3$ \\
T$_{d}$\,[K]                         &   $ 17.1\pm 0.4$      &   $ 16.9 \pm   0.5$ \\
$\beta_{d}$                          &  $ 1.5 \pm  0.1$      &   $ 1.5 \pm  0.1$ \\
A$_{sd} [10^{20}\,\mathrm{cm}^{-2}]$   & $  2.4  \pm   0.1$    &   $  2.0 \pm   0.1$  \\
$\Delta T_{\mathrm{CMB}}$\,[$\mu$K]     & $ 13.3  \pm   1.7$    &  $ 2.3 \pm  1.3$ \\
$ \chi^2_r$                           &      0.9                 &        1.9     \\
\hline
\end{tabular}
\label{tab:sed_fit_pars}
\end{table}
  
The difference in the fitted parameters between the CMB-subtracted
data and the un-subtracted maps is small and consistent with zero
within the uncertainties. The CMB component fitted in the
CMB-subtracted maps is consistent with zero. The fact that the
measured $\Delta T_{\mathrm{CMB}}=0.5\pm0.4\,\mu$K is consistent with
zero shows consistency within the two fits. The $\chi^2_r$ is higher
when using the CMB-subtracted maps. A reason for this is that the
error bars of the data points in this case are smaller, because the
fluctuations in the annular aperture are much smaller after
subtracting the CMB anisotropy. In this cloud, the CMB contribution is
significant on scales $\sim1^{\circ}$, but we show that it is well
quantified.

These SEDs show that there is significant AME present in this
cloud. If we assume the spinning dust component of the fit to be zero,
the overall CMB-subtracted fit is extremely poor, giving a
$\chi^2_r=40$, compared with the case when we include the spinning
dust component ($\chi^2_r=40$). The amplitude of the spinning dust
component is A$_{sd} =2.4\pm0.1$ and A$_{sd} =2.0\pm0.1$ for the
original and CMB-subtracted maps respectively. This corresponds to a
significance of 24$\sigma$ and 20$\sigma$ respectively, making
LDN\,1780 one of the clearest examples of AME on 1\dg\ angular
scales. In the analysis by \citet{planck_sd:13}, LDN\,1780 was not
detected as an AME source. We believe that the reason is that
LDN\,1780 does not appear as a conspicuous source in the original
maps, and only shows clearly after subtracting a CMB template, as its
location is coincident with a high value of the CMB anisotropy.

Another important aspect to highlight is the lack of emission from the
cloud in the low-frequency maps ($\nu \le 2.3$\,GHz). This means that
LDN\,1780 is a rising spectrum source at $\nu >5$\,GHz. 

\subsection{Peak location}
\label{sec:peak_location}
\begin{figure}
  \centering
  \includegraphics[angle=-90,width=0.47\textwidth]{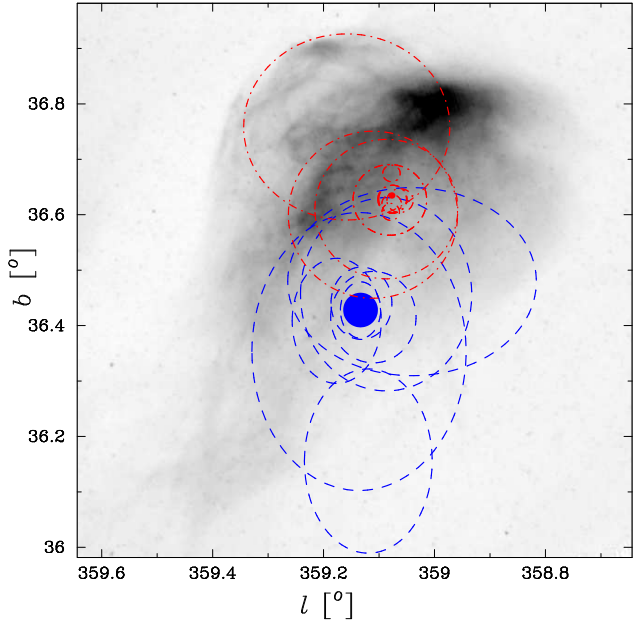}
  \caption{Ellipses centred at the location of the peak of the cloud
    for the WMAP, Planck and DIRBE maps. The size of the ellipses
    represent 1-$\sigma$ uncertainties, taken from Table
    \ref{tab:peak}. In {\it blue} are shown the frequencies between
    22.8\,GHz and 60.7\,GHz while in red the maps from 70.4 to
    2997\,GHz. The filled ellipses correspond to the averaged values
    shown also in Table \ref{tab:peak}.  On grey scale is the
    250\,$\mu$m Herschel map of LDN\,1780 in Galactic coordinates.  }
  \label{fig:peak_location}
\end{figure}

If we take a closer look to Fig. \ref{fig:l1780_maps_cmbsub}, we can
notice that the location of the peak of the cloud in the lower
frequencies (23--70\,GHz) is very close to the center of the
aperture. On the other hand, the cloud appears shifted north by a few
arcmin in the higher frequency maps (93--2997\,GHz). In order to
measure this shift, we measured the position of the peak of the cloud
in all the maps from 23 to 2997\,GHz using {\tt sextractor}
\citep{sextractor}. Table \ref{tab:peak} list the location and
uncertainties for the peak of the emission for all the maps between 23
and 2997\,GHz. In Fig. \ref{fig:peak_location} we plot these values
using coloured ellipses on top of the 250\,$\mu$m Herschel map
orientated in Galactic coordinates, where the {\it blue} ellipses
correspond to the maps from 22.8\,GHz to 60.7\,GHz, and the {\it red}
ellipses to the maps from 70.4\,GHz to 2997\,GHz. The averaged
position is also plotted for as a filled ellipse.

The location of the low-frequency (22.8--60.7\,GHz) peak is closer to
the peak of the IR emission originated from small grains
(e.g. 8\,$\mu$m, 12\,$\mu$m). This is most interesting as it is what
is expected from the spinning dust model. Moreover, this is also along
the direction of the local radiation field that illuminates the cloud,
which comes from the Galactic plane direction \citep{Witt:10}.  We
will explore further this morphological correlation using the CARMA
data in the next Section.

\begin{table}
  \centering
  \caption{Location of the peak of the cloud for WMAP, Planck and
    DIRBE maps in Galactic coordinates. We also list the averaged
    values for the maps in the range 70.4--2997\,GHz and
    22.8--60.7\,GHz.  }
  \begin{tabular}{lcccc}
    Map  &  Gal. Lon. [deg]  & Gal. Lat. [deg] \\
    \noalign{\hrule\vskip 6pt}
    DIRBE 2997& $  359.19 \pm     0.02$ & $   36.63\pm    0.02$ \\
    DIRBE 2141& $  359.17 \pm     0.02$ & $   36.67\pm    0.01$ \\
    DIRBE 1249& $  359.17 \pm     0.02$ & $   36.61\pm    0.02$ \\    
    Planck 857& $  359.17 \pm     0.02$ & $   36.62\pm    0.02$ \\
    Planck 545& $  359.18 \pm     0.03$ & $   36.63\pm    0.03$ \\
    Planck 353& $  359.17 \pm     0.03$ & $   36.63\pm    0.03$ \\
    Planck 217& $  359.17 \pm     0.07$ & $   36.63\pm    0.06$ \\
    Planck 143& $  359.16 \pm     0.13$ & $   36.61\pm    0.13$ \\
    Planck 100& $  359.13 \pm     0.15$ & $   36.60\pm    0.15$ \\
    WMAP 93.5&  $  359.07 \pm     0.19$ & $   36.76\pm    0.17$ \\
    Planck 70.4& $  359.10 \pm     0.19$ & $   36.35\pm    0.25$ \\
    \hline
    Averaged  & $ 359.077 \pm 0.002 $ & $ 36.635 \pm  0.002 $  \\
    \noalign{\hrule\vskip 6pt}
    WMAP 60.7& $  359.22 \pm     0.22$ & $   36.48\pm    0.17$ \\
    Planck 44.1& $  359.16 \pm     0.15$ & $   36.46\pm    0.17$ \\
    WMAP 40.7& $  359.12 \pm     0.12$ & $   36.16\pm    0.17$ \\
    WMAP 33& $  359.05 \pm     0.08$ & $   36.41\pm    0.11$ \\
    Planck 28.4& $  359.13 \pm     0.08$ & $   36.42\pm    0.08$ \\
    WMAP 22.8& $  359.11 \pm     0.05$ & $   36.44\pm    0.07$ \\
    \hline
    Averaged &    $ 359.13 \pm 0.04 $  &   $ 36.43  \pm 0.05 $ \\
    \hline
  \end{tabular}
  \label{tab:peak}
\end{table}

\section{Dust properties at 2$'$ resolution}
\label{sec:dust_properties}
Using the IR data available at angular resolutions similar to our
CARMA maps of 2$'$, we can obtain some physical properties of the
cloud, such as its temperature and column density. We do this by
fitting for the spectrum of the thermal dust emission (as defined in
Eq. \ref{eq:mod_bb}) in each pixel. For this fit, we use five data
points, at 70, 160, 250, 350, and 500\mic, which are dominated by
thermal dust.

\begin{figure}
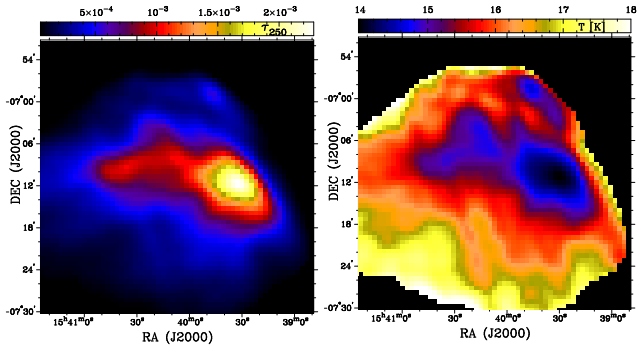

  \centering \newcommand{\widthfig}{0.235} \newcommand{\angfig}{-90}
  \includegraphics[angle=\angfig,width=\widthfig\textwidth]{figs/tau_map_midres_nosub.ps}
  \includegraphics[angle=\angfig,width=\widthfig\textwidth]{figs/temp_map_midres_nosub.ps}
    \caption{ Map of optical depth at 250\mic~({\it left}) and the
      dust temperature ({\it right}) for LDN\,1780. The coldest areas of
      the cloud, at $T_d\approx 14$\,K correspond to regions with
      larger optical depth. This is expected as the denser regions are
      more shielded from the radiation field.}
    \label{fig:l1780_tau_temp}
\end{figure}

Due to the smaller number of data points that we have here in
comparison to the previous fit at 1\dg scales (5 versus 20), we fix
the dust spectral index to $\beta_d=1.6$, a value similar to the one
measured in the 1\dg fit, and also, consistent with the values for the
diffuse medium measured by {\em Planck} \citep{planck_dust:13}. This
means that in this case our modified black-body fit only has two
parameters, the optical depth $\tau_{250}$ and the temperature of the
big grains. We have calculated the fit only where the signal-to-noise
ratio of the pixels is larger than 2. In Fig. \ref{fig:l1780_tau_temp}
we show the resulting map for the optical depth at 250\mic\ and for
the temperature of the dust. The colder regions corresponds to where
the optical depth is larger. This is expected as these regions are
more shielded from the ISRF.

From the optical depth map, we can obtain a hydrogen column density
map using the linear relation $\tau_{250}/N_H=
2.32\pm0.3\times10^{-25}$\,cm$^2$, as measured by
\citet{planck_extinction:11}.  From the temperature map, we can derive
a map for the radiation field. The radiation field $G_0$ can be estimated
using the following relation \citep{ysard:10a},
\begin{equation} 
  G_0=\left( \frac{T_d}{17.5\,[\mathrm{K}]} \right)^{\beta_d + 4},
  \label{eq:G0}
\end{equation}
where the spectral distribution of the radiation field is assumed to
have the standard shape, defined in
\citet{mathis:83}. \citet{planck_dust:13} has shown that this relation
might not hold in every environment, and variations in $T_d$ might be
due to variations in dust properties, such as grain structure or size
distribution. We will use this map in the following Section.

\subsection{IR correlations}

Here we investigate correlations with the IR data. We include the
\spitzer-IRAC map at 8\mic, which traces PAHs, as well as the
\spitzer-MIPS map at 24\mic, tracing VSGs. Similar analyses can be
found in the literature and they show different results in different
types of clouds and in different angular scales.  \citet{Scaife2010}
found in the LDN\,1246 cloud that the 8\,$\mu$m \spitzer\ map was the
closest to their 16\,GHz observations. \citet{Casassus2006} and
\citet{Tibbs2011} reported better correlations between radio data and
60\,$\mu$m. On large areas of the sky, the \planck\ team finds that
the FIR map correlates better with the AME template
\citep{Planck2016XXV}. On LDN\,1780, \citet{Vidal2011} found that the
31\,GHz data from the CBI was closer to IRAS 60\,$\mu$m. Using a
full-sky analysis, \citet{HensleyDraine2016} found that, on average,
the best correlation of AME is with the dust radiance map.

Here, to calculate the spatial correlations, we selected a rectangular
region of $25' \times 15'$ around the centre of the CARMA mosaic. We
smoothed all the maps to a common 2$'$ resolution, the same as the
CARMA map at 31\,GHz. We used Spearman's rank correlation coefficient,
$r_s$, to quantify the correlation of the two maps in a pixel-by-pixel
comparison. This coefficient has the advantage, over the traditional
Pearson correlation coefficient, in that the relation between the two
variables that are being compared does not have to be linear. A
Spearman's rank of $r_s =1$ will occur when the two quantities are
monotonically related, even if this relation is not linear. The
uncertainties in $r_s$ are estimated using 1000 Monte Carlo
simulations, calculated using the uncertainties in the maps.

\begin{figure}
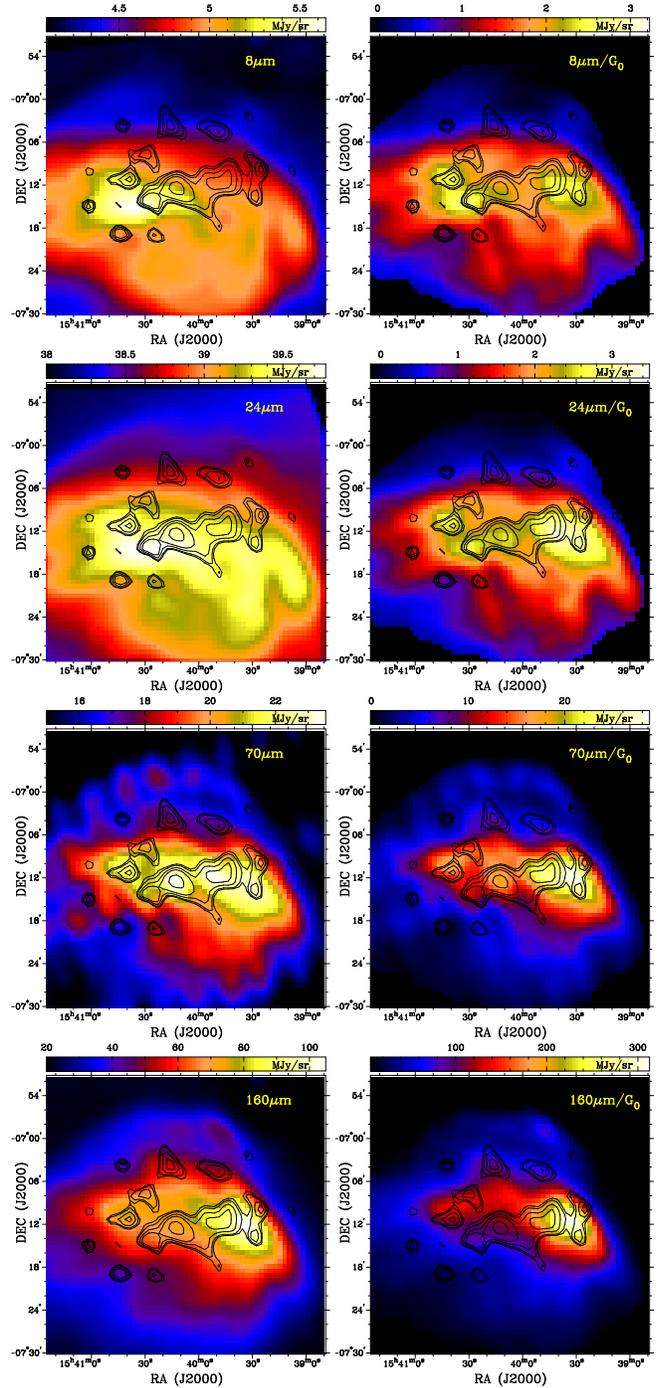
  
  \newcommand{\widthfig}{0.24} \newcommand{\angfig}{-90}
  \includegraphics[angle=\angfig,width=\widthfig\textwidth]{figs/irac_008_res_120_cont.ps}
  \includegraphics[angle=\angfig,width=\widthfig\textwidth]{figs/irac_008_res_120_bgsub_G0_corr_cont.ps}
  
  \includegraphics[angle=\angfig,width=\widthfig\textwidth]{figs/mips_024_res_120_cont.ps}
  \includegraphics[angle=\angfig,width=\widthfig\textwidth]{figs/mips_024_res_120_bgsub_G0_corr_cont.ps}
  
  \includegraphics[angle=\angfig,width=\widthfig\textwidth]{figs/mips_070_res_120_cont.ps}
  \includegraphics[angle=\angfig,width=\widthfig\textwidth]{figs/mips_070_res_120_bgsub_G0_corr_cont.ps}
  
  \includegraphics[angle=\angfig,width=\widthfig\textwidth]{figs/mips_160_res_120_cont.ps}
  \includegraphics[angle=\angfig,width=\widthfig\textwidth]{figs/mips_160_res_120_bgsub_G0_corr_cont.ps}
  \caption{NIR maps of LDN\,1780. The original maps are on the {\it left}
    and those corrected by the radiation field $G_0$ are on the {\it
      right}.  From {\it top} to {\it bottom}, the first row is 8\mic,
    the second 24\mic, the third 70\mic~and the fourth 160\mic.  The
    contours of the 31\,GHz CARMA map are overlaid on all the maps.
    The $G_0$-corrected 8\mic~and 24\mic~maps ({\it top} two) show a
    better correlation with the 31\,GHz data than the original
    8\mic~and 24\mic~maps. The opposite occurs for the 70\mic~and
    160\mic~({\it bottom} two). In this case, the correction for the
    ISRF results in a worse correlation with the 31\,GHz map compared
    to the original.  The quantitative values of the correlation are
    listed in Table \ref{tab:ldn_corr}.}
  \label{fig:ldn_8_24}
\end{figure}

We calculated $r_s$ between the 31\,GHz map and the IR templates at 8,
24, 70, 160, 250, 350 and 500\mic.  Because the IR emission of the
smallest grains depends on the radiation field, we also calculated the
correlation of the 31\,GHz data with the IR templates divided by the
radiation field map ($G_0\propto T_d^{5.6}$) to account for the
differences in the IRF across the cloud \citep{ysard:10a}. The maps
that are corrected by $G_0$ should be better tracers of the column
density of small grains than the original maps. In Table
\ref{tab:ldn_corr} we list the values of $r_s$ for the different IR
templates for both the original IR maps, the versions corrected by
$G_0$ and the ratio between both quantities.  Among the original maps,
the best correlation is between the 31\,GHz map and
70\mic~template. The NIR maps at 8 and 24\mic~present the lowest
correlation coefficient, and the maps at 160, 250, 350 and
500\mic~show similar $r_s$ as expected, as these maps are tracing the
same population of large grains. After dividing the IR maps by the
$G_0$ template, the correlation with the NIR maps at 8 and
24\mic~improves by a factor of 2-3. This increase is significant and
can be appreciated even by eye. In Fig. \ref{fig:ldn_8_24} we show on
the left column the original 8, 24, 70 and 160\mic\ maps of the
cloud. On the right column, we show the same maps after being divided
by the $G_0$ map. The $G_0$-corrected 8 and 24\mic\ maps present a
morphology closer to the 31\,GHz black contours. The correlation with
the longer wavelength maps ($\ge 70$\mic) degrades, but not
significantly after the $G_0$ correction.

The fact that the correlation improves significantly (by a factor 2.7
and 2.2, see Table \ref{tab:ldn_corr}) after the correction for the
radiation field illumination of the small grains, traced both by the
8\mic\ and 24\mic\ maps suggests that the emission seen at 31\,GHz
might be produced not only by the PAH's (traced by the 8\mic\ map) but
also by small and warmer grains more exposed to the external radiation
field, traced by the 24\mic\ map.

\subsection{Spinning dust modelling}
\label{sec:spdust_modelling} 
The peak of the 31\,GHz emission in LDN\,1780 is not coincident with
the regions with higher column density. This implies a larger radio
emissivity from the less dense regions, which can be due to either due
to a lack of small grains (e.g. due to coagulation of small grains to
big grains) or due to local enhancement of the environmental
conditions that trigger the spinning dust emission. Here we
investigate if such variations in emissivity can be explained using a
spinning dust model.

l

\begin{table}
\centering
\caption{Spearman's rank, $r_s$, between the 31\,GHz map and the
  different IR templates. The column on the centre shows the
  correlation value for the IR maps after they have been divided by
  the radiation field map, in order to account for variations in the
  illumination of the grains across the cloud. The column on the right
  shows the ratio between both quantities. }
\begin{tabular}{cccc}
\hline
\hline
 Wavelength & $r_s$    &   $r_s[G_0]$  & $r_s[G_0]/r_s$  \\
 $[\micron]$ & \\
\hline
$  8$ & $ 0.14 \pm  0.06$ & $ 0.38 \pm  0.07$    & $  2.7 \pm   0.5$ \\
$ 24$ & $ 0.21 \pm  0.06$ & $ 0.46 \pm  0.06$    & $  2.2 \pm   0.3$ \\
$ 70$ & $ 0.49 \pm  0.07$ & $ 0.45 \pm  0.07$    & $  0.9 \pm   0.2$ \\
$160$ & $ 0.36 \pm  0.07$ & $ 0.31 \pm  0.07$    & $  0.9 \pm   0.3$ \\
$250$ & $ 0.35 \pm  0.06$ & $ 0.31 \pm  0.07$    & $  0.9 \pm   0.3$ \\
$350$ & $ 0.34 \pm  0.06$ & $ 0.30 \pm  0.07$    & $  0.9 \pm   0.3$ \\
$500$ & $ 0.34 \pm  0.06$ & $ 0.30 \pm  0.06$    & $  0.9 \pm   0.3$ \\
\hline
 \end{tabular}
\label{tab:ldn_corr}
\end{table}

\begin{figure}
  \centering
  \includegraphics[angle=-90,width=0.5\textwidth]{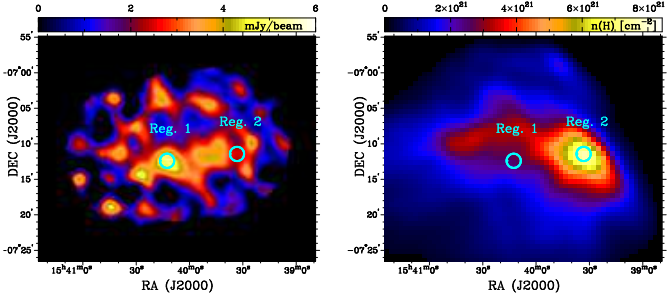}
  \caption{{\it Left:} 31\,GHz CARMA map of LDN\,1780. {\it Right:}
    hydrogen column density map obtained from the IR data in Section
    \ref{sec:dust_properties}. Region 1 corresponds to the peak of the
    CARMA map while region 2 in centred at the peak of the column
    density map on the {\it right}. Both circular regions have 2$'$
    diameter. }
  \label{fig:emm_regs}
\end{figure}

We compare the emissivity at the peak of the 31\,GHz map, with the
emissivity of the region with largest column density of the cloud. In
Fig. \ref{fig:emm_regs} we show these two regions and in Table
\ref{tab:emm_regs} we list average values over a 2$'$ diameter
aperture for the column density, dust temperature, and the relative
intensity of the ISRF obtained from the maps produced in
Sec. \ref{sec:dust_properties}. We also list the flux densities at
31\,GHz in the 2$'$ aperture and the ratio of the flux with the mean
hydrogen column density.

\begin{table}
\centering
\caption[Parameters of the regions shown in
  Fig. \ref{fig:emm_regs}.]{Parameters of the regions shown in
  Fig. \ref{fig:emm_regs}. The column density, dust temperature and
  radiation field intensity $G_0$ are obtained from the map produced
  in Sec. \ref{sec:dust_properties}. The fluxes at 31\,GHz are
  integrated over a 2$'$ diameter aperture. Their location is
  shown in Fig. \ref{fig:emm_regs}.}
\begin{tabular}{lccccc}
\hline
\hline
Region & $N(H)$    & $T_d$ & $G_0$ & $S_{31}$ & $S_{31}/N(H)$ \\
       &[$\times 10^{21}$\,cm$^{-2}$] & [K]   &       &  mJy    &   $\times 10^{-24}$\,[Jy\,cm$^{-2}$]  \\  
\hline
1 & 2.4   &  16.6   &   0.7  & 4.5 & 18.7 \\
2 & 7.3   &  15.0   &   0.4  & 2.4 & 3.2 \\
\hline
 \end{tabular}
\label{tab:emm_regs}
\end{table}

Region 1 shows a 31\,GHz emissivity which is $18.7/3.2=5.8$\,times
larger than that of Region No. 2. We will see if the {\sc SPDUST}
package from \citet{ali:09,silsbee:11} can produce emissivities
different by a factor $\sim 5.8$ within this cloud, with plausible
physical conditions.

In {\sc SPDUST}, there are seven input parameters that are related to
the environmental conditions of the emitting region. These are,
\begin{enumerate}
\item Total hydrogen number density $n_H$.
\item Gas temperature $T$. 
\item Intensity of the radiation field relative to the average interstellar radiation field $G_0$.
\item Hydrogen ionization fraction $x_H \equiv n_{H^+} /n_H$.
\item Ionized carbon fractional abundance $x_C \equiv n_C+ /n_H$.
\item Molecular hydrogen fractional abundance $y \equiv 2n(H_2)/n_H$.
\item ``line'' parameter: Parameters that define the grain size
  distribution. It corresponds to the number line number of Table 1 of
  \citet{weingartner:01}.
\end{enumerate}

Another critical parameter is the shape of the grain size
distribution.  In LDN\,1780, the differences in the IR morphology of
the cloud can be explained by a difference in the type of grains
across the cloud. \citet{weingartner:01} describe parametrizations of
the grain size distribution, where the main dust components, silicates
and carbonaceous grains are described. The grain size distribution of
silicate grains has a power-law shape for the smallest grains, which
are probably the ones responsible for some of the spinning dust
emission.  Carbonaceous grains on the other hand, show a more
complicated distribution, with two ``bumps'' in the small-size
regime. These local peaks are the small PAHs, large molecules with
less than $\sim 10^3$ atoms. A larger abundance of C also increases
the proportion of PAHs. PAHs are also expected to be important
spinning dust emitters.

The number of PAHs in the {\sc SPDUST} model, which is characterised
by the relative size of the bumps in the grains size distribution,
produces large differences in the emissivity at radio frequencies.  We
have modified the {\sc SPDUST} package to allow modifications of the
b$_{\rm c}$ parameter, which defines the relative size of the PAH
bumps in the grain size distribution, for a fixed total carbon
abundance. A similar analysis has recently been done by
\citet{Tibbs2016}.

The {\sc SPDUST} code has been used by many authors to compare the AME
emissivity with radio data. Normally, most parameters are kept fixed
to standard values for different astrophysical environments (e.g. cold
neutral medium, warm ionised medium). Here we would like to constrain
the range of some parameters using additional datasets of LDN\,1780.
Different combinations of the {\sc SPDUST} parameters can produce
similar output spectra. Also, some of these parameters are strongly
correlated. To tackle these complications, we use an exhaustive
approach where we run {\sc SPDUST} over a grid of parameters, for a
total of 10$^{7}$ runs. Table \ref{tab:spdust_params} lists the range
and the spacing for each parameter. 

\begin{table}
\centering
\caption{Range of the parameters used for defining the grid for
  running {\sc SPDUST}. The parameters are: the hydrogen column
  density, gas temperature, intensity of the radiation field,
  hydrogen, carbon and H$_2$ fractional abundances. $b_c$ quantifies
  the proportion of PAH grains in the dust.  }
\begin{tabular}{lccccc}
\hline
\hline
Parameter & min     & max &  Steps & Type  \\
\hline
$n_H$      & 0.1 & $10^{5}$ &     10 & log\\
$T$        & 10 &  $10^{5}$ &     10 & log\\
$\chi$     & $10^{-4}$ & 3000 &   10 & asinh\\
$x_{\rm H}$  & $10^{-4}$ & 1 &      10 & asinh\\
$x_{\rm C}$  & $10^{-4}$ & 1 &     10 & asinh\\
$y$        & $10^{-4}$ & 1 &     10 & asinh\\
$b_{\rm c}$  & 0        & 1 &     10 & linear\\
\hline
\end{tabular}
\label{tab:spdust_params}
\end{table}

From each run of {\sc SPDUST} we recover the peak frequency, the peak
emissivity and also the parameters that define a fourth order
polynomial fit to the {\sc SPDUST} spectrum. The polynomial fit is
calculated around the peak of the spectrum. In order to explore the
results from the \spdust\ runs, we first use observational constraints
from ancillary sources in some of the physical parameters in LDN\,1780.

\citet{mattila:79} observing neutral hydrogen and OH with the 100-m
Effelsberg radio telescope found that the kinetic temperature of
hydrogen was in the range $ T_k=40-56$\,K. They also quote a mean
value for the total density of the gas of
$n=1.8 x 10^3$\,cm$^{-3}$\!\!. \citet{laureijs:95} finds an average total
density of 10$^{3}$\,cm$^{-3}$, while \citet{toth:95} obtains a lower
value of $0.6\times10^{3}$\,cm$^{-3}$. These two works also show that
the cloud is in virial equilibrium and presents an $r^{-2}$ density
profile.

We can assume that Region No. 1, which corresponds to the peak of the
31\,GHz map, has a density equal to the average density of the cloud
of 1000\,cm$^{-3}$\!\!. This is reasonable as the column density at
that point has a near-average value over the cloud (see left panel of
Fig. \ref{fig:emm_regs}). As this point is about half-way to the
border of the cloud, and we know that the density profile of the cloud
presents an $r^{-2}$ dependency, we can estimate the value of the
highest density of the cloud to be $0.5^{-2}= 4$ times larger than the
average. We can also assume that the coldest region of the cloud will
be the one with higher column density (this region also has the lowest
dust temperature, as we shown in Section
\ref{sec:dust_properties}). We therefore assign to Region No. 2, the
lowest gas temperature allowed by the work of \citet{mattila:79},
$T=40$\,K. The largest temperature that \citet{mattila:79} predicts
for the cloud, $T=56$\,K is assigned to Region 1.
For the radiation field, we use the values that we calculated
previously. Region 2, at the peak of the column density, is also
coincident with a peak in the $^{13}$CO map from \citet{toth:95}, this
implies that the ionization fraction in this region is very close to
zero, due to the molecular nature of the gas at this position. For the
ionization fraction, we use the value that \citet{DL98b} defines for
the cold neutral medium (CNM), as listed in Table
\ref{tab:sp_par}. The carbon ionization fractions for the two regions
are also taken from the conditions for MC and CNM in \citet{DL98b}. We
list the parameters for Regions 1 and 2 in Table \ref{tab:sp_par}.  We
note that the absolute value of these parameters is not as important
as the ratio between them, as we want to compare the ratio of the
emissivities produced by {\sc SPDUST}.

\begin{table}
\centering
\caption{{\sc SPDUST} parameters of the regions shown in
  Fig. \ref{fig:emm_regs}. The column density, dust temperature and
  radiation field intensity $G_0$ are obtained from the maps produced
  in Section \ref{sec:dust_properties}. ``line'' corresponds to the
  parameters that define the grain size distribution, defined in Table
  1 of \citet{weingartner:01}, the parameters listed in line 7 are the
  favored by \citet{weingartner:01} for the Milky Way.  The locations
  of Reg 1 and Reg 2 are shown in Fig. \ref{fig:emm_regs}.}
\begin{tabular}{lrrccccc}
\hline
\hline
Region & $n(H)$ & $T$ & $G_0$ & $x_H$ & $x_C$ & $y$ & ``line''\\
      &cm$^{-3}$ & [K] &       &      &  & &  WD2001             \\  
\hline
MC      & 300    & 20   & 0.01 & 0 & 0.0001 & 0.99 & 7 \\
CNM     & 30     & 100  & 1 & 0.0012 &0.0003  & 0  & 7 \\
Reg 1   & 1000   & 56   & 0.7 & 0.0012 & 0.0003 & 0.5 & 7 \\
Reg 2   & 4000   & 40   & 0.4 & 0 & 0.0001 & 0.99 & 7 \\
\hline
 \end{tabular}
\label{tab:sp_par}
\end{table}

\begin{figure}
  \centering
  \includegraphics[angle=0,width=0.48\textwidth]{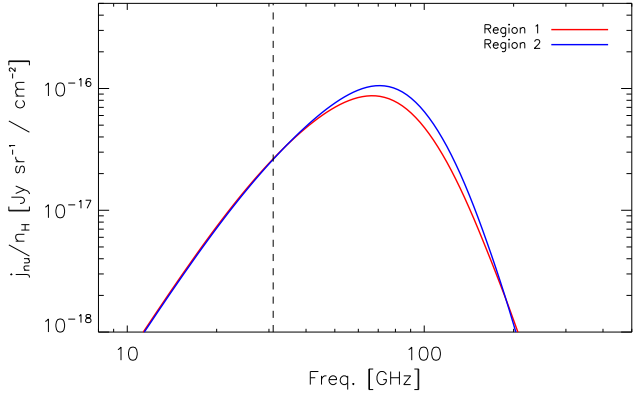}
  \caption{{\sc SPDUST} spectra for regions 1 and 2, using the
    parameters listed in Table \ref{tab:sp_par}.  The vertical dashed
    line in all the plots marks the frequency of our CARMA data,
    31\,GHz.}
  \label{fig:sp_spectra_regs}
\end{figure}

In Fig. \ref{fig:sp_spectra_regs} we show the resulting spectra for
regions 1 and 2, using the parameters listed in Table
\ref{tab:sp_par}. The difference in the parameters produces a
difference in emissivity that is almost zero at 31\,GHz and only 21\%
at the peak of each spectrum, around 75\,GHz. This is not unexpected due
to the small variation in the parameters from region 1 to region 2.
This shows that variation in the environmental conditions are not
sufficient to explain the emissivity differences observed in the
cloud.

\begin{figure}
  \centering
  \includegraphics[angle=0,width=0.5\textwidth]{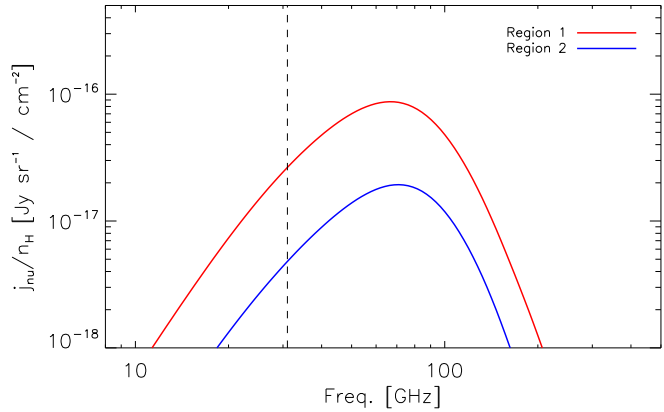}
  \caption{{\sc SPDUST} spectra for regions 1 and 2, using the
    parameters listed in Table \ref{tab:sp_par}, but this time,
    changing the ``line'' parameter, to change the grain size
    distribution by increasing the total carbon abundance. In this
    case, the ratio of the two spectra at 31\,GHz (showed with the
    dashed line) is 5.5, close to the measured 5.8 from the CARMA
    data.}
  \label{fig:sp_spectra_regs_line}
\end{figure}

The number of PAHs in the {\sc SPDUST} model, which is characterised
by the relative size of the bumps in the grain size distribution,
produces large differences in the emissivity at radio frequencies.  If
we allow this value to change across the cloud, we can reproduce the
observed differences in radio emissivity. In
Fig. \ref{fig:sp_spectra_regs_line} we show the {\sc SPDUST} spectra
for Region 1 and 2 using the same parameters of Table
\ref{tab:sp_par}, but this time changing the ``line'' parameter to
represent a variation in carbon abundance by a factor 6. By doing
this, the emissivity ratio at 31\,GHz is 5.5, close to the measured
5.8 from the CARMA data.

A factor of six in the total carbon abundance is difficult to explain
given the proximity of the two regions in the same cloud. An
alternative to such a drastic change in the chemical composition of
the cloud is to modify the relative size of the PAH bumps in the grain
size distribution, for a fixed total carbon abundance. The values
quoted by \citet{weingartner:01}, of 0.75 and 0.25 for the amplitude
of the large and small peaks represent a best-fit value to a number of
different clouds, so they will likely differ from cloud to cloud. 

It is clear that, given the degrees of freedom of {\sc SPDUST}, the
observed differences in emissivity across the cloud are compatible
with spinning dust. Nevertheless, we have shown that if we constrain
the parameter space to be compatible with the physical properties of
the cloud found in the literature, the variations in 31\,GHz
emissivity observed in the CARMA data cannot be explained only by
environmental variations.  We explored the parameter space within the
values listed in Table \ref{tab:sp_par} that are compatible with the
observed emissivity for Regions 1 and 2 listed in Table
\ref{tab:emm_regs}. We find that the {\em only} models compatible with
the observed emissivities present a different grain size distribution
($b_{\rm c}$ parameter). In is important to note that
\citet{Ridderstad2006} reached the same conclusion, in requiring a
significant variation in the grain size distribution along the E-W
axis of LDN\,1780, but based on the study of infrared data and
radiative transfer modelling.

These differences in the grain properties across the cloud can be
expected, as there are differences in the IR morphology of the
cloud. In this scenario, the denser region of the cloud, which shows
low radio emission has a smaller proportion of the smallest PAHs, and
this fraction is $\sim 9$ times larger at the peak of the 31\,GHz
emission. The origin for this difference might be related to the
coagulation of the smallest grains in the denser and colder region of
LDN\,1780.

\section{Conclusions}
\label{sec:conclusions}
Following the detection of AME in the LDN\,1780 translucent cloud
presented in \citet{Vidal2011} using data from the Cosmic Background
Imager (CBI), we performed follow-up observations at 31\,GHz using the
CARMA 3.5-m array. These new data have an angular resolution of $2'$,
3 times better than that of the CBI. We measured the correlation
between the 31\,GHz data and different IR templates. We found that the
best correlation occurs with MIPS $70\,{\mu}$m, with a Spearman's rank
$r_S=0.49\pm0.07$. This confirms what we found using the CBI data in
\citet{Vidal2011}, where the radio data correlated better with a
60\,$\mu$m IR map. The correlation between the CARMA data and
\spitzer\ $8\micron$ and \spitzer\ 24\,$\mu$m, which traces PAHs and
VSGs, is poor. Here, $r_S=0.14\pm0.06$ and $r_S=0.21\pm0.06$ for
8\,$\mu$m and 24\,$\mu$m respectively. These two correlation values
increase significantly when correcting the IR maps by the ISRF,
yielding $r_S=0.38\pm0.07$ and $r_S=0.46\pm0.06$ for the corrected
8\,$\mu$m and 24\,$\mu$m templates respectively. This is important as
these ISRF corrected templates should be better tracers of the PAHs
and VSGs column density, as opposed to the uncorrected maps, in which
the emission is proportional also to the incoming radiation field.

We constructed an SED of LDN\,1780 on 1\dg scales between 0.408\,GHz
and 2997\,GHz, including \planck\ data. The region of the cloud is
dominated by the CMB anisotropy between 23\,GHz and 217\,GHz. After
subtraction of the CMB emission, the presence of AME is very clear and
it is well fitted using a spinning dust model. AME is detected with a
significance $> 20\sigma$. On these angular scales, there is a
significant shift of the peak of the cloud between the emission at low
frequencies (23--70\,GHz) versus the emission at higher frequencies
(93--2997\,GHz). This means that the AME in this cloud does not
originate at the same location that the thermal dust emission.

In the 31\,GHz CARMA maps, there are differences in the emissivity
along the cloud. These differences are compatible with the spinning
dust model. The spinning dust emission depends on the physical
parameters of the dust grain and also on environmental conditions of
the cloud, such as density, temperature and ISRF. Some of these
parameters for LDN\,1780 are known from the literature, so we fixed
them and concluded that there must be variations in the grain size
distribution along the cloud, with and E--W enhancement of the PAHs
population. Only by doing this the model can reproduce the observed
factor $\sim 6$ difference in the AME emissivity.

Given the large number of free parameters that the spinning dust
models have, it is not difficult to account for AME variations in
different environments. This is something to keep in mind when trying
to interpret the observations. It is clear that a greater number of
observations are required in order to fully constrain the spinning
dust models. Multi-wavelength observations are needed in order to
constrain parameters independently. In particular, separating
different grain populations. High angular resolution observations of
AME sources using current and future instruments (VLA, ALMA, ngVLA,
SKA) will help greatly in this respect.

\section*{Acknowledgments}

We thank Anthony Banday for very useful comments on this work. We
thank Justin Jonas for allowing us the use of the 2.3\,GHz map. MV
acknowledges support from FONDECYT through grant 3160750. CD
acknowledges funding from an STFC Advanced Fellowship, STFC
Consolidated Grant (ST/L000768/1), and an ERC Starting (Consolidator)
Grant (no.~307209). We acknowledge the use of the Legacy Archive for
Microwave Background Data Analysis (LAMBDA). Support for LAMBDA is
provided by the NASA Office of Space Science. Some of the work of this
paper was done using routines from the IDL Astronomy User's
Library\footnote{http://idlastro.gsfc.nasa.gov/}. Some of the results
in this paper have been derived using the HEALPix \citep{gorsky:05}
package.  Support for CARMA construction was derived from the Gordon
and Betty Moore Foundation, the Kenneth T. and Eileen L. Norris
Foundation, the James S. McDonnell Foundation, the Associates of the
California Institute of Technology, the University of Chicago, the
states of California, Illinois, and Maryland, and the National Science
Foundation.  CARMA development and operations were supported by the
National Science Foundation under a cooperative agreement, and by the
CARMA partner universities.

\label{lastpage}

\bibliographystyle{mn2e}
\bibliography{refs} 

\bsp

\end{document}